\documentclass[twocolumn,           
               showpacs,            
               preprintnumbers,     
               aps,                 
               prd,                 
               a4paper,             
                   superscriptaddress,
               nofootinbib,         
               tightenlines,        
               amsmath,amssymb
               floats,floatfix      
               ]{revtex4}

\usepackage{graphicx,epsfig}
\usepackage{subfigure}
\usepackage{latexsym}
\usepackage{amsmath,amssymb}        
\usepackage[draft=false]{hyperref}
\usepackage{enumerate}
\usepackage[usenames]{xcolor}

\usepackage{multirow}

\newcommand{\be}{\begin{equation}}
\newcommand{\ee}{\end{equation}}

\newcommand{\kmax}{k_\mathrm{max}}
\newcommand{\kmin}{k_\mathrm{min}}
\newcommand{\Pgal}{P_\mathrm{gal}}
\newcommand{\PGiggleZ}{P_\mathrm{GiggleZ}}
\newcommand{\Pdamp}{P_\mathrm{damped}}
\newcommand{\Phf}{P_\mathrm{hf}}
\newcommand{\Plin}{P_\mathrm{lin}}
\newcommand{\Pnw}{P_\mathrm{nw}}
\newcommand{\Phfnw}{P_\mathrm{hf,nw}}

\newcommand{\Pm}{P_m}
\newcommand{\Pcon}{P_\mathrm{con}}
\newcommand{\Pobs}{P_\mathrm{obs}}
\newcommand{\trial}{\mathrm{trial}}
\newcommand{\fid}{\mathrm{fid}}
\newcommand{\zeff}{z_\mathrm{eff}}
\newcommand{\ascl}{a_\mathrm{scl}}

\newcommand{\ombh}{\Omega_\mathrm{b} h^2}
\newcommand{\omch}{\Omega_\mathrm{CDM} h^2}

\newcommand{\Mpc}{\textrm{Mpc}}
\newcommand{\Gpc}{\textrm{Gpc}}

\newcommand{\s}{\mathrm{s}}
\newcommand{\km}{\mathrm{km}}

\begin{document}



\title{The WiggleZ Dark Energy Survey: \\ Final data release and cosmological results}
\author{David Parkinson}\email{d.parkinson@uq.edu.au} 
\affiliation{School of Mathematics and Physics, University of Queensland, Brisbane, QLD 4072, Australia} 
\author{Signe Riemer-S\o rensen}
\affiliation{School of Mathematics and Physics, University of Queensland, Brisbane, QLD 4072, Australia}
\author{Chris Blake}
\affiliation{Centre for Astrophysics \& Supercomputing, Swinburne University of Technology, P.O. Box 218, Hawthorn, VIC 3122, Australia}  
\author{Gregory B.  Poole}
\affiliation{Centre for Astrophysics \& Supercomputing, Swinburne University of Technology, P.O. Box 218, Hawthorn, VIC 3122, Australia}  
\author{Tamara M. Davis}
\affiliation{School of Mathematics and Physics, University of Queensland, Brisbane, QLD 4072, Australia}
\author{Sarah Brough}
\affiliation{Australian Astronomical Observatory, P.O. Box 296, Epping, NSW 1710, Australia}
\author{Matthew Colless}
\affiliation{Australian Astronomical Observatory, P.O. Box 296, Epping, NSW 1710, Australia}
\author{Carlos Contreras}
\affiliation{Centre for Astrophysics \& Supercomputing, Swinburne University of Technology, P.O. Box 218, Hawthorn, VIC 3122, Australia}  
\author{Warrick Couch}
\affiliation{Centre for Astrophysics \& Supercomputing, Swinburne University of Technology, P.O. Box 218, Hawthorn, VIC 3122, Australia}  
\author{Scott Croom}
\affiliation{Sydney Institute for Astronomy, School of Physics, University of Sydney, Sydney, NSW 2006, Australia}
\author{Darren Croton}
\affiliation{Centre for Astrophysics \& Supercomputing, Swinburne University of Technology, P.O. Box 218, Hawthorn, VIC 3122, Australia}  
\author{Michael J.\ Drinkwater}
\affiliation{School of Mathematics and Physics, University of Queensland, Brisbane, QLD 4072, Australia}
\author{Karl Forster}
\affiliation{California Institute of Technology, MC 278-17, 1200 East California Boulevard, Pasadena, CA 91125, United States }
\author{David Gilbank}
\affiliation{South African Astronomical Observatory, P.O. Box 9, Observatory, 7935, South Africa}
\author{Mike Gladders}
\affiliation{Department of Astronomy and Astrophysics, University of Chicago, 5640 South Ellis Avenue, Chicago, IL 60637, United States}
\author{Karl Glazebrook}
\affiliation{Centre for Astrophysics \& Supercomputing, Swinburne University of Technology, P.O. Box 218, Hawthorn, VIC 3122, Australia}  
\author{Ben Jelliffe}
\affiliation{Sydney Institute for Astronomy, School of Physics, University of Sydney, NSW 2006, Australia}
\author{Russell J.\ Jurek}
\affiliation{CSIRO Astronomy \& Space Sciences, Australia Telescope National Facility, Epping, NSW 1710, Australia}
\author{I-hui Li}
\affiliation{Centre for Astrophysics \& Supercomputing, Swinburne University of Technology, P.O. Box 218, Hawthorn, VIC 3122, Australia}  
\author{ Barry Madore}
\affiliation{Observatories of the Carnegie Institute of Washington, 813 Santa Barbara St., Pasadena, CA 91101, United States}
\author{D.\ Christopher Martin}
\affiliation{California Institute of Technology, MC 278-17, 1200 East California Boulevard, Pasadena, CA 91125, United States }
\author{Kevin Pimbblet}
\affiliation{School of Physics, Monash University, Clayton, VIC 3800, Australia}
\author{Michael Pracy}
\affiliation{Sydney Institute for Astronomy, School of Physics, University of Sydney, NSW 2006, Australia}
\author{Rob Sharp}
 \affiliation{Australian Astronomical Observatory, P.O. Box 296, Epping, NSW 1710, Australia}
\affiliation{Research School of Astronomy \& Astrophysics, Australian National University, Weston Creek, ACT 2611, Australia}
\author{Emily Wisnioski}
\affiliation{Centre for Astrophysics \& Supercomputing, Swinburne University of Technology, P.O. Box 218, Hawthorn, VIC 3122, Australia}  
\author{David Woods}
\affiliation{Department of Physics \& Astronomy, University of British Columbia, 6224 Agricultural Road, Vancouver, BC V6T 1Z1, Canada}
\author{Ted K.\ Wyder}
\affiliation{California Institute of Technology, MC 278-17, 1200 East California Boulevard, Pasadena, CA 91125, United States }
\author{H.K.C. Yee}
\affiliation{Department of Astronomy and Astrophysics, University of Toronto, 50 St.\ George Street, Toronto, ON M5S 3H4, Canada}
\date{\today} 


 

\begin{abstract} 
This paper presents cosmological results from the final data release of the WiggleZ Dark Energy Survey.  We perform full analyses of different cosmological models using the WiggleZ power spectra measured at z=0.22, 0.41, 0.60, and 0.78, combined with other cosmological datasets. The limiting factor in this analysis is the theoretical modelling of the galaxy power spectrum, including non-linearities, galaxy bias, and redshift-space distortions.  In this paper we assess several different methods for modelling the theoretical power spectrum, testing them against the Gigaparsec WiggleZ simulations (GiggleZ). We fit for a base set of 6 cosmological parameters, \{$\ombh$, $\Omega_{\rm CDM}h^2$, $H_0$, $\tau$, $A_s$, $n_{\rm s}$\}, and $5$ supplementary parameters \{$n_{\rm run}$, $r$, $w$, $\Omega_k$, $\sum m_{\nu}$\}.  In combination with the Cosmic Microwave Background (CMB), our results are consistent with the $\Lambda$CDM concordance cosmology, with a measurement of the matter density of $\Omega_m =0.29 \pm0.016$ and amplitude of fluctuations $\sigma_8=0.825\pm0.017$. Using WiggleZ data with CMB and other distance and matter power spectra data,  we find no evidence for any of the extension parameters being inconsistent with their $\Lambda$CDM model values. The power spectra data and theoretical modelling tools are available for use as a module for CosmoMC, which we here make publicly available at \url{http://smp.uq.edu.au/wigglez-data}. We also release the data and random catalogues used to construct the baryon acoustic oscillation correlation function.
\end{abstract}

\maketitle

 
\section{Introduction} 
The aim of the WiggleZ Dark Energy Survey has been to measure the large scale structure of the Universe and use that to learn about the properties of dark energy.  
Here we present our final data release and cosmology results, using the full large scale structure power spectrum.  

The Universe is filled with structure, forming under gravitational collapse. We observe this structure through light, emitted when the gas is compressed and heated. Cosmology is the study of this light, from stars in distant galaxies, to cosmic microwaves emitted by the Big Bang -- and inferences are made as to the structure that it traces. We have built up a complex picture of structure formation, where it is seeded as tiny density fluctuations and grows under gravitational collapse to form the `cosmic web' of galaxies and clusters that we observe in the late-time Universe. The cosmological model of acausal-seeded, hierarchical structure formation in a geometrically flat, late-time accelerating universe has been well tested using precision cosmological datasets such as measurements of the Cosmic Microwave Background (CMB) by the COsmic Background Explorer (COBE) and Wilkinson Microwave Anisotropy Probe (WMAP), the expansion history of the Universe using standard candles and rulers such as Type-Ia Supernovae (SNIa) and Baryon Acoustic Oscillations (BAO), and the late Universe matter power spectra ($P(k)$) and growth of structure. The name of this `concordance' cosmological model is $\Lambda$CDM.

The definition of a cosmological model is one that describes the whole Universe, and possesses a number of tuneable parameters. In the past, cosmological models were incommensurable, in that it was almost impossible to judge one in the light of another (such as the Big Bang model compared to the Steady State model). The advent of precision cosmological data has allowed $\Lambda$CDM to emerge as the predominant model, but it must always be checked to be consistent with new data as it becomes available. New alternatives/extensions must also be tested, to see if they make better predictions than the standard model, which would indicate that it should be updated.

The concordance cosmological model is described by six free parameters: the Hubble parameter today, $H_0$ (where the dimensionless $h=H_0/100 {\rm km s}^{-1}{\rm Mpc}^{-1}$), the physical baryon density, $\ombh$, the physical Cold Dark Matter density, $\omch$, the optical depth to reionisation, $\tau$, the amplitude of the primordial scalar density perturbations on scales of $k=0.05 h^{-1}\Mpc$, $A_s$, and the spectral index of the primordial power spectrum of density perturbations, $n_s$. Extensions to this standard model involve adding one or more extra parameters such as curvature $\Omega_k$, the equation of state of the dark energy, $w$, the mass of the neutrinos, $\sum m_\nu$, the running of the spectral index of the primordial power spectrum of density perturbations, $n_{\rm run}$ ($= d\ln n_s/d \ln k$),  and the amplitude of the primordial tensor perturbations (generated by primordial gravitational waves from the end of inflation), $r$. All of these represent some extra physical effects that have been suggested theoretically and would contribute to the expansion and/or structure formation of the Universe, but may not be detectable using current cosmological data. 

The advantage of a complete model of structure formation is that it allows us to combine information from the early and late Universe to test cosmological models. This has been done by the CfA Redshift Survey \citep{Huchra:1983}, the APM Galaxy Survey \citep{Maddox:1990}, VIMOS-VLT Deep Survey (VVDS) \citep{lefevre:2005}, the Two-Degree Field Galaxy Redshift Survey (2dFGRS) \citep{colless:2001} and the Sloan Digital Sky Survey (SDSS) \citep{York:2000}.
 
In this paper we present analyses of the cosmological models using  data from the WiggleZ Dark Energy Survey \citep{Drinkwater:2010}. WiggleZ is a large-scale galaxy redshift survey conducted with the AAOmega multi-object spectrograph on the Anglo-Australian Telescope at Siding Springs Observatory. WiggleZ was designed to study the effect of dark energy on the  expansion history of the Universe and on the growth of cosmological structures across an unprecedented period of cosmic history. We have already presented our measurements of several important features such as the baryon acoustic oscillation scale \citep{blakedavis11,blakekazin11}, the growth of structure \citep{Blake:2011growth}, the Alcock-Paczynski effect, \citep{Blake:2011ap}, and the mass of the neutrino \citep{riemersorensen:2011}.


An important step in these types of analysis is generating accurate predictions from theory for what the large-scale structure in the Universe should look like.  In practice that usually means predicting the shape of the galaxy power spectrum or correlation function.  Using only linear theory, we can easily make predictions using perturbation theory. However, the relationship between the observed galaxy power spectrum and the linear perturbation theory prediction for the matter power spectrum is complicated by three physical effects: non-linear structure formation, galaxy bias, and redshift space-distortions.

{\it Non-linear structure formation:}   When density perturbations become large enough they are no longer well described by first order perturbation theory.  For testing cosmological models using the CMB power spectrum, the linear theory prediction is accurate enough when compared to the data (at least at the moment), as the amplitude of the density fluctuations at $z \sim 1100$ are very small. For the matter power spectrum at low-redshift this is not the case, as some non-linear evolution of the density field has taken place.  The amplitude of this non-linear evolution can be estimated by numerical simulations \citep{Jennings:2010,Bird:2011,Marulli:2011}, and then applied as a correction to the linear prediction (e.g., Halofit \citep{Smith:2003}).

{\it Galaxy bias:} This is not the end of the story, since we observe the distribution of galaxies, but our theory predicts the distribution of matter, which includes non-luminous, non-baryonic dark matter.   The galaxies are slightly decoupled from the matter, through the complex baryonic physics of star and galaxy formation. The simplest coupling to be assumed is the idea of the linear `bias' \citep{Kaiser:1984}, an overall, shape-independent amplitude scaling from the matter power spectrum to the galaxy power spectrum. This bias parameter is different for different populations of galaxies, and relates directly to the galaxy-formation history of the population being sampled. Because of this, we also expect the bias to evolve with redshift and environment (e.g., density where the galaxies are found), which would introduce scale-dependence to the bias. Using numerical simulations of galaxies, it should be possible to predict the bias for a given population of galaxies, and the technology has advanced to such a state that this is becoming possible. 

{\it Redshift space distortions:}
Another complication arises because all our observed distances are inferred from measurements of redshifts (and angular separations). 
However, galaxy redshifts do not solely come from the expansion of the Universe, but also local velocities induced by gravitational over-densities. These peculiar velocities introduce redshift-space distortions into the clustering power spectrum (as measured using the WiggleZ data in \citep{Blake:2011growth,Blake:2011ap}), and will have a measurable effect on scales small enough for peculiar velocities to be significant.  The effect of redshift space distortions can be detected by observing differences in the clustering pattern measured perpendicular to and parallel to the line-of-sight. 

A key part of the analysis in this paper is a detailed exploration of the different approaches we have used to successfully model the power spectrum in terms of the cosmological parameters we are constraining.  
We have tested seven different approaches for our analysis. The models are fairly similar at low values of the comoving wavenumber $k$ (large scales), where the large-scale clustering can be treated as linear. There the theory is quite robust, and there is little difference between the predictions. However the difference between the models starts to increase for $k>0.2\,h\,\Mpc^{-1}$, which significantly affects the outcome of the fitting.  As we discuss in section \ref{sec:systematics}, we are fitting up to $k_{\rm max} = 0.3\, h{\rm Mpc}^{-1}$, as we found this to be the optimal compromise between robust modelling and data inclusion.

With this paper, we also make public our data, as well as a code to implement the preferred power spectrum modelling method. The power spectrum data (measurements, covariance matrices and window functions) we make available along with the modelling code as a module for the CosmoMC cosmological analysis package \citep{Lewis:2002}. We also make available the data that was used to construct the baryon acoustic peak correlation function measurements reported in \citet{blakekazin11}.

The structure of the paper is as follows. In Section \ref{sec:Data} we describe the WiggleZ Dark Energy Survey, the data that is available, the selection functions, the $N$-body simulations (GiggleZ), and the random catalogues --- in other words, all the raw material you need in order to perform the analysis in this and our other papers.   We then describe in Section \ref{sec:generating_pk} how we extracted the observed power spectrum from this data.  In Section \ref{sec:modelling} we describe how we model the galaxy power spectrum theoretically, starting with the linear matter power spectrum and making additions to account for extra physical effects. In particular in this section we discuss the different formalisms for dealing with non-linear structure formation, galaxy bias, and redshift space distortions, and select a particular model to perform our analyses.  In Section \ref{sec:cosmology} we conduct statistical analyses of a number of different cosmological models, combining the WiggleZ data with independent observations of the CMB, BAO, LSS, and SNIa. 
Finally, we summarise our results in Section \ref{sec:conclusions}.

\section{The WiggleZ Dark Energy Survey}
\label{sec:Data}

\subsection{Data}

The WiggleZ Dark Energy Survey was designed to detect the BAO scale at higher redshifts than was possible with previous datasets. Galaxies were initially selected by a joint selection from optical galaxy surveys (from the Sloan Digital Sky Survey \citep{York:2000} in the Northern Galactic Cap, and from the Red Sequence Cluster Survey 2 (RCS2) \citep{Gilbank:2011} in the Southern Galactic Cap) and from ultraviolet imaging by the Galaxy Evolution Explorer satellite (GALEX). A number of magnitude and colour cuts were made to select high-redshift star-forming galaxies with prominent emission lines, and then spectra of these galaxies were taken by the AAOmega spectrograph \citep{Sharp:2006} in 1-hour exposures at the Anglo-Australian Telescope. In total $238,000$ galaxy redshifts were measured by the survey.

\begin{figure}
	\includegraphics[width=0.99\columnwidth]{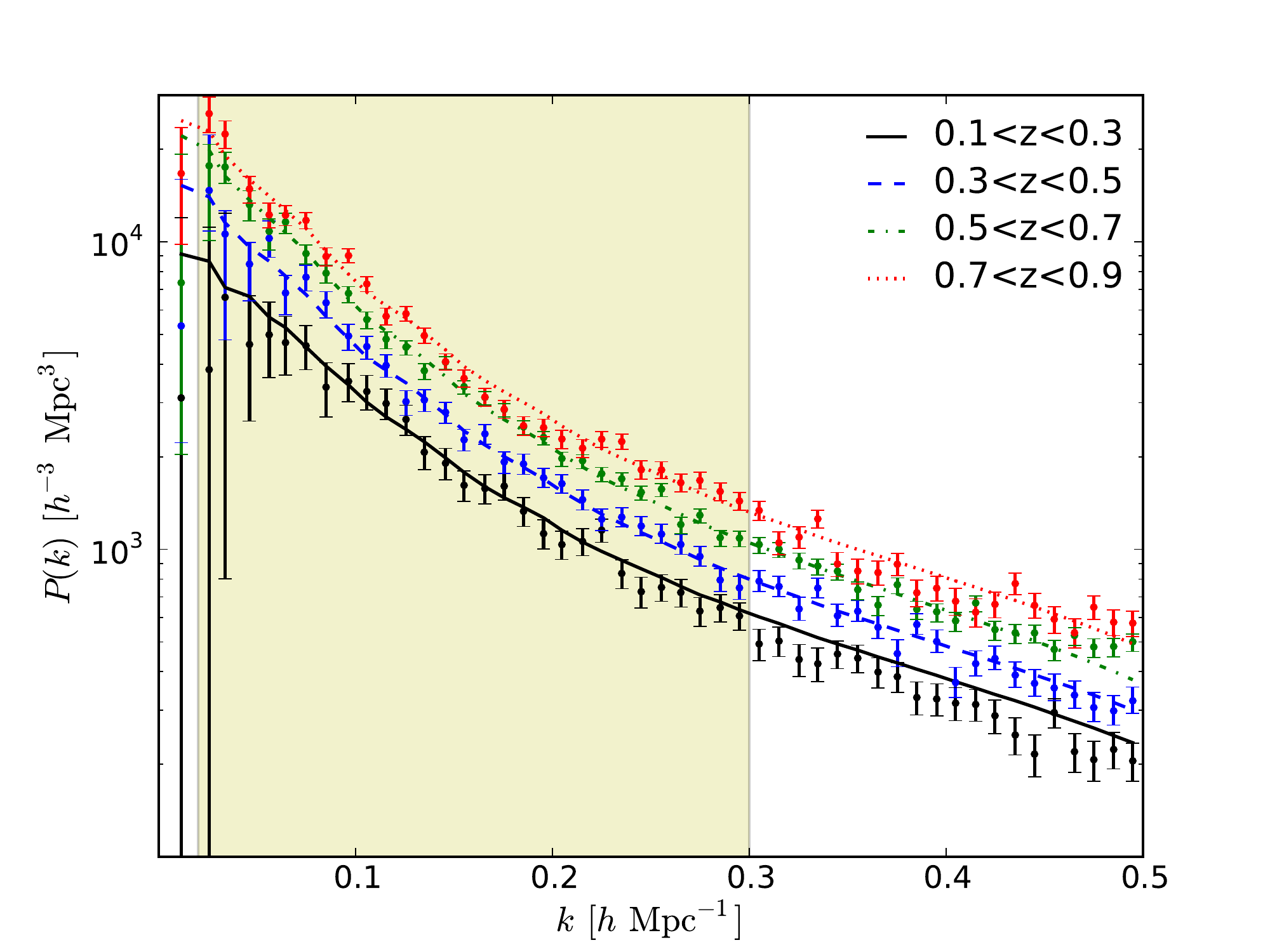}
	\caption{(color online). The WiggleZ power spectrum in the four redshift bins, weighted averaged over the seven regions. We artificially offset the redshift bins in the vertical direction to distinguish them (with the lowest redshift bin at the bottom moving up to highest at the top). At each redshift the line gives the best fit flat-$\Lambda$CDM cosmology prediction, convolved with the window function for each region and then averaged over the regions. The shaded region gives the range in wavenumber of data we used for the analysis. 
	}
	\label{fig:wigglez_pk_bins}
\end{figure}

These galaxies were used to map seven regions of the sky with a total volume of $1\,\Gpc^3$ in the redshift range $z<1$ \citep{Drinkwater:2010}. We split the data into four redshift bins  with ranges $0.1 < z < 0.3$, $0.3< z < 0.5$, $0.5 < z < 0.7$ and $0.7 < z < 0.9$. The number of galaxies included in each redshift slice is listed in Table \ref{tabgals2}. We determined the effective redshift of the power spectrum estimate for each slice by weighting each pixel in the selection function by its contribution to the power spectrum error, as given by Eq. (13) in \citep{blakedavis11}. The effective redshifts of these bins is evaluated to be $\zeff=[0.22, 0.41, 0.6, 0.78]$. The power spectra, $\Pobs$, and covariance matrices, $C$, are measured in $\Delta k = 0.01\,h$\,Mpc$^{-1}$ bins using the optimal-weighting scheme proposed by \citet{Feldman:1994} for a fiducial cosmological model with matter density $\Omega_m = 0.27$ \citep{Blake:2010a}. These power spectra and covariance matrices, along with the window functions for the power spectrum, we make available as part of our cosmology analysis package.\footnote{The package contains the data combined with a module for CosmoMC, and is available at {\tt http://smp.uq.edu.au/wigglez-data}.}

\begin{figure*}
	\includegraphics[width=1.79\columnwidth]{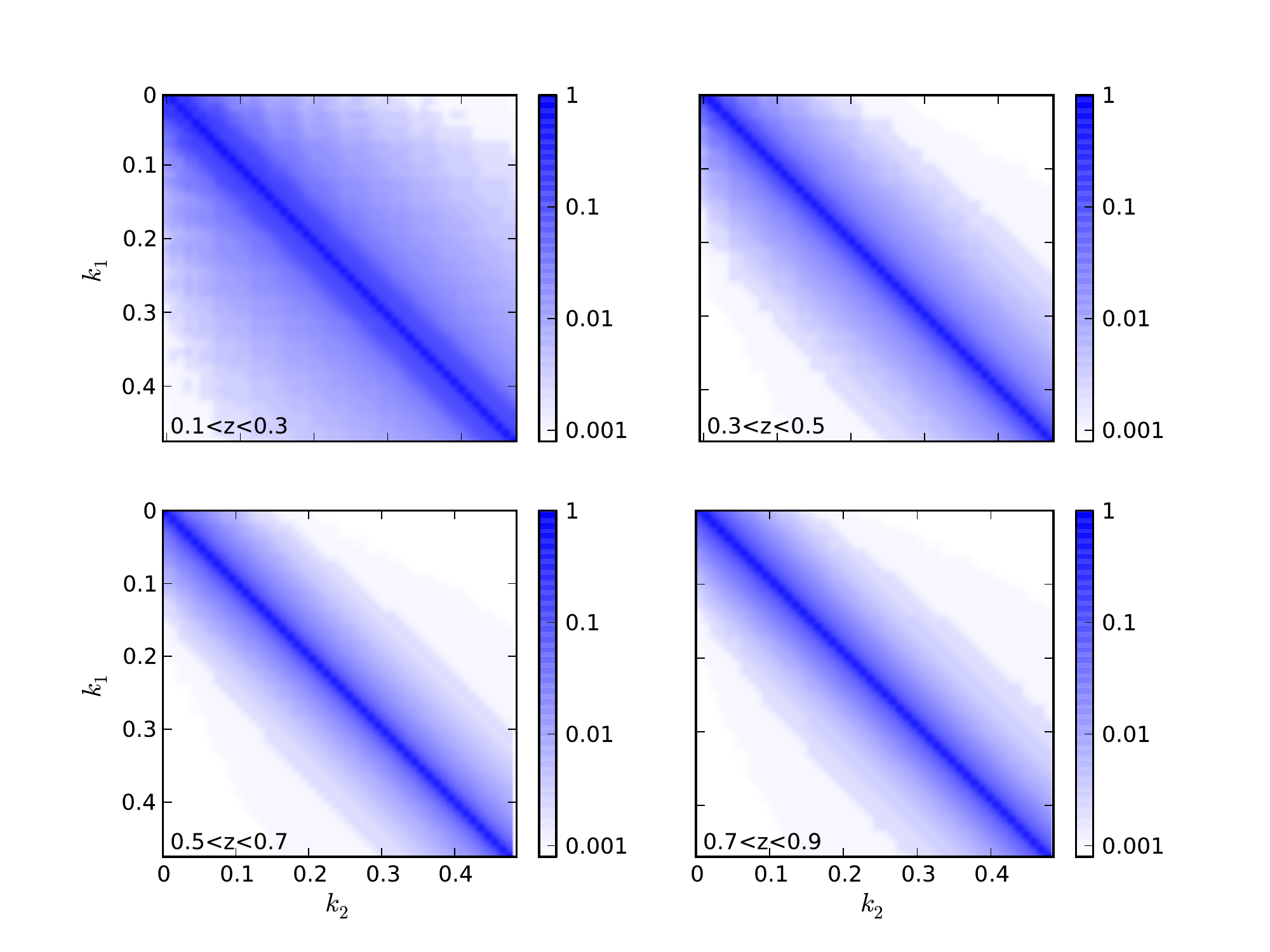}
	\caption{(color online). The WiggleZ covariance matrix in the four redshift bins, as a weighted average over the seven regions.}
	\label{fig:wigglez_cov_matrix}
\end{figure*}

The region-averaged power spectrum for each of the redshift bins, and the best fit flat $\Lambda$CDM cosmology curve with these window function effects taken into account, are shown in Fig. \ref{fig:wigglez_pk_bins} and the covariance matrices in Fig. \ref{fig:wigglez_cov_matrix}.

\begin{table}
\begin{center}
\caption{Number of galaxies used in the power spectrum estimation,
divided by survey region and redshift slice.}
\label{tabgals2}
\begin{tabular}{ccccc}
\hline
Region & $0.1 < z < 0.3$ & $0.3 < z < 0.5$ & $0.5 < z < 0.7$ & $0.7 < z < 0.9$ \\
\hline
9-hr & 2671 & 7438 & 11294 & 5189 \\
11-hr & 3446 & 8118 & 11766 & 5910 \\
15-hr & 3578 & 10942 & 17928 & 8943 \\
22-hr & 5782 & 7728 & 8924 & 5337 \\
0-hr & 2814 & 2986 & 5667 & 3689 \\
1-hr & 2654 & 2596 & 5016 & 3144 \\
3-hr & 2973 & 3522 & 5982 & 4315 \\
\hline
\end{tabular}
\end{center}
\end{table}

The methodology used to construct the WiggleZ survey selection function and random catalogues is described in \citet{Blake:2010a}.  Monte Carlo realizations of catalogues are produced following the angular and redshift dependences of the selection function for each region. This process models several effects including the variation of the GALEX target density with dust and exposure time, the incompleteness in the current redshift catalogue, the variation of that incompleteness imprinted across each 2-degree field by constraints on the positioning of fibres and throughput variations with fibre position, and the dependence of the galaxy redshift distribution on optical magnitude.

\subsubsection{BAO Data}

In conjunction with this paper, we make available the data that we used to construct the baryon acoustic peak correlation function measurements reported in \citet{blakekazin11}. This dataset consists of measurements in three overlapping redshift slices $0.2 < z < 0.6$, $0.4 < z < 0.8$ and $0.6 < z < 1.0$ for the 9-hr, 11-hr, 15-hr, 22-hr, 1-hr and 3-hr survey regions. For each region and redshift slice we make available the data catalogue and 10 random catalogues.  We also provide the resulting correlation function and covariance matrix for each measurement (derived from lognormal realizations). The number of galaxies included in each data subset is listed in Table \ref{tabgals}. (Note that in the BAO measurement we do not use the 0-hr region, due to incompleteness concerns). This sample is constructed from the full WiggleZ redshift catalogue in each of these regions by applying cuts for redshift and contiguity. 

Note that this 'Large-Scale Structure catalogue' is a sub-set of the full redshift catalogue, but no matching random catalogues exist for the full redshift catalogue, only for this LSS sample. This only covers the sample used for the BAO measurement, and cannot be used for computing the full galaxy power spectrum and its errors.

\begin{table}
\begin{center}
\caption{Number of galaxies divided by survey region and redshift slice
  used in the baryon acoustic peak correlation function analysis reported
  by \citet{blakekazin11}.}
\label{tabgals}
\begin{tabular}{cccc}
\hline
Region & $0.2 < z < 0.6$ & $0.4 < z < 0.8$ & $0.6 < z < 1.0$ \\
\hline
9-hr & 15128 & 18978 & 11424 \\
11-hr & 19202 & 23940 & 15064 \\
15-hr & 22309 & 30015 & 19471 \\
22-hr & 15884 & 16146 & 11024 \\
1-hr & 6927 & 9437 & 7880 \\
3-hr & 8000 & 10241 & 8756 \\
\hline
\end{tabular}
\end{center}
\end{table}

\subsection{The GiggleZ simulation}
The Gigaparsec WiggleZ Survey (GiggleZ) simulations \citep{Poole:2012} were designed to probe the low-mass haloes traced by WiggleZ galaxies, whilst providing an equivalent survey volume allowing the measurement of power spectrum modes with $k = 0.01\,\text{--}\,0.5\,h\, \Mpc^{-1}$. They provide a powerful means for testing and calibrating our modeling algorithms. The main simulation we use is a $2160^3$ particle dark matter simulation run in a periodic box of side $1 h^{-1}\, \Gpc$. The resulting particle mass of this simulation is $7.5 \times 10^9 h^{-1} \, M_\odot$ which permits us to resolve bound systems with masses $\sim 1.5 \times 10^{11} h^{-1}\, M_\odot$, facilitating studies of halos with clustering biases near unity such as WiggleZ galaxies. The initial conditions of the simulation were constructed using a CAMB\footnote{\url{camb.info}} power spectrum based on a flat $\Lambda$CDM cosmology using the best fit WMAP 5-year values with $\{\Omega_m, \Omega_{\rm b}, h, \sigma_8, n_s\} = \{0.273, 0.0456, 0.705, 0.812, 0.960\}$.

Bound structures in the dark matter catalogue were identified using Subfind \citep{Springel:2001}, which uses a friends-of-friends (FoF) scheme followed by a substructure analysis to identify bound overdensities within each FoF halo.  We use the Subfind substructures for our analysis, taking the value of each halo's maximum circular velocity as a proxy for mass. In order to generate mock WiggleZ catalogues at a particular redshift we selected the range of halo groupings with large-scale clustering bias closest to the WiggleZ sample under analysis, and applied the WiggleZ selection function to the mock dataset.

By comparing simulated and observed power spectra, we found that over the range of scales and halo masses relevant for this analysis, the galaxy bias is scale-independent to within 1\% \citep{Poole:2012}. 


\section{Taking fiducial model and window function into account}\label{sec:generating_pk}
The method we use for generating a power spectrum from our measured positions of galaxies has been detailed in \citet{Blake:2010a}, and here we expand it to the entire WiggleZ sample.   Crucially, the observed power spectrum is naturally convolved with the survey window function.  (The window function quantifies the holes and incompleteness in the data that are due to the observing strategy rather than any real large scale structure.)   In addition, when generating this observed power spectrum we have used a fiducial cosmological model to convert observed angles and redshifts into distances.  Before the observed galaxy power spectrum can be compared to theoretical models, we must take those factors into account.  

The fiducial model and window function are accounted for by convolving the theoretical power spectrum with the survey window function, and scaling all distances with respect to the fiducial model
\begin{equation}
\Pcon^i(k) = \sum_j \frac{W_{ij}(k) \Pgal^j(k/\ascl)}{a^3_{scl}} \, .
\end{equation}
Details of the window function, $W_{ij}(k)$, can be found in section \ref{sec:Data} and \citet{Blake:2010a}. The scaling, $\ascl$, takes into account that the observed galaxy redshift-space positions are converted to real space positions using a fiducial cosmology. It is calculated as \citep{Tegmark:2006,Reid:2009,Swanson:2010}
\begin{equation}
\ascl^3 = \frac{[D_A(\zeff)]^2H_{\mathrm{fid}}(\zeff)}{[D_{A,\mathrm{fid}}(\zeff)]^2H(\zeff)}, \,
\end{equation}
where $D_A$ is the angular diameter distance and $H$ is the Hubble parameter evaluated for the effective redshift, $\zeff$, of the galaxy sample. A subscript `fid' means this is the value in the fiducial model, while no subscript means this is the value in the model being tested.  For a $\Lambda$CDM Universe $D_A(z)$ is given by
\begin{equation}
D_A(z) = \frac{c}{H_0(1+z)}\frac{1}{|\Omega_k|^{1/2}}S\left\{|\Omega_k|^{1/2}\int_0^z\frac{dz'}{H(z')/H_0}\right\} \, ,
\end{equation}
where $S$ is $\sin(x)$, $x$ or $\sinh(x)$ depending on whether the Universe is closed, flat or open respectively, and $H(z)$ given by the Friedman equation,
\begin{equation} \label{eqn:H}
\frac{H^2(z)}{H_0^2} = \Omega_{r,0}(1+z)^4+\Omega_{m,0}(1+z)^3 + \Omega_{k,0}(1+z)^2+\Omega_{\Lambda,0}   \, ,
\end{equation}
which can be integrated numerically.

\section{Modelling the Galaxy Power Spectrum}\label{sec:modelling}

As mentioned in the introduction, the limiting factor in extracting cosmological constraints from a galaxy power spectrum lies in accurately modelling the theoretical power spectrum in the cosmological model being tested.  A major part of the analysis of this paper has been testing several different approaches to modelling the power spectrum.  In testing this modelling we relied heavily on comparison to $N$-body simulations (GiggleZ) to validate our treatment of non-linear structure formation, galaxy bias, and redshift space distortions.  

In terms of these modelling challenges the WiggleZ Dark Energy Survey has several potential advantages over previous surveys: 

{\it Higher redshift:} The effect of non-linear structure formation increases with time as density perturbations grow. For the distant galaxies probed by WiggleZ, the contamination from non-linearities is smaller than for previous surveys. 

{\it Lower scale-dependence in bias:} The relationship (bias) between the observed galaxy distribution and the dark matter distribution depends on the observed galaxy type. Previous studies \cite[e.g.][]{Percival:2009,Reid:2010} measured red galaxies, which tend to cluster in the center of dark matter halos, whereas the star-forming blue WiggleZ galaxies avoid the densest regions. This leads to a lower overall bias, which makes WiggleZ less susceptible to any possible systematics that could arise from a scale-dependence of the bias.  

{\it Low non-linear pairwise velocities:}  The WiggleZ galaxies have less significant fingers-of-god effects (due to lower non-linear pairwise velocities) because these low-mass tracers avoid massive clusters.

\subsection{Seven approaches to modelling $P(k)$}
\label{sec:Approaches}

Here we describe seven different approaches to modelling the power spectrum.  They differ primarily in the way they treat non-linearities and bias. In the next section we will test these approaches by how well they reproduce the input cosmology of the GiggleZ simulation.

\subsubsection{Model A: Linear}
We use CAMB to calculate the linear matter power spectrum $\Pm(k)$ at redshift $z$ for each set of cosmological parameters. 
We assume a linear bias $b$ given by:
\begin{equation} \label{eqn:Kaiser}
\Pgal(k) = b^2\Pm(k) \, .
\end{equation}
Since the linear bias s constant with respect to $k$, we analytically marginalise over $b$ \cite{Lewis:2002},
\begin{equation}
\label{eqn:biasmarge}
b^2 = \frac{\sum_{j,k}\Pcon^j C_{jk}^{-1}\Pobs^k}{\sum_{j,k} \Pcon^jC_{jk}^{-1}\Pcon^k} \,
\end{equation}
where $\Pcon(k)$ is the convolved theoretical power spectrum, and $\Pobs(k)$ is the observed power spectrum with the covariance matrix $C$.

\subsubsection{Model B: Halofit}
The Halofit model was developed by \citet{Smith:2003}, calibrated from simulations made by the Virgo consortium. It is an improvement on the then commonly used Peacock-Dodds formula \cite{Peacock:1994}, as it allowed for the modification of dark matter halos by continuing mergers. It is still only a simple empirical scaling however, and doesn't attempt to include the effects of redshift-space distortions or  galaxy bias.

We use the CAMB Halofit module to calculate the non-linear matter power spectrum $\Phf (k)$ at redshift $z$. The bias is treated the same way as for the linear model.

\subsubsection{Model C: Simulation inspired fitting formulae} \label{model:jennings}

Modelling the non-linear redshift space distortions of the galaxy power spectra is difficult, and one approach is to use large scale $N$-body simulations. In \citet{Jennings:2010}, the authors studied such effects in their simulations, and produced fitting formula for the density, velocity divergence and cross-power spectra in the non-linear regime that provides a good fit to the simulations (up to $k \sim 0.2 h\, {\rm Mpc}^{-1}$). We briefly outline our interpretation/application of  these formulae as follows.

The angle averaged redshift-space power spectrum of matter can be written as \citep{Scoccimarro:2004}
\begin{eqnarray}
P_m(k,z) &=& A_0(k)b^2(z)P_{\delta\delta}(k,z) \\ 
		&& + 2 A_2(k) f b P_{\delta\theta}(k,z) + A_4(k) f^2 P_{\theta\theta}(k,z)  \, , \nonumber
\end{eqnarray}
where $P_{\delta\delta}(k)$ is the non-linear density power spectrum, $P_{\theta\theta}(k)$ is the power spectrum of the velocity divergence field often referred to as the velocity power spectrum, and $P_{\delta\theta}(k)$ is the cross-power spectrum.
We use Halofit to calculate $P_{\delta\delta}(k,z)$. The velocity and cross-power spectrum are calculated using the fitting formulae of \citet[][Eq. (15)]{Jennings:2010}
\begin{equation}\label{eqn:jennings}
P_{xy}(k,z=0) = \frac{\alpha_0\sqrt{P_{\delta\delta}(k,z=0)}+\alpha_1P^2_{\delta\delta}(k,z=0)}{\alpha_2+\alpha_3P_{\delta\delta}(k,z=0)} \, ,
\end{equation}
where $P_{\delta\delta}(k,z=0)$ is the non-linear matter power spectrum at $z=0$. The values of $\alpha$ are given in Table \ref{tab:alphas} for $(xy)=(\theta\theta,\delta\theta)$.

\begin{table}[!h]
\centering
\caption{\label{tab:alphas} Best fit values of $\alpha$ in Eq. \ref{eqn:jennings} from \citet{Jennings:2010}}
\begin{tabular}{l|l|l}
			&$P_{\theta\theta}$ & $P_{\theta\delta}$ \\ \hline
$\alpha_0$	& -12462.1	& -12288.7\\
$\alpha_1$	& 0.839		& 1.43\\
$\alpha_2$	& 1446.6		& 1367.7\\
$\alpha_3$	& 0.806		& 1.54\\
\end{tabular}
\end{table}

The $P_{\theta\theta}(k)$ and $P_{\delta\theta}(k)$ must be scaled to redshift \citep[][Eq. (17)]{Jennings:2010},
\begin{equation}
C(z=0,z) = \frac{D(z=0)+D^2(z=0)+D^3(z=0)}{D(z)+D^2(z)+D^3(z)} \, ,
\end{equation}
where $D$ is the growth factor. For a flat universe with a cosmological constant, the growth factor can be approximated as
\begin{equation}
D(z) = \frac{H(z)}{H_0} \int_z^\infty \frac{dz' (1+z')}{H^3(z')} \left(\int_0^\infty \frac{dz'(1+z')}{H^3(z')}\right)^{-1} \, ,
\end{equation}
where $H(z)$ is given by Eq. \ref{eqn:H}. The power spectrum at redshift $z$ is then given by \citep[][Eq. (17)]{Jennings:2010}
\begin{equation} \label{eqn:Pxy}
P_{xy}(k,z) = \frac{P_{xy}(k,z=0)-P_{\delta\delta}(k,z=0)}{C(z=0,z)^2}+P_{\delta\delta}(k,z)
\end{equation}

The coefficients, $A_n$, are calculated as \citep{Scoccimarro:2004,Swanson:2010}
\begin{equation}\label{An}
A_n(k) \equiv \frac{1}{2}\int_{-1}^1 d\mu \mu^n \exp(-(f\mu k\sigma_v)^2) \, ,
\end{equation}
where $\mu = \hat{k}\cdot\hat{z}$ is the cosine of the angle between the wave vector, $\hat{k}$, and the direction of the line of sight, $z$. The growth rate, $f$, is given by $f=d\ln D/d\ln a \approx \Omega_m(z)^{0.55}$ with $a$ being the scale factor, and the one-dimensional velocity dispersion, $\sigma_v$, given by
\begin{equation}
\sigma_v^2 \equiv \frac{1}{3} \int \frac{d^3k'}{(2\pi)^3}\frac{P_{\theta\theta}(k')}{k'^2} \, .
\end{equation}
$P_{\theta\theta}$ is assumed to be spherically symmetric so the integral becomes
\begin{equation} \label{eqn:sigmav}
\sigma_v^2 = \frac{2}{3} \frac{1}{(2\pi)^2} \int d k' P_{\theta\theta}(k')\, .
\end{equation}

We marginalise numerically over $b$.

\subsubsection{Model D: Simulation inspired fitting formula with zero pairwise velocity}
The fitting formulae used in model C were derived for dark matter particles and not for halos. Setting all galaxy velocity dispersions to zero in model C provides a better fit to the GiggleZ halo catalogues, and we have treated this special case as a separate model.

\subsubsection{Model E: Empirical redshift space distortion damping} \label{model:RSD}
Non-linear structure formation will lead to increased peculiar galaxy velocities at low redshift (sometimes denoted non-linear velocities), leading to a damping in the observed structure, which can be described by the empirical model \citep{Peacock:1994}:
\begin{eqnarray}
\Pgal(k) &=& b_r^2 \Pm(k)\int_0^1\frac{(1+\beta\mu^2)^2}{1+(k\sigma_v\mu)^2}d\mu \\
               &=& \Pm(k)\left[ b_r^2 \int_0^1\frac{d\mu}{1+(k\sigma_v\mu)^2}  \right. \\ \nonumber
               && + 2b_rf\int_0^1\frac{\mu^2 d\mu}{1+(k\sigma_v\mu)^2} \left. +f^2\int_0^1\frac{\mu^4d\mu}{1+(k\sigma_v\mu)^2} \right] \, , \nonumber
\end{eqnarray}
where $\mu$ is the angle to the line of sight (defined as in Eq. \ref{An}), and $\sigma_v$ is the pairwise galaxy velocity dispersion in units of $\Mpc/h$. Here the damping due to the pairwise velocity dispersion is a Lorentzian, rather than a Gaussian. The last equality is obtained using $\beta=f/b_r$ with $f= \Omega_m(z)^{0.55}$. The velocity dispersion can then be calculated theoretically (e.g., from Eq. \ref{eqn:sigmav} with $P_{\theta\theta}(k)=f^2P_{\delta\delta}(k)$). The standard in the literature, which we follow here, is to set $P_{\theta\theta}(k)=f^2\Plin(k)$. 

Setting $\sigma_v=0\, \Mpc/h$ we recover the linear bias case given in Eq. \ref{eqn:Kaiser}, with the linear bias related to $b_r$ by a simple numerical factor, $b = b_r \sqrt{\int_0^1(1+\beta\mu^2)^2 d\mu}$. Here we numerically marginalise over $b_r$.

\subsubsection{Model F: $N$-body simulation calibrated}
All the non-linear effects are present in the GiggleZ simulation for a fiducial cosmology. Apart from the non-linear structure formation which we get from Halofit, we assume that any other effect is either negligible or can be scaled with cosmology. This is a simplified version of the method presented in \citet{Reid:2010}, who considered three modifications of the linear power spectrum, treated independently. These modifications are: the damping of the BAO, the change in the broad shape of the power spectrum because of non-linear structure formation, and the bias because we observe galaxies in haloes in redshift space rather than the real space matter distribution. Here we simplify it by substituting the complex scale-dependent bias of the Luminous Red Galaxies (LRGs) with the linear bias of WiggleZ galaxies.

For the BAO damped power spectrum we use the following form
\begin{equation}
\label{eqn:trialdamped}
\Pdamp^\trial(k) = P_{\rm lin}^\trial(k)f_{\mathrm{damp}}(k)+\Pnw^\trial(k)(1-f_{\mathrm{damp}})\,,
\end{equation}
where $P_{\rm lin}$ is the linear power spectrum, and $\Pnw$ is a `no-wiggles' power spectrum. Here the damping factor is $f_\mathrm{damp}(k)=\exp(-(k\sigma_v)^2)$, with $\sigma_v$ given by Eq. \ref{eqn:sigmav} and $P_{\theta\theta}(k)=\Plin(k)$. The no-wiggles power spectrum (without the acoustic peaks) we get by spline-fitting the output power spectrum from CAMB following the approach of \citet{Reid:2010} and \citet{Swanson:2010}. The exact positions of the nodes do not affect the final result as long as they are consistent for all sets of trial parameters.

For the change in broadband power due to non-linear structure growth we use the Halofit model, applied as a  ratio to the no-wiggles power spectrum (since Eq. \ref{eqn:trialdamped} has already taken into account the effect on the BAO peak of non-linear structure formation). This ratio term is given by
\begin{equation}
r_{\rm hf} = \frac{\Phfnw(k)}{\Pnw(k)}\,.
\end{equation}

Finally, we need to model the bias that comes from observing galaxies in halos. For this we make use of the GiggleZ simulations. $\PGiggleZ^\fid(k)$ is given by a fifth-order polynomial fit to the power spectrum of a set of halos in the GiggleZ simulation chosen to match the clustering amplitude of the WiggleZ galaxies (details of this fitting formula can be found in Appendix \ref{sec:gigglezfitting}),
\begin{equation}
\PGiggleZ^\fid(k) = P_{\rm{poly}}(\fid)(k)\Pdamp^\fid(k)/\Pnw^\fid(k) \, .
\end{equation}
where ``fid'' refers to the fiducial cosmology for the GiggleZ simulation.

Combining all these terms together, we get (for a given trial cosmology)
\begin{equation}
\Pgal^\trial(k) = b^2\Pdamp^\trial(k) \frac{\Phfnw^\trial(k)}{\Pnw^\trial(k)}\frac{\Pnw^\fid(k)}{\Phfnw^\fid(k)}\frac{\PGiggleZ^\fid(k)}{\Pdamp^\fid(k)} \, ,
\end{equation} 
and re-aranging we get
\begin{equation} \label{eqn:Pgaltrial}
\Pgal^\trial(k) = b^2\Phfnw^\trial(k)\frac{\Pdamp^\trial(k)}{\Pnw^\trial(k)}\frac{P_{\rm{poly}}^\fid(k)}{\Phfnw^\fid(k)}\, ,
\end{equation}
where $b^2$ is the linear scaling (related to galaxy bias). We marginalize analytically over $b$. The second factor in Eq. \ref{eqn:Pgaltrial} represents the smooth power spectrum of the trial cosmology. The third term defines the acoustic peaks and their broadening caused by the bulk flow motion of galaxies from their initial positions in the density field, and the fourth factor describes all the additional non-linear effects from the $N$-body simulation.

\subsubsection{Model G: $N$-body simulation calibrated without damping}

The approach in model F was derived for a particular data set \citep{Reid:2010}, with several assumptions that went into the simulation and modelling that my not be present in WiggleZ, or we have included and that were not included in \citet[Reid:2010]. One particular case we tested was 
to see if introducing a strong damping may in fact lead to biased parameter values. Removing all damped power spectra in model F (so $\sigma_v=0$ and $f_{\mathrm{damp}}(k)=1$ in Eq. \ref{eqn:trialdamped}) simplifies it to:
\begin{equation} \label{eqn:Pgaltrialnodamp}
\Pgal^\trial(k) = b^2\Phf^\trial(k)\frac{P_{\rm{poly}}^\fid(k)}{\Phf^\fid(k)}\, .
\end{equation}
We marginalize analytically over $b$. 

We found this model provides a better fit to the GiggleZ halo catalogues, recovering the input fiducial cosmology more accurately than in model F, and we have treated this special case as a separate model.

\renewcommand \thesubsubsection{\arabic{subsubsection}}

\subsection{Bias}

The GiggleZ simulations show that the effect of scale dependent bias is less than 1\% for haloes with large-scale clustering amplitude matching that of the WiggleZ sample, over the considered $k$-interval, and consequently we only consider scale-independent bias \citep{Poole:2012}. Where possible we analytically marginalise over this bias, and otherwise numerically, as denoted for each of the approaches. 

The bias is required to be identical for all windows at the same redshift, so when we marginalise we assume a single parameter that specifies the bias at that redshift. We allow the bias to vary between redshifts, since the survey colour and magnitude cuts cause the galaxy populations observed in our sample to be different at different redshifts.

\subsection{A comparison of the approaches}
\label{sec:comparison}

\begin{figure}
	\includegraphics[width=0.99\columnwidth]{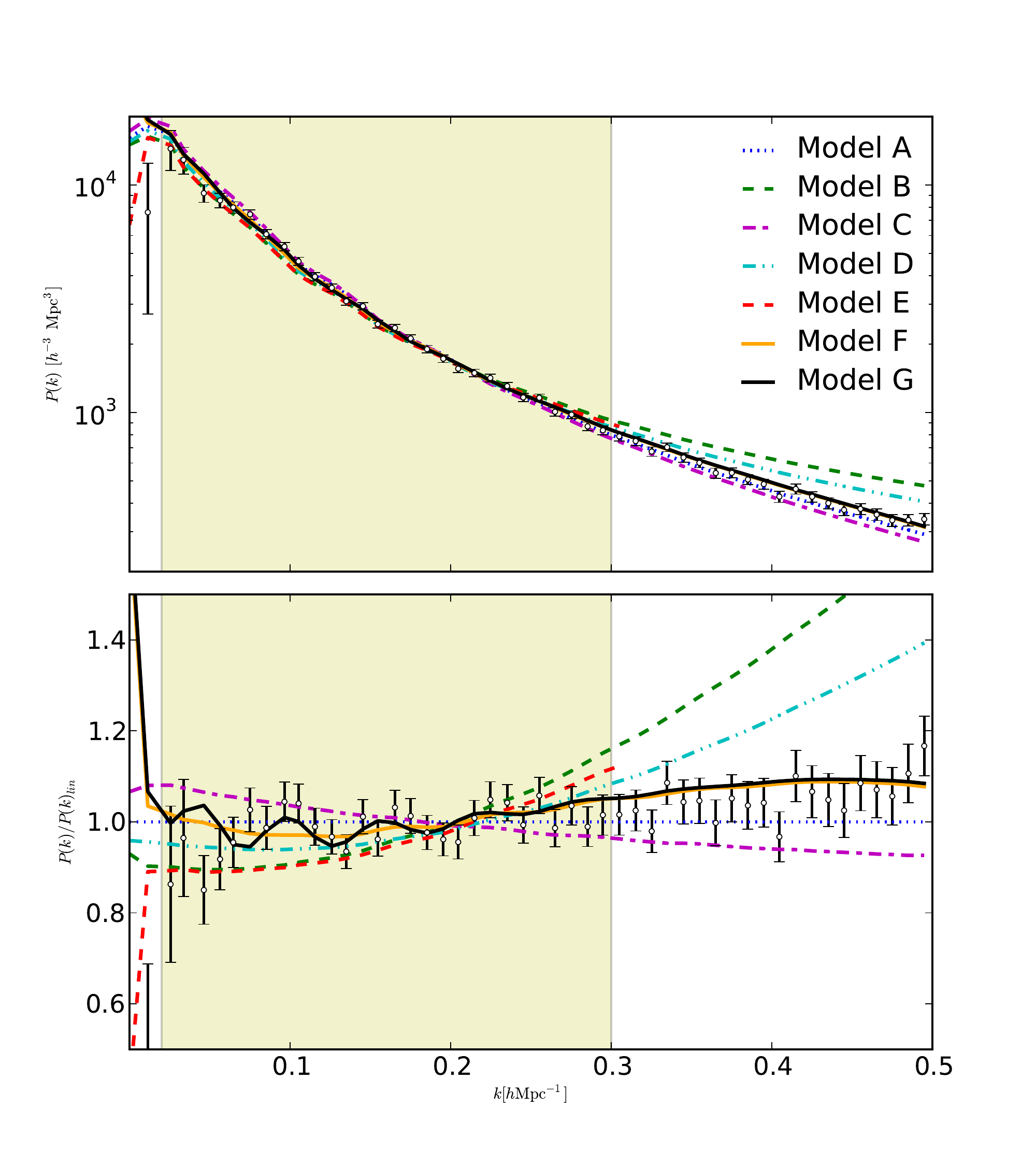}
	\caption{(color online). ({\it top}): A weighted average of the WiggleZ power spectra in the survey regions and redshifts, and the seven models described in section \ref{sec:Approaches} for the best fit cosmology of the best model (G). ({\it bottom}): The same models, but now the power spectrum has been rated to the linear prediction. In both plots the shaded region from $k=0.02\,h\,\Mpc^{-1}$ to $k=0.3\,h\,\Mpc^{-1}$ is the range of data we are using in our analysis. This shows the divergence of models at large $k$ and the reason why careful modelling is necessary.}
	\label{fig:pk}
\end{figure}

The power spectra for a fixed cosmology for models A-G are shown in Fig. \ref{fig:pk}. The different models tend to diverge for large $k$ ($k>0.2\,h\,\Mpc^{-1}$), and these different predictions will have a significant effect on fitting the cosmological parameters. 

We tested the different approaches by fitting to the $z=0.6$ power spectrum of the GiggleZ simulation\footnote{This comparison was summarised in \citet{riemersorensen:2011}, but we repeat it here for the purposes of ongoing discussion, and the inclusion of model F.}. For this we used two sets of parameter grids: $\Omega_m$ versus $f_b$ (where $f_b=\Omega_{\rm b}/\Omega_m$ is the baryon fraction), and $\Omega_m$ versus $n_s$ both with the remaining parameters fixed to the GiggleZ fiducial cosmology values. We chose these grids because the parameters are tightly related to the shape of the power spectrum. In both cases we obtain very similar conclusions so here we only present the results of $\Omega_m$ versus $f_b$. 

\begin{figure}[h]
\centering
	\includegraphics[width=0.99\columnwidth]{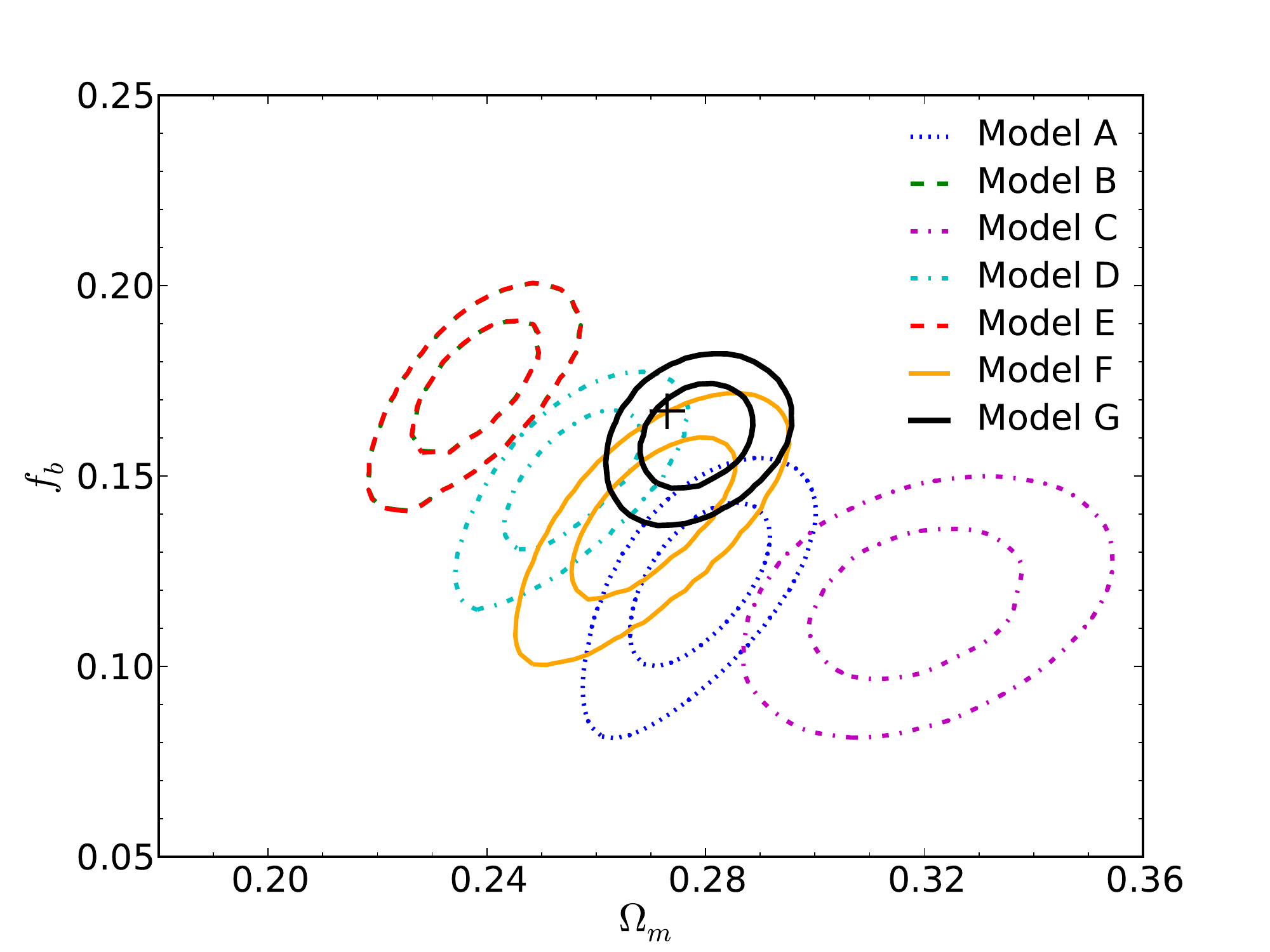}
	\caption{(color online). 1 and $2\sigma$ contours for in the \{$\Omega_m$, $f_b$\} plane for fits to the GiggleZ power spectra using model A-G. All other cosmological parameters are held fixed at the GiggleZ fiducial cosmology (WMAP 5-year best fit values). The GiggleZ parameter values are marked by a cross. The red ellipses (model E) lie directly on top of the green ones (model B), and so are not visible. This shows that the choice of model has a significant effect on the cosmological constraints from the power spectrum. Our preferred model is the Model G, which successfully reproduces the input parameters of the GiggleZ simulation.}
	\label{fig:GiggleZgrid}
\end{figure}

\begin{figure}
\centering
	\includegraphics[width=0.99\columnwidth]{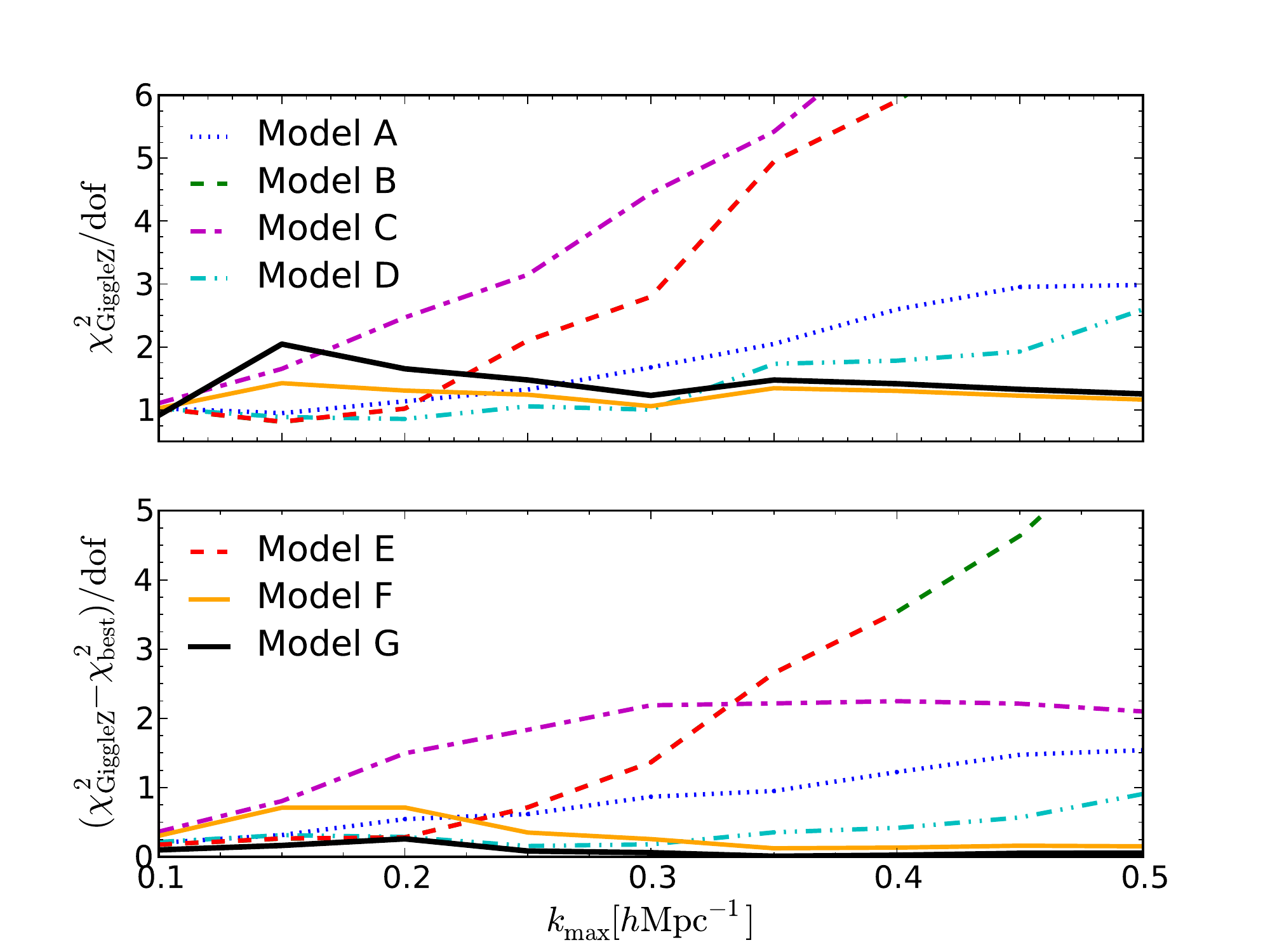}
	\caption{({\it Top}) Reduced $\chi^2$ of models fitted to the $N$-body simulation halo catalogue for the GiggleZ fiducial cosmology values. In absence of systematic errors the models should recover the input cosmology with $\chi^2/\mathrm{dof} \simeq1$. ({\it Bottom}) Difference in reduced $\chi^2$ values when using the GiggleZ fiducial cosmological parameters and the best fit values.}
	\label{fig:GiggleZchi2}
\end{figure}

The ability to recover the input model is shown in Fig. \ref{fig:GiggleZgrid}. For $\kmax<0.2\,h\,\Mpc^{-1}$ most of the models produce a good fit, whereas for $\kmax=0.3\,h\,\Mpc^{-1}$ model B, C and E give reduced $\chi^2$ values above $1.5$. Models B \& C were calibrated to be good fits up to $k=0.2\,h\,\Mpc^{-1}$, so this is not surprising. The upper panel of Fig. \ref{fig:GiggleZchi2} shows the $\chi^2$ for the fiducial GiggleZ cosmological parameters, which is a measure of how well the models recover the input parameters. The lower panel shows the difference between $\chi^2$ of the GiggleZ values and the best fit, indicating how far the best fit is from the input values. We assume that the $N$-body simulation, which provides a complete census of the relevant non-linear effects, yields the most accurate clustering model. In this sense the good performance of model G (figures \ref{fig:GiggleZgrid} and \ref{fig:GiggleZchi2}) is a consistency check, and the variations of results produced by the other models are due to the breakdown in their performances compared to the simulation.


It is clear that model G demonstrates the best ``fit and recover" performance, and consequently we have chosen this model to be the default model of the WiggleZ CosmoMC module.

\subsection{Systematics}
\label{sec:systematics}

In measuring the power spectrum from the WiggleZ data a number of choices were made in modelling certain unknown factors, which may impact on the resulting data product and cosmological parameters. These were the nature of the radial selection function, $N(z)$, and the fiducial cosmological parameters used to generate the power spectra. We test a number of possible systematics to see how much they impact on the resulting cosmological parameter constraints.

\begin{figure}[h]
\centering
	\includegraphics[width=1\columnwidth]{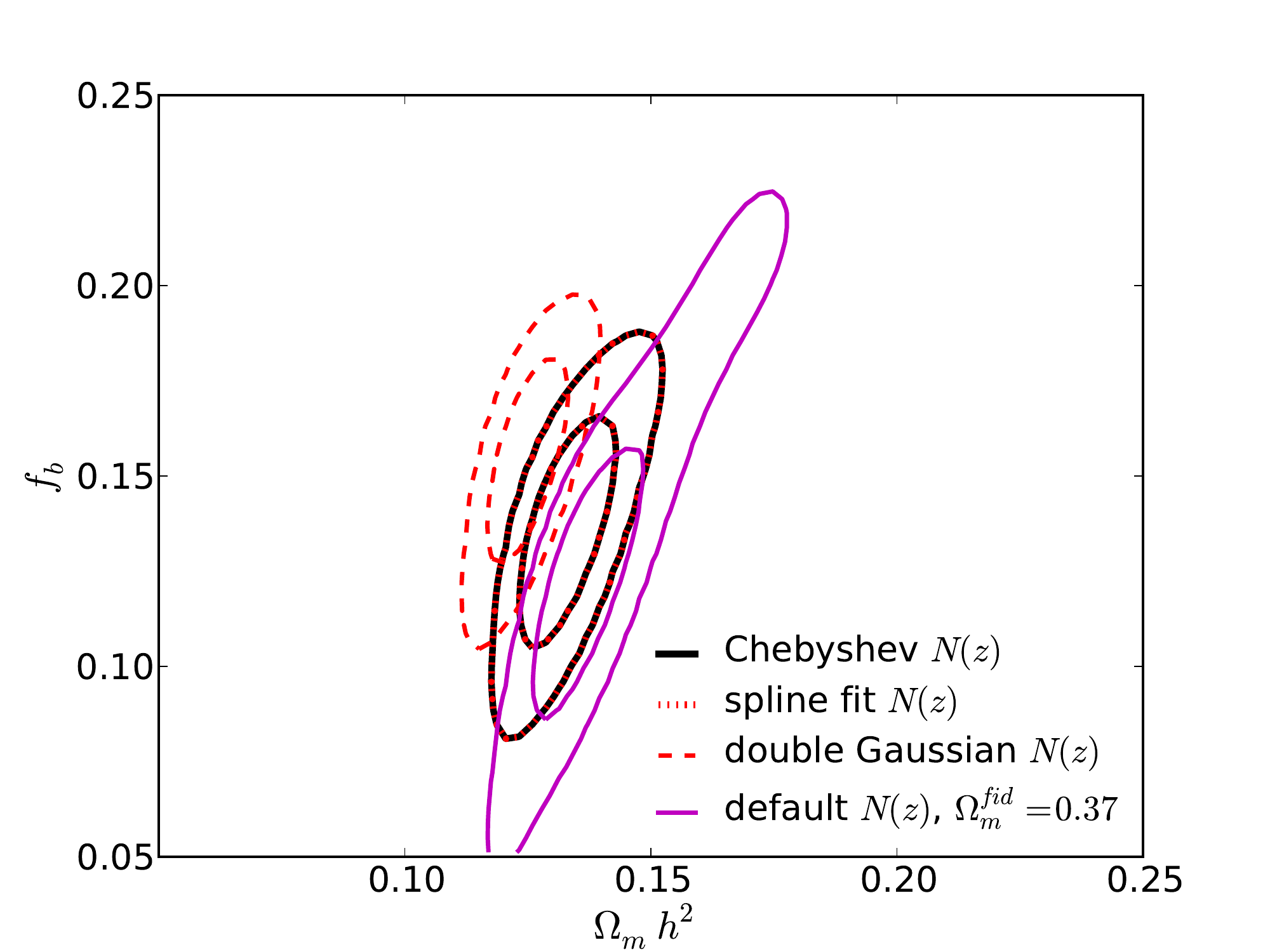}
	\caption{(color online). 1 and $2\sigma$ contours for in the \{$\Omega_m$, $f_b$\} plane for fits to the WiggleZ power spectra, checking for systematic errors using a number of different data analyses. 
	The red dotted lines  (spline fit) lie directly on top of the black (Chebyshev), and so are not visible.}
	\label{fig:systematics}
\end{figure}

Firstly there is the choice of radial selection function, $N(z)$. The number of galaxies in the WiggleZ sample will vary not just as a function of angular position on the sky, but also as a function of redshift, because the survey colour and magnitude cuts cause the observed galaxy populations to be different at different redshifts. It is difficult to calibrate this directly, so we assume some smooth function form for $N(z)$. We fit the same $N(z)$ function to each region of the sky, but a different one for each observing priority band (WiggleZ observations were prioritised by magnitude, with apparently-faint galaxies given highest priority and each band representing an equal interval of the range $20.0 \le r \le 22.5$). The default choice for the function is a Chebyshev polynomial, but alternatives include a double-peaked Gaussian, and a cubic-spline fit. In Fig. \ref{fig:systematics} we show the effect of these choices on the recovered joint cosmological constraints on the physical matter density $\Omega_m h^2$ and the baryon fraction $f_b$ (keeping the remaining parameters fixed at the WMAP7 best fit values). We see the effect on the recovered cosmological parameters from the choice of $N(z)$ is small, with the error ellipses being consistent between each other at the two-sigma level. This is as expected given that the data are identical. Since there is no obvious large systematic, our choice of $N(z)$ is motivated by our initial confidence in which of these models will work best. The double Gaussian has too few parameters to describe real variations, whereas the spline fit has too many parameters and may over fit, removing real fluctuations. The Chebyshev polynomial (whose order is truncated as driven by the data) is a compromise between these effects.

Secondly there is the choice of fiducial cosmological parameters in converting the redshifts to distances. We include the Alcock-Paczynski scaling factor when computing the predicted power spectrum in the likelihood calculation, as given in section \ref{sec:generating_pk}. We check the effect of assuming the wrong cosmology by  re-measuring the power spectra using a fiducial matter density of $\Omega_m^{\rm fid} = 0.37$.

From Fig. \ref{fig:systematics} it is clear that the choice of fiducial cosmological parameter values has a small effect on the best fit values, though the size of the error ellipse is expanded.

\begin{figure}[h]
\centering
	\includegraphics[width=1\columnwidth]{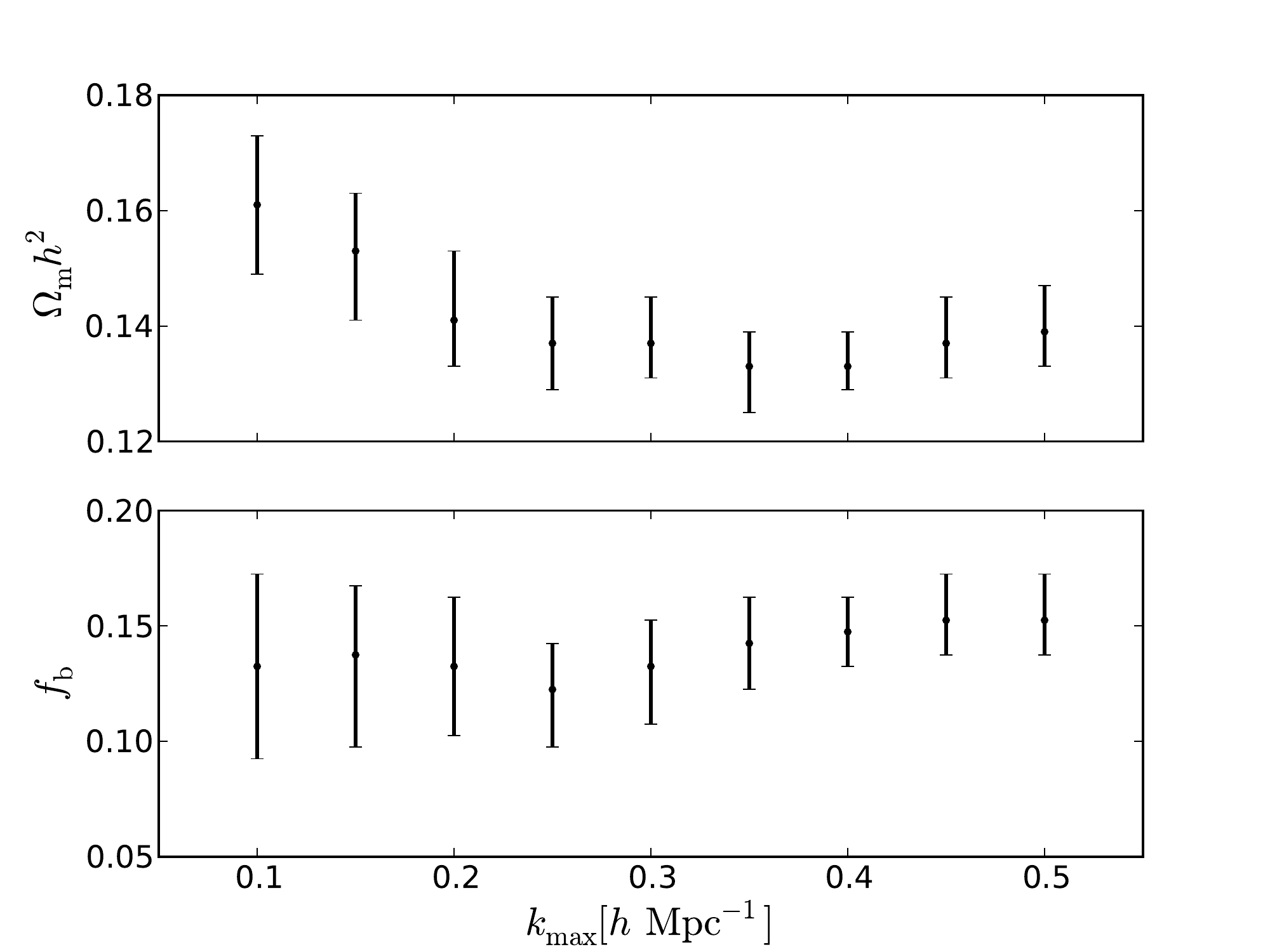}
	\caption{(color online). Measurement of $\Omega_m h^2$ (top) and $f_{\rm b}$ (bottom) as a function of $k_{\rm max}$ for the WiggleZ data,  with all other cosmological parameters fixed to the WMAP 7-year best fit values, with those two being measured simultaneously.}
	\label{fig:kmax}
\end{figure}

There is also the choice of the range of $k$-values to fit the data to. Throughout our discussion of these models, and further parameter analysis, we have fixed the lower limit of power spectrum fitting to be $\kmin=0.02\,h\,\Mpc^{-1}$, which corresponds to the largest modes observed in each of the WiggleZ regions (the final results are not very sensitive to the exact value). The upper limit on the fitting range was fixed to $\kmax=0.3\,h\,\Mpc^{-1}$ after running a series of flat $\Lambda$CDM fits to WMAP + WiggleZ varying $\kmax$. As seen in Fig. \ref{fig:kmax}, the uncertainties decrease until  $k_{\rm max} = 0.3\,h\,\Mpc^{-1}$ and then start to increase again (due to the effect of systematics), before decreasing to roughly the same value at $\kmax = 0.5\,h\,\Mpc^{-1}$. Furthermore, the mean values of the parameter probability distributions are roughly consistent in the range of maximum $k$ considered, similar to what we saw in section \ref{sec:comparison} when making comparisons to simulations. Our choice of $k_{\rm max} = 0.3\,h\,\Mpc^{-1}$ is then a compromise between minimising the parameter errors, and trusting that our simulations are accurate down to those small scales. 

\subsection{WiggleZ power spectrum fit}

\begin{figure}[h]
\centering
	\includegraphics[width=0.49\textwidth]{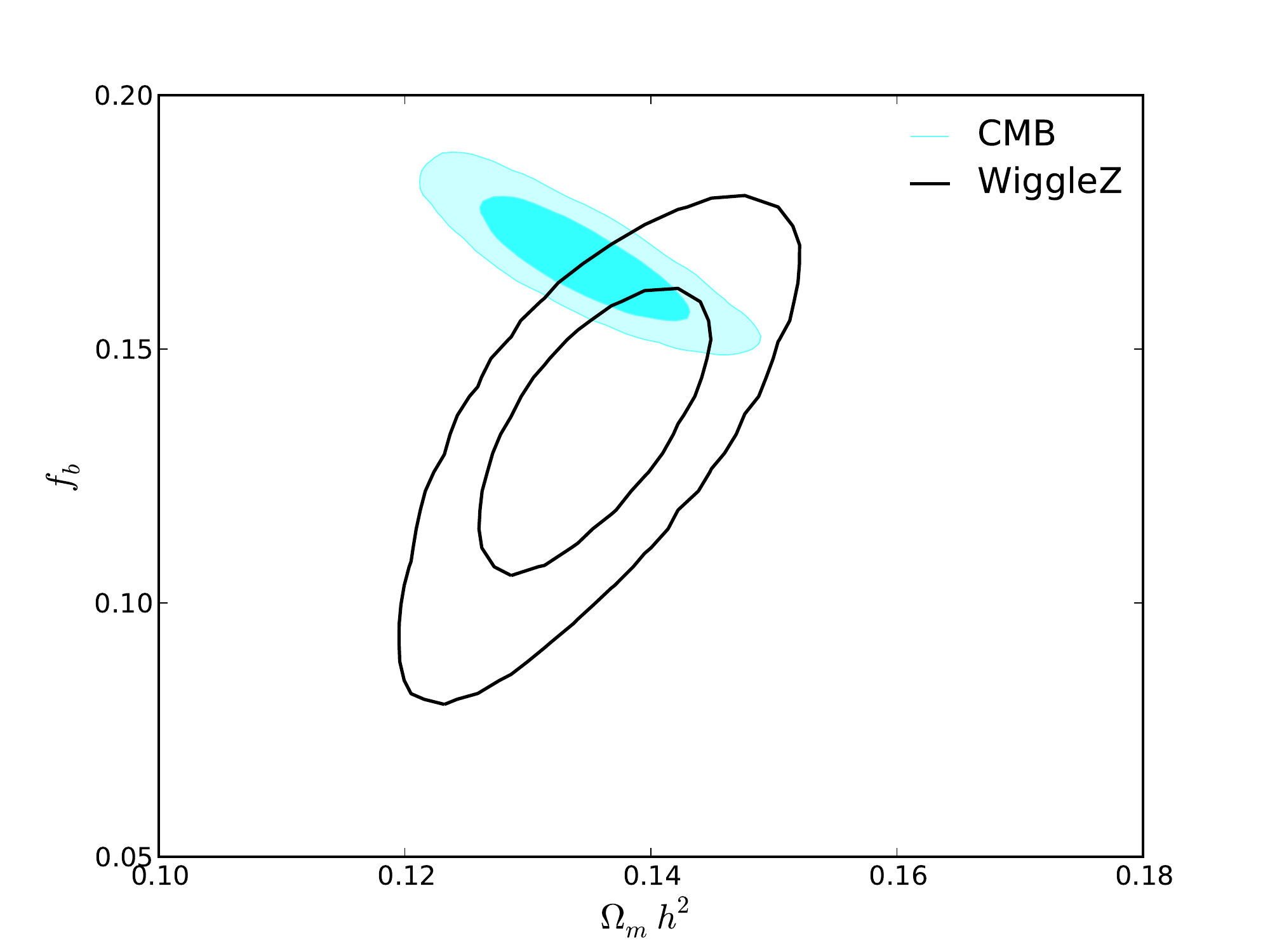}
	\caption{(color online). The probability contours in the \{$\Omega_m h^2$, $f_b$\} plane for WiggleZ alone (black solid line), with all other cosmological parameter fixed to the WMAP 7-year best fit values. We also plot the contours from the WMAP data for a $\Lambda$CDM model, marginalising over all other parameters (blue filled contour). We see good agreement between the two data sets, though the WiggleZ data does favour slightly smaller values of the baryon fraction. }
	\label{fig:2D_wigglez_only}
\end{figure}

To check the consistency between the cosmological parameters derived from WMAP and WiggleZ, we fitted the WiggleZ power spectra alone varying the physical matter density, $\Omega_m h^2$, and the baryon fraction, $f_b$, keeping the remaining parameters fixed at the WMAP7 best fit values. The results are shown in Fig. \ref{fig:2D_wigglez_only}, where we also show the constraints from the CMB (WMAP 7-year) in comparison. The two data sets give consistent cosmological parameters in this model. There is a tendency for WiggleZ alone to favour slightly smaller values of the baryon fraction in comparison to the CMB data alone, but not at any statistically significant level.  

\section{Cosmological Constraints}
\label{sec:cosmology} 

In this section we will discuss the different cosmological parameters and combinations of parameters (models) to be tested, and present the results of our analysis.

\subsection{Cosmological Models}


The approach we take in this paper is to start with the flat $\Lambda$CDM model as our benchmark. We first check the consistency of our new WiggleZ $P(k)$ data with previous data sets, within the flat $\Lambda$CDM model. The basic model has five parameters, as described in the introduction and summarised in table \ref{table:params}.


To extend the $\Lambda$CDM model, we consider parameters whose existence is required or suggested by either external datasets (for example neutrino oscillation experiments demonstrating that the neutrinos have mass) or from theoretical predictions (e.g., primordial tensor perturbations generated by inflation). While there is a wealth of possible extension parameters to choose from, we limit ourselves to these five: the equation of state of the dark energy $w$ (assumed to be a constant), the curvature, the running of the primordial density power spectrum, the tensor to scalar ratio, and the density fraction of massive neutrinos.

The standard set and extension parameters are given in table \ref{table:params}. The different models we will consider will consist of the base parameter set plus one or more extra parameters, by themselves or in combination with others. These are shown in table \ref{table:models}.

\begin{table}
\begin{tabular}{llc} \hline
Cosmological Parameter & symbol  \\ \hline
{\it Base parameters (for Flat $\Lambda$CDM)} & \\
Physical baryon density & $\Omega_{\rm b} h^2$   \\
Physical CDM density & $\Omega_{\rm CDM} h^2$  \\
Angular size of the sound horizon  & \multirow{2}{*}{$\theta$} \\
at decoupling & \\
Optical depth of reionization & $\tau$ \\
Scalar spectral index & $n_s$\\
Amplitude of the scalar perturbations & $A_s$ \\
Sunyaev-Zel'dovich (SZ) template normalization & $A_{\rm SZ}$ \\ \hline
{\it Extension parameters} & \\
Dark energy Equation of state & $w$  \\
Curvature & $\Omega_k$  \\
Neutrino density fraction & $f_{\nu}$ \\
running of the spectral index & $n_{\rm run}$ \\
Tensor to scalar amplitude ratio & $r$ \\ \hline
{\it Derived parameters} &  \\
Matter density & $\Omega_m$ \\
Dark energy density & $\Omega_\Lambda$ or $\Omega_{\rm DE}$ \\
Hubble parameter & $H_0$ or $h$ \\
Matter power dispersion at $8 h^{-1}$Mpc & $\sigma_8$ \\ 
Combined mass of the neutrino species & $\sum m_{\nu}$\\
\hline
\end{tabular}
\caption{\label{table:params}Cosmological parameters to be varied.}
\end{table}

\begin{table}
\begin{tabular}{lll} \hline
Section & Model & Extra Parameter(s) \\ \hline
\ref{sec:lcdm} & Flat $\Lambda$CDM & none  \\
\ref{sec:wcdm} & Flat $w$CDM  & $w$ \\
\ref{sec:kcdm} & $\Lambda$CDM & $\Omega_k$ \\
\ref{sec:owcdm} & $w$CDM & $\Omega_k$ \& $w$ \\
\ref{sec:mnu} & massive neutrinos & $f_\nu$ \\
\ref{sec:running} & running spectrum & $n_{\rm run}$ \\
\ref{sec:grav} & gravitational waves & $r$ \\
\hline
\end{tabular}
\caption{\label{table:models}Cosmological models to be analysed.}
\end{table}

If a parameter is not varied in some model, and cannot be derived from the other parameters that are being varied, then it is held fixed at some default value. These default values are zero, except in the case of the equation of state of the dark energy, where the default value is $w=-1$ (i.e. a cosmological constant).

\subsection{Parameters and the physics being constrained}

The cosmological observations that we make constrain different physics, and so it is important, especially with many-parameter models, to understand exactly which data are contributing to constraints on which parameter or parameters. The major cosmological datasets are:

{\it CMB}: The CMB is probably the most powerful tool available to a cosmologist. The features of the CMB power spectrum are the Sachs-Wolfe plateau which measures $A_s$, and gives weak constraints on the dark energy parameters $\Omega_{DE}$ and $w$ through the Integrated Sachs Wolfe effectl the position of the acoustic peaks which function as a standard ruler measuring $\Omega_m$, $\Omega_\Lambda$, and $H_0$; the relative height of the acoustic peaks which measure $\Omega_m$ and $\Omega_{\rm b}$; and the damping tail, which gives some weak constraint on $\tau$. There is also the polarisation power spectrum, which gives stronger constraints on $\tau$. Finally, as the CMB is measuring the amplitude of density fluctuations over a range of scales it provides a constraint on the spectral index $n_s$. For the analysis presented here we have used the CMB observations from the WMAP 7-year data release \citep{Komatsu:2011}.

{\it BAO}: The Baryon Acoustic Oscillations are a standard ruler, and so give constraints on those parameters that relate to distance, $\Omega_m$, $\Omega_{\rm DE}$, and $w$. Their size is set in co-moving coordinates, so it is not possible to use them to measure the Hubble parameter today ($H_0$) without reference to some physical size, such as from the CMB. Despite their name, the size of the ruler depends on the baryon density only weakly, and since our measurement of $\Omega_{\rm b} h^2$ from the CMB is already very accurate, this parameter is usually fixed when using BAO data. For the analysis presented here we have used the BAO scale as measured by SDSS \citep{Percival:2007, Percival:2009}. 

Note that we do not include the WiggleZ BAO measurements, as they come from the same galaxy sample as our $P(k)$ measurement, and so we are being conservative in the light of potential covariance. We compare the BAO results from WiggleZ with the results from the WiggleZ $P(k)$ in a couple of cases of interest (curvature and the dark energy equation of state), and find fairly good agreement in the cosmological constraints from the two datasets.

\begin{figure*}
\centering
	\includegraphics[width=1.85\columnwidth]{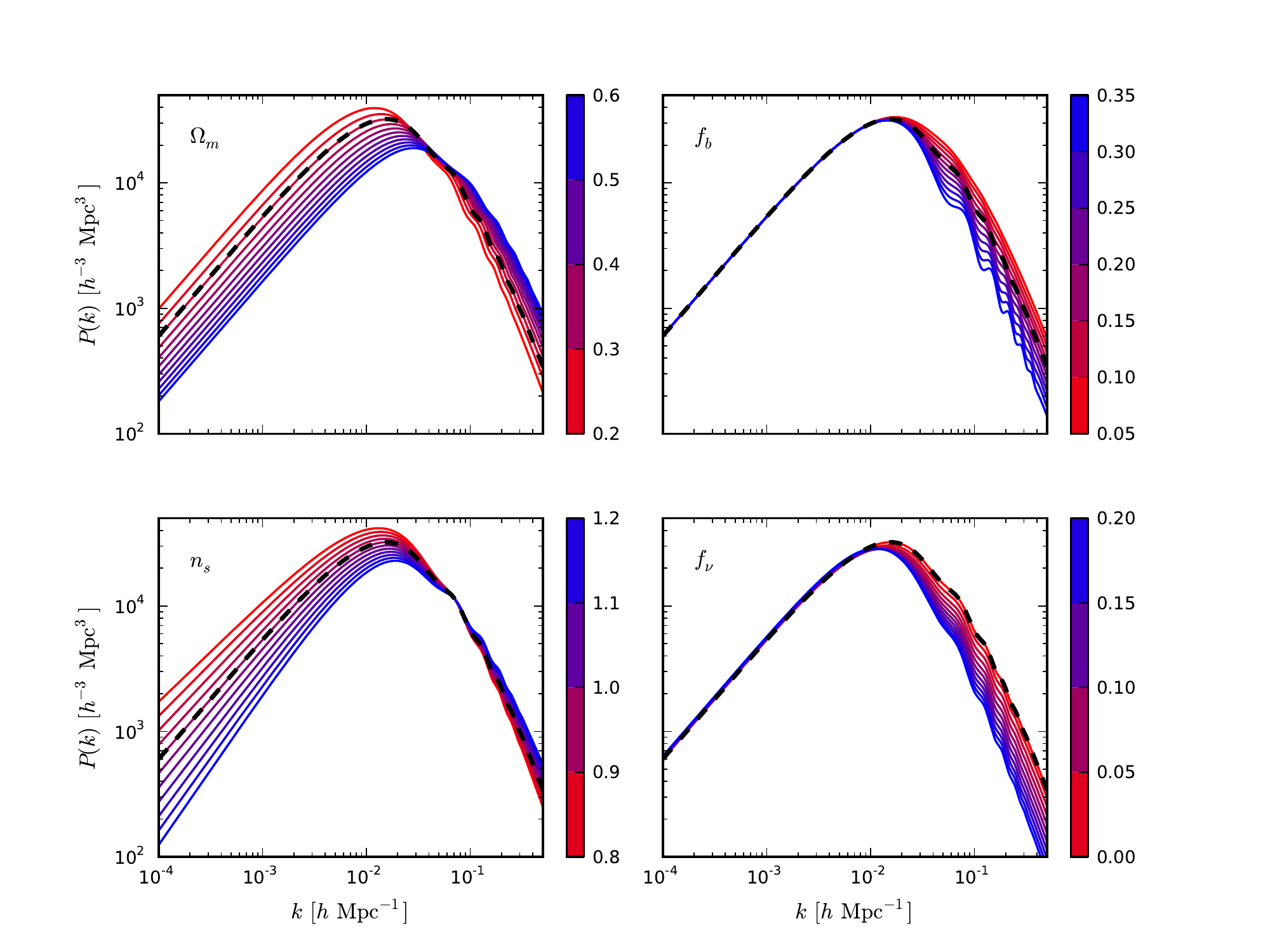}
	\caption{\label{fig:pkcosmoparams}(color online). The matter power spectrum plotted for different values of the cosmological parameters. For each plot we fix the other parameters to the fiducial cosmological values while varying only that particular one. The colour bar indicates the value  of the parameter. The dashed black curve is the power spectrum for the fiducial cosmology, and is the same in all four plots. }
	
\end{figure*}

{\it LSS}: The large-scale structure of matter gives information about the matter density $\Omega_m$ through the broadband shape of the power spectrum, with some modulation by baryons $\Omega_{\rm b}$ or neutrinos $f_{\nu}$, and the amplitude of fluctuations $\sigma_8$ (though this measurement is somewhat degenerate with the linear bias). It also contains the BAO peaks, and so in principle can be used to give the same distance information as BAO. However, when measuring the full power spectrum the overall shape is more important than the wiggles, so the BAO features have only a small effect on the cosmological constants for the full power spectrum. 
The LSS data used here is SDSS LRG sample \citep{Reid:2009}, and the WiggleZ $P(k)$ measurement in this paper.  (Note that again, because of previously mentioned covariance between the two data sets, we never combined the SDSS LSS and BAO data in the same analysis). In Fig. \ref{fig:pkcosmoparams}, we show the effect of varying the matter density, baryon fraction, spectral index and neutrino fraction on the matter power spectrum.

{\it SNIa}: Type-Ia Supernovae are standard candles, and so also give constraints on the parameters that control the cosmological distances, $\Omega_m$, $\Omega_{\rm DE}$ and $w$. Their absolute magnitude is not known, and is degenerate with the Hubble parameter, so this combination is normally marginalised over when using this data set. We have used the SNLS data set \citep{Guy:2010,Conley:2011,Sullivan:2011}.

{\it ${H_0}$}: Direct measurements of the Hubble parameter from local standard candles or rulers also play a key role, establishing a normalising `ruler' in the cosmological distance ladder. Here we have used the $H_0$ data point given in \citep{Riess:2009}.

\subsection{Analysis method}\label{sec:wigglezcosmomc}

The most commonly used approach in cosmology is a Bayesian analysis of the the cosmological data. By Bayes' theorem, the posterior probability distribution of a parameter, $\theta$, given a model, $M$, and a dataset, $D$, is given by
\be
P(\theta|D,M) = \frac{P(D|\theta,M)P(\theta|M)}{P(D|M)} = \frac{\mathcal{L}(\theta)\pi({\theta})}{E}
\ee
where $P(\theta|M)$ (or $\pi(\theta)$) is the `prior' parameter distribution, and $P(D|\theta,M)$ (or $\mathcal{L}$) is the `likelihood\footnote{The normalising coefficient $P(D|M)$ (or $E$) is the `model likelihood' or `evidence', and can be used to evaluate the probability of the model as a whole (for examples of this use in cosmology, see \citep{Beltran:2005,Bridges:2006,Mukherjee:2006,Parkinson:2006,Trotta:2007,Valiviita:2009,Kilbinger:2010}).}.' The different models we test are given in Table \ref{table:models}. In this paper we focus on the posterior distributions (parameter constraints) of the different models for different data combinations, and set aside comparisons between the models, except in view of their consistency with the data. 

For the parameter priors, we assume a uniform probability distribution, with limits given in Table \ref{table:priors}. These prior ranges are chosen such that they are much larger than the expected posterior ranges (with certain exceptions, e.g., neutrino mass), so that we do not not miss regions of interest in the parameter space, and do not necessarily correspond to any physical intuition.

\begin{table}
\begin{tabular}{lcc} \hline
Parameter & minimum & maximum  \\ \hline
$\Omega_{\rm b} h^2$ & 0.005   & 0.34 \\
$\Omega_{\rm CDM} h^2$& 0.01  & 1.5 \\
100$\theta$&  0.5 & 10 \\
$\tau$&  0.01 & 0.5 \\
$n_s$&  0.5 & 1.5 \\
$\log(10^{10} A_s)$& 2.7   & 4 \\
$A_{\rm SZ}$&  0 & 2 \\
$w$&  -3  & 0 \\
$\Omega_k$ & -0.3   & 0.3 \\
$f_\nu$ &  0 & 1 \\
$n_{\rm run}$ & -0.2  & 0.2 \\
$r$ & 0  & 1 \\\hline
\end{tabular}
\caption{\label{table:priors}Prior ranges on the cosmological parameters. We assume uniform probability distribution between these ranges.}
\end{table}

Due to the complexity of the datasets under consideration it is impossible to solve for the individual parameter posteriors analytically. Instead we must proceed by sampling of the parameter space, and averaging (`marginalisation') of the likelihoods over the parameter ranges. For this we used the CosmoMC cosmological analysis package \citep{Lewis:2002}. CosmoMC is a Markov Chain Monte Carlo package that compute the likelihoods of all the cosmological datasets we use. We sample the parameter space using the Metropolis-Hastings algorithm, which comes as standard in the package.

We have created an additional WiggleZ power spectrum module for the code, that includes the modelling details of Model G. The module computes the likelihood of each WiggleZ redshift bin independently, but computes contributions from all regions simultaneously. It does this in order to marginalise over the linear bias $b$, using Eq. \ref{eqn:biasmarge}, assuming the bias is the same for every region at a given redshift. We make the power spectrum data and the module available with this paper.

\subsection{Results}
\subsubsection{Flat $\Lambda$CDM} \label{sec:lcdm}

We initially investigate how combining the WiggleZ power spectrum data with the other cosmological data sets constrains parameters in the concordance $\Lambda$CDM model. The WiggleZ power spectrum data alone are not able to constrain all the parameters, so we add the WMAP 7-year CMB data. We then considered the effect of adding extra datasets to the CMB +  WiggleZ pairing.

\begin{figure}
\centering
	\includegraphics[width=0.99\columnwidth]{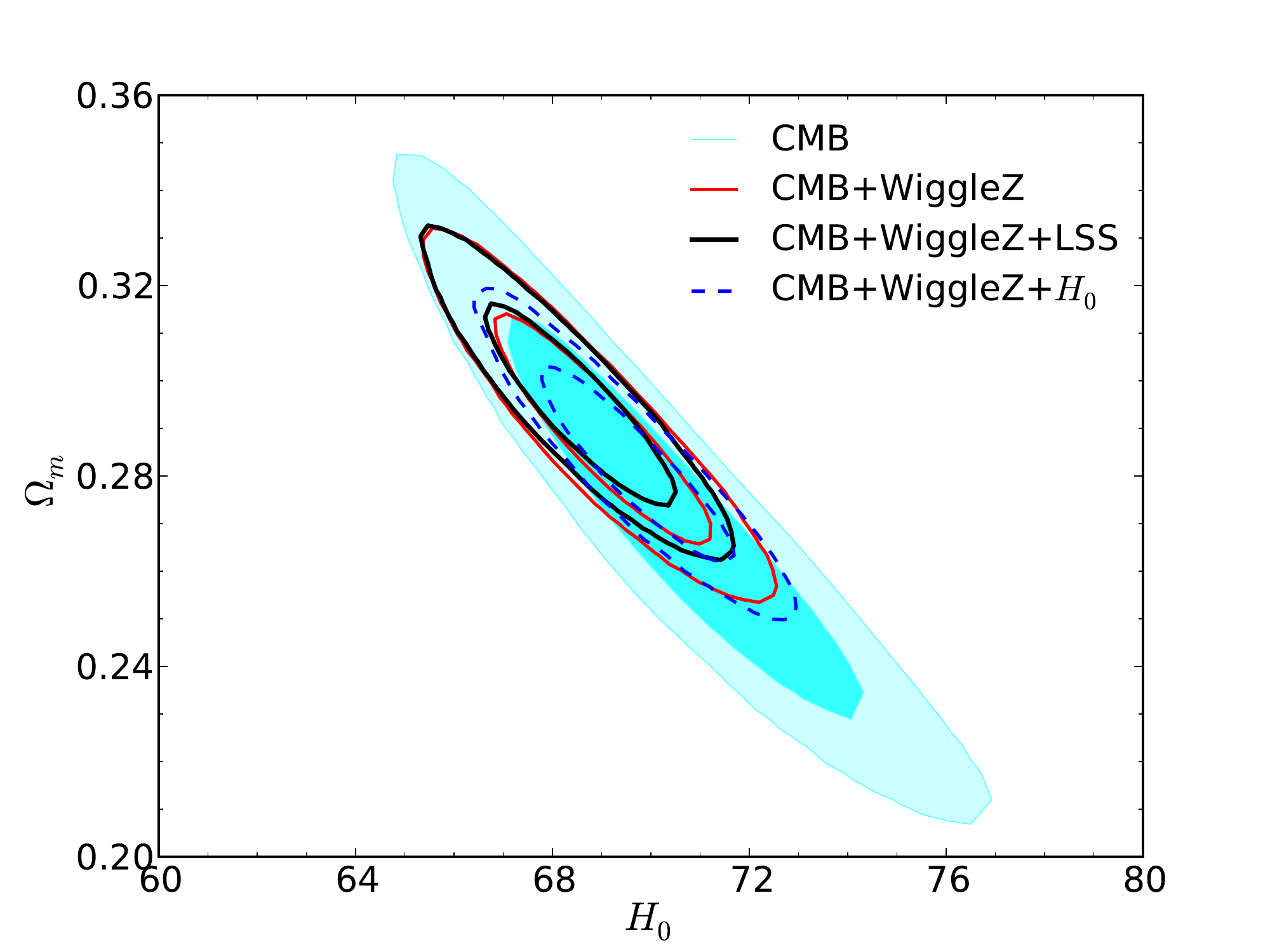}
	\caption{\label{fig:LCDM_om_H0_2D}(color online). The two-dimensional constraints in the \{$\Omega_m$, $H_0$\} plane for a flat $\Lambda$CDM model, using a number of different datasets (see legend). The WiggleZ data are consistent with the CMB data (from WMAP), reducing the size of the error ellipse, but pushing the preferred value of $\Omega_m$ upwards. }
	
\end{figure}

\begin{figure}
\centering
	\includegraphics[width=0.99\columnwidth]{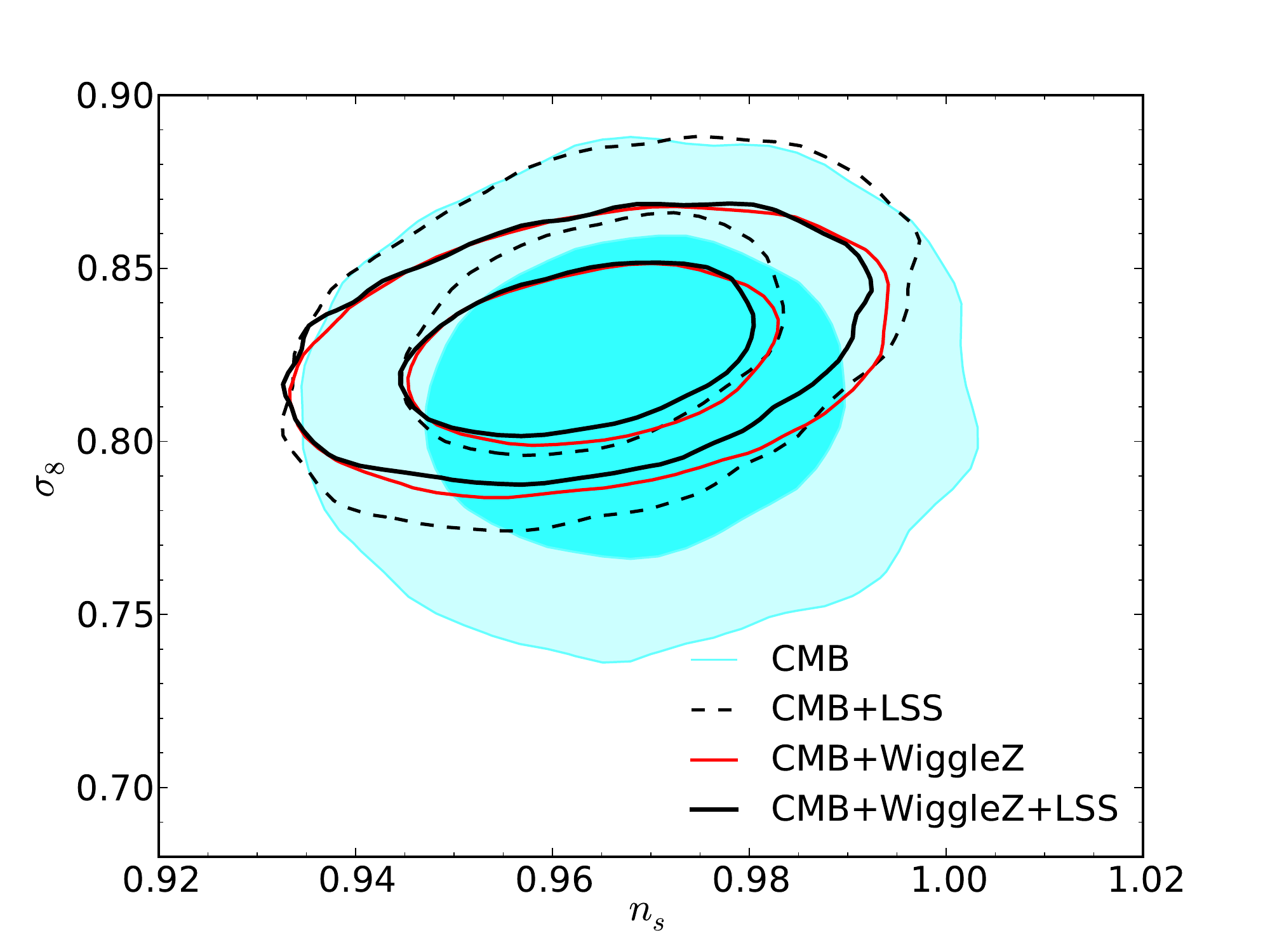}
	\caption{\label{fig:LCDM_ns_sig8_2D}(color online). The two-dimensional constraints in the \{$\sigma_8$, $n_s$\} plane for a flat $\Lambda$CDM model, using  a number of different data combinations (see legend). We find that combination of LSS data (from either WiggleZ or SDSS)  with the CMB data (WMAP) only improves the constraints on the spectral index $n_s$ marginally, but improves the constraints on the amplitude of fluctuations $\sigma_8$ by a large amount.}
	
\end{figure}

The marginalised constraints are given in table \ref{table:distanceconstraints}. The two-dimensional parameter constraint contours for $\{ \Omega_m,~H_0\}$ are shown in Fig. \ref{fig:LCDM_om_H0_2D} and $\{ n_s,~ \sigma_8\}$ are shown in Fig. \ref{fig:LCDM_ns_sig8_2D}.

We find that adding the WMAP CMB data to the WiggleZ $P(k)$ data considerably shrinks the size of the confidence contours, as well as allowing us to constrain other parameters we would not be able to using only the WiggleZ data (such as $\tau$). The contours are smaller than both individual datasets by themselves, and the WiggleZ data are complimentary to WMAP, narrowing the confidence contours of the total matter density around $\Omega_m =0.29$ (and since the distance to recombination is well constrained in this model, preferring values of $h=0.689$). In terms of the parameters that govern the primordial power spectrum, $n_s$, and $\sigma_8$, we see from Fig. \ref{fig:LCDM_ns_sig8_2D} that the WiggleZ data improves constraints only on the amplitude of fluctuations, but does not help with the spectral index (a similar effect is seen using CMB + LSS data). This is due to the fact that the constraint from WMAP on $n_s$ is already very good, and the large scale measurements of the matter power spectra at late times do not offer enough accuracy over a large enough range of scales to be effective.


\begin{table*}
\begin{tabular}{llccccc} \hline
Model & Parameter &  CMB + WiggleZ & + $H_0$ & + SN-Ia &  + BAO		& + $H_0$ + BAO \\ \hline
Flat $\Lambda$CDM  & $ 100\Omega_{\rm b} h^2 $ & $ 2.238 \pm 0.052 $ & $ 2.255 \pm 0.050 $ & $ 2.240 \pm 0.053 $ & $ 2.239 \pm 0.050 $ & $ 2.253 \pm 0.050 $ \\ 
& $ \Omega_{\rm CDM} h^2 $ & $ 0.1153 \pm 0.0027 $ & $ 0.1145 \pm 0.0026 $ & $ 0.1150 \pm 0.0028 $ & $ 0.1152 \pm 0.0024 $ & $ 0.1146\pm 0.0024 $ \\ 
& $ 100\theta $ & $ 1.039 \pm 0.002 $ & $ 1.040 \pm 0.002 $ & $ 1.039 \pm 0.003 $ & $ 1.039 \pm 0.002 $ & $ 1.039 \pm0.002 $ \\ 
& $ \tau $ & $ 0.083 \pm 0.014 $ & $ 0.084 \pm 0.014 $ & $ 0.083 \pm 0.014 $ & $ 0.083 \pm 0.014 $ & $ 0.084 \pm0.014 $ \\ 
& $ n_s $ & $ 0.964 \pm 0.012 $ & $ 0.968 \pm 0.012 $ & $ 0.965 \pm 0.013 $ & $ 0.964 \pm 0.012 $ & $ 0.968\pm 0.011 $ \\ 
& $ \log(10^{10} A_s) $ & $ 3.084 \pm 0.029 $ & $ 3.086 \pm 0.029 $ & $ 3.085 \pm 0.030 $ & $ 3.083 \pm 0.029 $ & $ 3.086\pm 0.029 $ \\ 
& $ \Omega_m $ & $ 0.290 \pm 0.016 $ & $ 0.283 \pm 0.014 $ & $ 0.288 \pm 0.017 $ & $ 0.289 \pm 0.013 $ & $ 0.284\pm 0.012 $ \\ 
& $ H_0 [\km\, \s^{-1}\, \Mpc ^{-1}] $ & $ 68.9 \pm 1.4 $ & $ 69.6 \pm 1.3 $ & $ 69.1 \pm 1.6 $ & $ 69.0 \pm 1.2 $ & $ 69.5 \pm 1.2 $ \\ 
& $ \sigma_8 $ & $ 0.825 \pm 0.017 $ & $ 0.825 \pm 0.017 $ & $ 0.825 \pm 0.017 $ & $ 0.825 \pm 0.017 $ & $ 0.825\pm 0.017 $ \\

\hline
Flat $w$CDM &  $ 100\Omega_b h^2 $ & $ 2.265 \pm 0.062 $ & $ 2.253 \pm 0.057 $ & $ 2.228 \pm 0.055 $ & $ 2.247 \pm 0.056 $ & $ 2.253 \pm 0.056 $ \\ 
& $ \Omega_{DM} h^2 $ & $ 0.1164 \pm 0.0036 $ & $ 0.1146 \pm 0.0030 $ & $ 0.1157 \pm 0.0030 $ & $ 0.1147 \pm 0.0029 $ & $ 0.1148 \pm0.0030 $ \\ 
& $ 100\theta $ & $ 1.039 \pm 0.003 $ & $ 1.039 \pm 0.003 $ & $ 1.038 \pm 0.003 $ & $ 1.039 \pm 0.003 $ & $ 1.039 \pm0.003 $ \\ 
& $ \tau $ & $ 0.084 \pm 0.015 $ & $ 0.084 \pm 0.014 $ & $ 0.082 \pm 0.014 $ & $ 0.084 \pm 0.014 $ & $ 0.084\pm 0.014 $ \\ 
& $ n_s $ & $ 0.975 \pm 0.019 $ & $ 0.968 \pm 0.014 $ & $ 0.962 \pm 0.014 $ & $ 0.967 \pm 0.014 $ & $ 0.968\pm 0.014 $ \\ 
& $ log[10^{10} $ & $ 3.096 \pm 0.031 $ & $ 3.086 \pm 0.030 $ & $ 3.082 \pm 0.029 $ & $ 3.085 \pm 0.030 $ & $ 3.086\pm 0.030 $ \\ 
& $ w $ & $ -0.525 \pm 0.293 $ & $ -1.007 \pm 0.084 $ & $ -1.062 \pm 0.072 $ & $ -0.973 \pm 0.086 $ & $ -1.008\pm 0.085 $ \\ 
& $ \Omega_m $ & $ 0.487 \pm 0.132 $ & $ 0.283 \pm 0.018 $ & $ 0.844 \pm 0.028 $ & $ 0.294 \pm 0.018 $ & $ 0.284 \pm0.018 $ \\ 
& $ H_0 $ & $ 55.2 \pm 8.4 $ & $ 69.7 \pm 2.1 $ & $ 70.5 \pm 2.3 $ & $ 68.4 \pm 2.0 $ & $ 69.7\pm 2.1 $ \\ 
& $ \sigma_8 $ & $ 0.664 \pm 0.110 $ & $ 0.826 \pm 0.032 $ & $ 0.844 \pm 0.028 $ & $ 0.815 \pm 0.033 $ & $ 0.827 \pm0.032 $ \\ 
\hline
$\Lambda$CDM & $100\Omega_{\rm b} h^2$  		& $2.215  \pm 0.055 \,$	&  $2.263  \pm 0.054 \,$		&	$2.256\pm0.054\,$		& $2.252  \pm 0.054 \,$ 		&  $2.262  \pm 0.052 \,$ \\
&$\Omega_{\rm CDM} h^2$  		& $0.1118 \pm 0.0039 \,$	& $0.1162 \pm 0.0039  \,$	&	$0.114\pm0.0042\,$			& $0.1150 \pm 0.0038 \,$	&  $0.1161 \pm 0.0038 \,$\\
&$100\theta$ 				& $1.038 \pm 0.003\,$		& $1.040\pm0.003 \,$		&	$1.040\pm0.003\,$			& $1.040 \pm 0.003\,$		& $1.040 \pm 0.003\,$ \\
& $\tau$ 				& $0.086 \pm 0.014\,$		& $0.088\pm 0.015\,$		&	$0.089\pm0.014$			& $0.088 \pm 0.015\,$		& $0.088 \pm 0.014\,$ \\
&$n_s$ 				& $0.958 \pm  0.013 \,$		& $0.970 \pm 0.013$		&	$0.969\pm0.013$		& $0.968 \pm  0.013 \,$		&  $0.969 \pm  0.013 \,$ \\
&$\log(10^{10} A_s)$		& $3.072\pm  0.031 \,$		& $3.101\pm 0.031 \,$		&$3.096\pm0.031\,$	& $3.096\pm  0.031 \,$		& $3.101\pm  0.030 \,$\\
&$\Omega_m$ 			& $0.454 \pm 0.058 \,$		& $0.287\pm 0.029\,$		&$0.303\pm0.038$	& $0.302 \pm 0.020 \,$		& $0.288\pm0.016\,$\\
&$\Omega_k$			& $-0.046 \pm  0.017$		& $0.001 \pm 0.008\,$		&$-0.005\pm0.012\,$	& $-0.004 \pm 0.006 \,$	& $0.000\pm 0.005 \,$ \\
&$H_0\, [\km\, \s^{-1}\, \Mpc ^{-1}]$& $54.65 \pm  3.8 \,$ 	& $69.86\pm3.6 \,$		& $67.7\pm4.7$	& $67.6 \pm  2.3 \,$ 		& $69.9\pm 3.6\,$\\
&$\sigma_8$ 			& $0.782 \pm  0.024\,$		& $0.838\pm0.023 \,$		& $0.825\pm0.026$				& $0.829 \pm 0.022\,$		& $0.838\pm0.023\,$\\

\hline	
	
$w$CDM & $ 100\Omega_{\rm b} h^2 $ & $ 2.231 \pm 0.058 $ & $ 2.248 \pm 0.059 $ & $ 2.205 \pm 0.055 $ & $ 2.244 \pm 0.057 $ & $ 2.246 \pm 0.056 $ \\ 
& $ \Omega_{\rm CDM} h^2 $ & $ 0.1117 \pm 0.0038 $ & $ 0.1133 \pm 0.0045 $ & $ 0.1094 \pm 0.0041 $ & $ 0.1130 \pm 0.0041 $ & $ 0.1131 \pm  0.0040 $ \\ 
& $ 100\theta $ & $ 1.039 \pm 0.003 $ & $ 1.039 \pm 0.003 $ & $ 1.038 \pm 0.003 $ & $ 1.039 \pm 0.003 $ & $ 1.039 \pm  0.003 $ \\ 
& $ \tau $ & $ 0.086 \pm 0.015 $ & $ 0.084 \pm 0.014 $ & $ 0.083 \pm 0.014 $ & $ 0.084 \pm 0.014 $ & $ 0.084 \pm  0.014 $ \\ 
& $ n_s $ & $ 0.964 \pm 0.015 $ & $ 0.966 \pm 0.014 $ & $ 0.956 \pm 0.013 $ & $ 0.966 \pm 0.014 $ & $ 0.966 \pm  0.014 $ \\ 
& $ \log(10^{10} A_s)$ & $ 3.075 \pm 0.032 $ & $ 3.080 \pm 0.034 $ & $ 3.054 \pm 0.032 $ & $ 3.079 \pm 0.032 $ & $ 3.079 \pm  0.031 $ \\ 
& $ w $ & $ -0.502 \pm 0.160 $ & $ -1.102 \pm 0.219 $ & $ -1.215 \pm 0.117 $ & $ -1.060 \pm 0.177 $ & $ -1.113 \pm 0.169 $ \\ 
& $ \Omega_m $ & $ 0.614 \pm 0.061 $ & $ 0.283 \pm 0.018 $ & $ 0.354 \pm 0.041 $ & $ 0.289 \pm 0.020 $ & $ 0.278 \pm 0.017 $ \\ 
& $ \Omega_k $ & $ -0.052 \pm 0.018 $ & $ -0.005 \pm 0.012 $ & $ -0.032 \pm 0.015 $ & $ -0.005 \pm 0.010 $ & $ -0.005\pm 0.009 $ \\ 
& $ H_0\, [\km\, \s^{-1}\, \Mpc ^{-1}] $ & $ 46.9 \pm 2.6 $ & $ 69.4 \pm 2.1 $ &  $ 61.3 \pm 4.2 $  & $ 68.5 \pm 2.1 $ & $ 69.9 \pm  1.8 $ \\ 
& $ \sigma_8 $ & $ 0.654 \pm 0.051 $ & $ 0.836 \pm 0.040 $ & $ 0.818 \pm 0.028 $ & $ 0.826 \pm 0.040 $ & $ 0.842 \pm  0.037 $ \\
\hline
\end{tabular}
\caption{\label{table:distanceconstraints}Cosmological parameter constraints combining WiggleZ data with other cosmological datasets, for different cosmological models.}
\end{table*}

\subsubsection{Flat $w$CDM} \label{sec:wcdm}

Here we consider models where the equation of state of the dark energy, $w$, is allowed to be different from $-1$ but kept constant with time. Again we only consider flat models.

\begin{figure}
\centering
	\includegraphics[width=0.99\columnwidth]{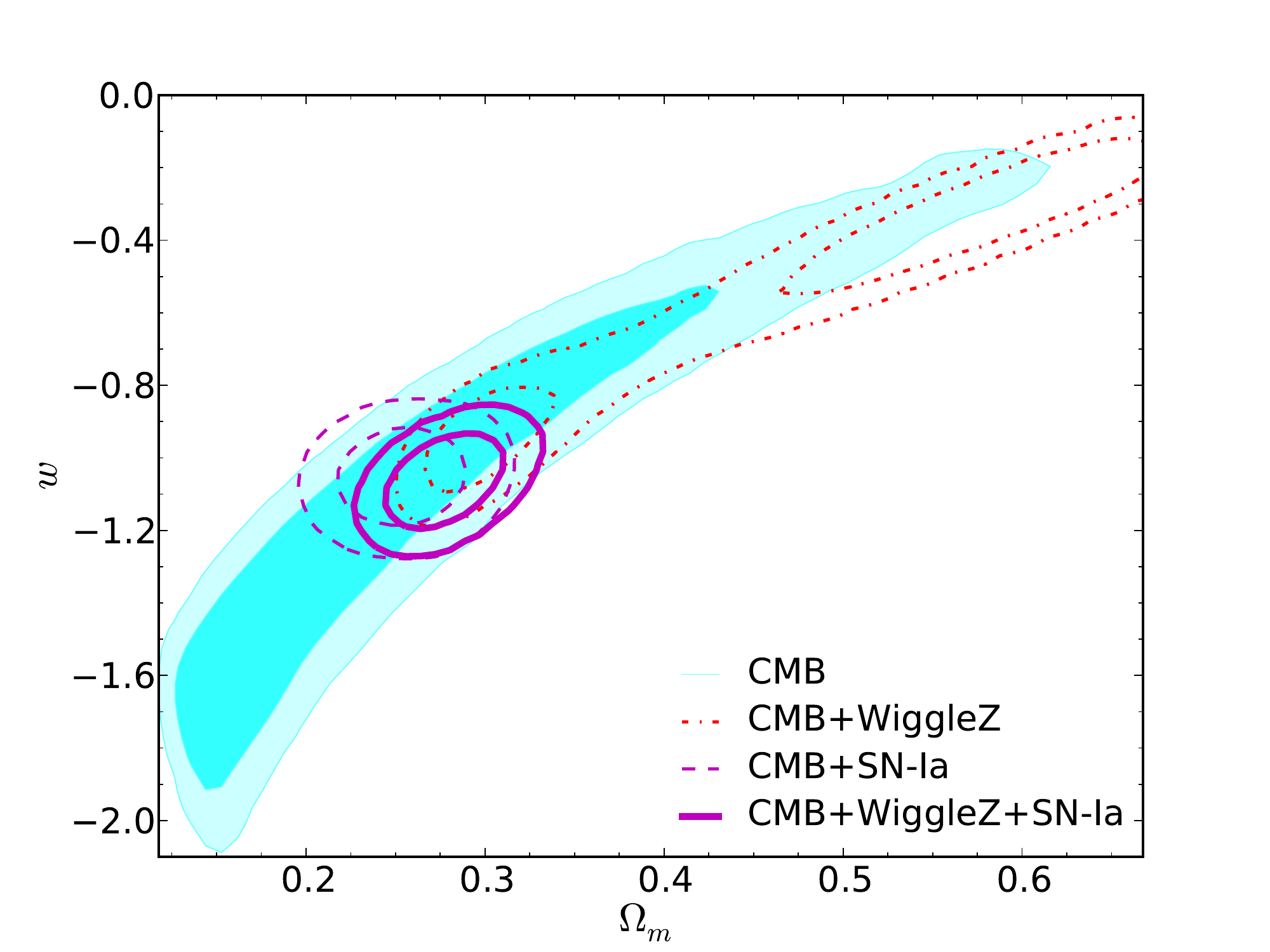}
	\caption{\label{fig:wCDM_om_w_2D}(color online). The two-dimensional constraints in the \{$\Omega_m$, $w$\} plane for a flat $w$CDM model (see legend). The supernovae data (solid dark blue) is far more powerful at measuring distances, having many more data-points spread over a range of redshifts, but the addition of the WiggleZ data (thick red) has an effect on the best fit $\Omega_m$ value, pushing it slightly higher.}
\end{figure}

In addition to CMB + WiggleZ, we also compare constraints on the constant equation of state $w$ using the WiggleZ $P(k)$ with the SNLS Type-Ia Supernova compilation, and then combining the two. The supernova data are far more powerful at measuring distances, having many more data-points spread over a range of redshifts, and so perform much better than the WiggleZ $P(k)$, as shown in Fig. \ref{fig:wCDM_om_w_2D}.

The marginalised constraints are given in table \ref{table:distanceconstraints}.

\subsubsection{$\Lambda$CDM} \label{sec:kcdm}

\begin{figure}
\centering
	\includegraphics[width=0.99\columnwidth]{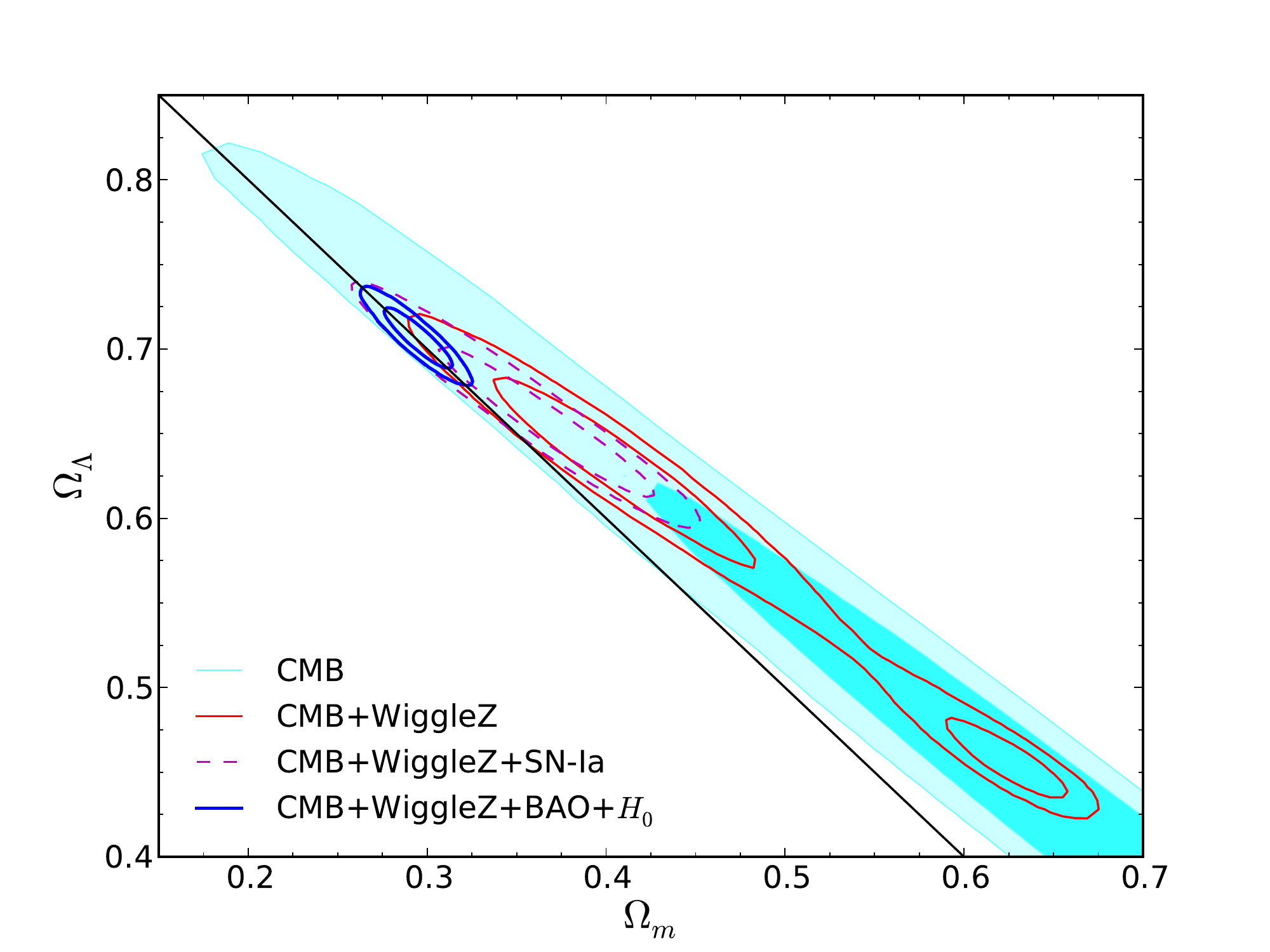}
	\caption{\label{fig:kCDM2D}(color online). The two-dimensional constraints in the \{$\Omega_m$, $\Omega_{\Lambda}$\} plane for a $\Lambda$CDM model, for a number of different data compilations (see legend). The straight black line gives the location of $\Omega_m+\Omega_{\Lambda}=1$, i.e. a flat universe. The WiggleZ data are consistent with the CMB data (from WMAP) with a peak likelihood close to flatness, but have a second likelihood peak for larger values of $\Omega_m$, and away from the flatness line.}
\end{figure}

Here we consider models where the curvature is allowed to be different from flat, allowing $\Omega_k$ to be non-zero. The marginalised constraints are given in table \ref{table:distanceconstraints}. The two-dimensional parameter constraint contours for  $\{\Omega_m,~\Omega_\Lambda\}$ are shown in Fig. \ref{fig:kCDM2D}. 

\begin{figure}
\centering
	\includegraphics[width=0.99\columnwidth]{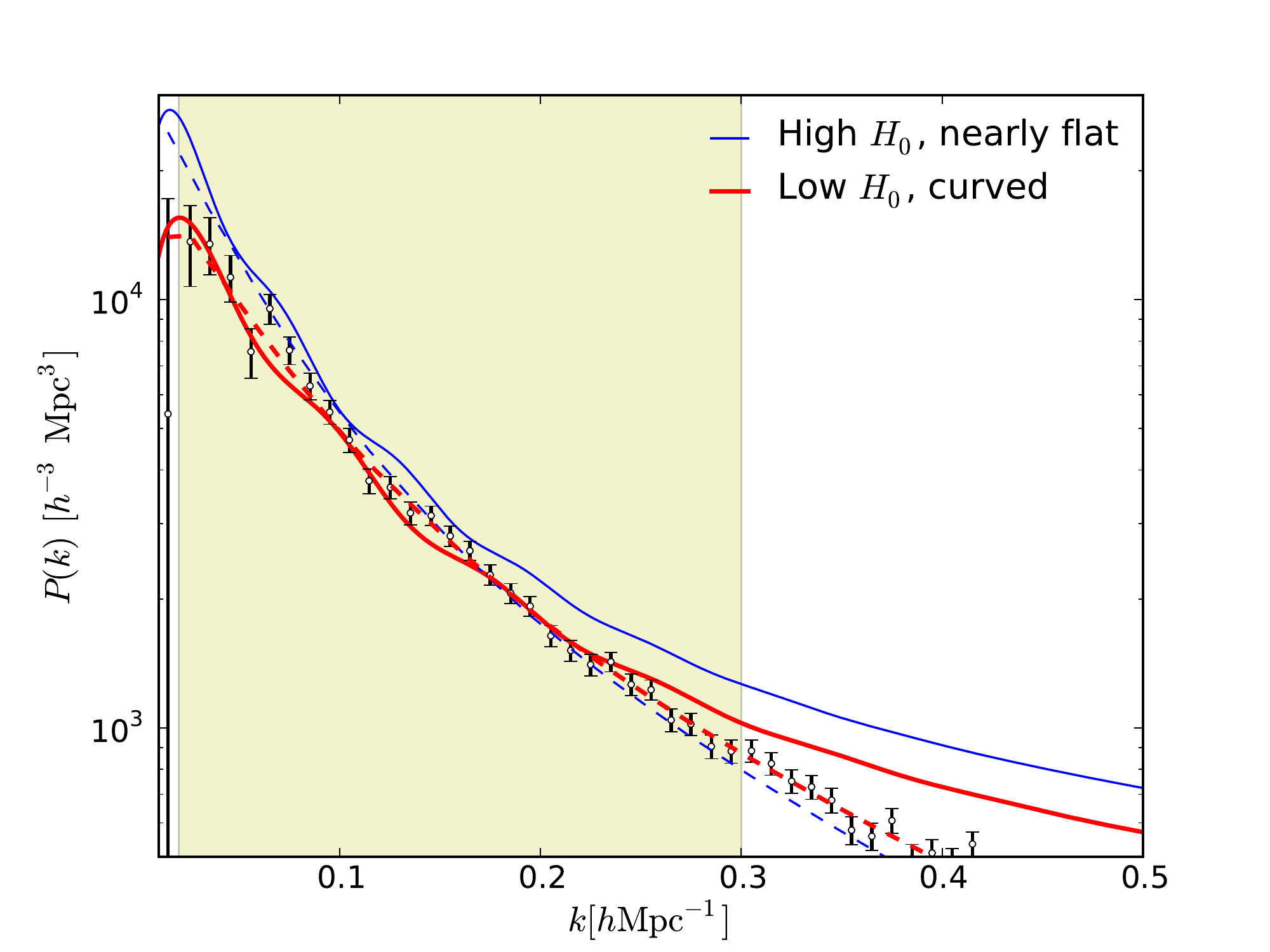}
	\caption{ \label{fig:curved_models_pk} (color online). The matter power spectrum for two different sets of cosmological parameters, both of which are good fits to the $\Lambda$CDM model using CMB + WiggleZ data. The solid curves give the predicted linear $P(k)$. The dashed curves give are the same $P(k)$ after the non-linear effects and window functions have been applied, to compare to the data at $z=0.6$ (black error bars).}
\end{figure}

We find two, slightly separated, high-likelihood peaks. One of these is close to flatness, with a matter density $\Omega_m = 0.29$ and Hubble parameter $h =0.69$. The other is slightly further away, with matter density $\Omega_m =0.52  $ and Hubble parameter  $h = 0.50$. These are allowed by the CMB due to the extra dynamical freedom obtained by letting the Universe move away from flatness. The high matter density peak is not ruled out by including the WiggleZ data, as both of these parameter sets have the same broad-spectrum shape. We show this by plotting the relevant power spectrum for the two peaks in Fig. \ref{fig:curved_models_pk}. Here we show both the initial linear power spectra and final prediction power spectra (convolved with the region-averaged window function at $z=0.6$) in comparison with the data, and show that both models are good fits. However, if we were able to precisely determine the position of the turnover in the matter power spectrum, we would be able to rule out one of these models. (For further discussion of using the WiggleZ data to determine the turnover position and using the turnover to measure the matter density, $\Omega_m$, see reference \citep{Poole2012}.) We can also rule out the high $\Omega_m$, low $H_0$ peak using other data, such as distance data from SN-Ia or BAO, or $H_0$ measurements. This is shown in Fig. \ref{fig:kCDM2D}, where the second peak vanishes once the other data sets are applied.

\begin{figure}
\centering
	\includegraphics[width=0.99\columnwidth]{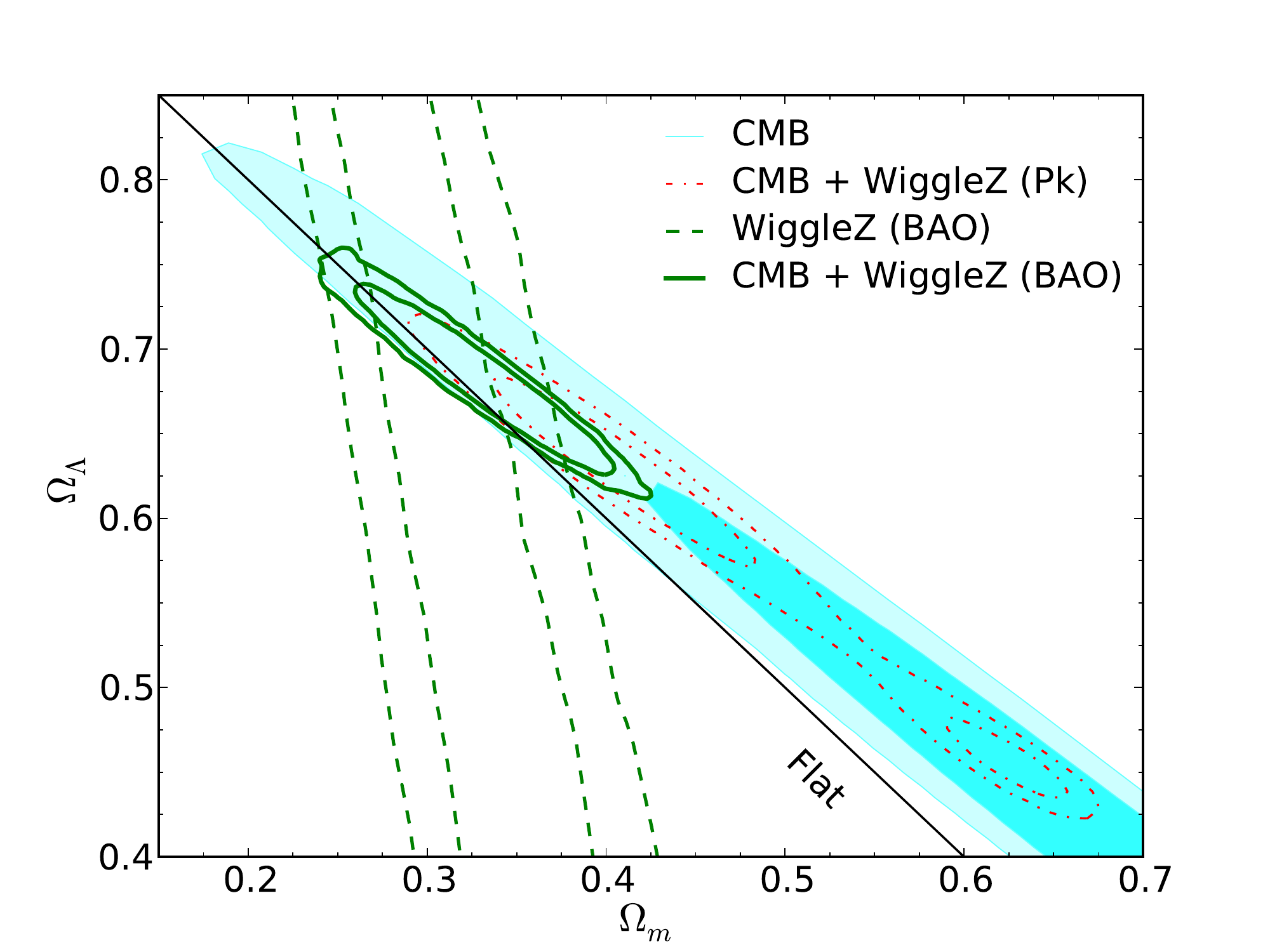}
	\caption{\label{fig:kCDM2D_compare}(color online). The two-dimensional constraints in the \{$\Omega_m$, $\Omega_{\Lambda}$\} plane for a $\Lambda$CDM model, comparing the WiggleZ $P(k)$ data with the WiggleZ BAO data (see legend). The straight black line gives the location of $\Omega_m+\Omega_{\Lambda}=1$, i.e., a flat universe. The WiggleZ $P(k)$ data are consistent with the BAO from WiggleZ, for the peak likelihood close to flatness, but not for the second peak at high $\Omega_m$ values.}
	
\end{figure}

We also compare the constraints in the  \{$\Omega_m$, $\Omega_{\Lambda}$\} plane between the WiggleZ $P(k)$ data and the WiggleZ BAO, as shown in Fig. \ref{fig:kCDM2D_compare}. We see that only one of the high likelihood peaks in the $P(k)$ result is favoured by the WiggleZ BAO measurement.

\subsubsection{$w$CDM} \label{sec:owcdm}

\begin{figure*}
\centering
	\includegraphics[width=0.99\columnwidth]{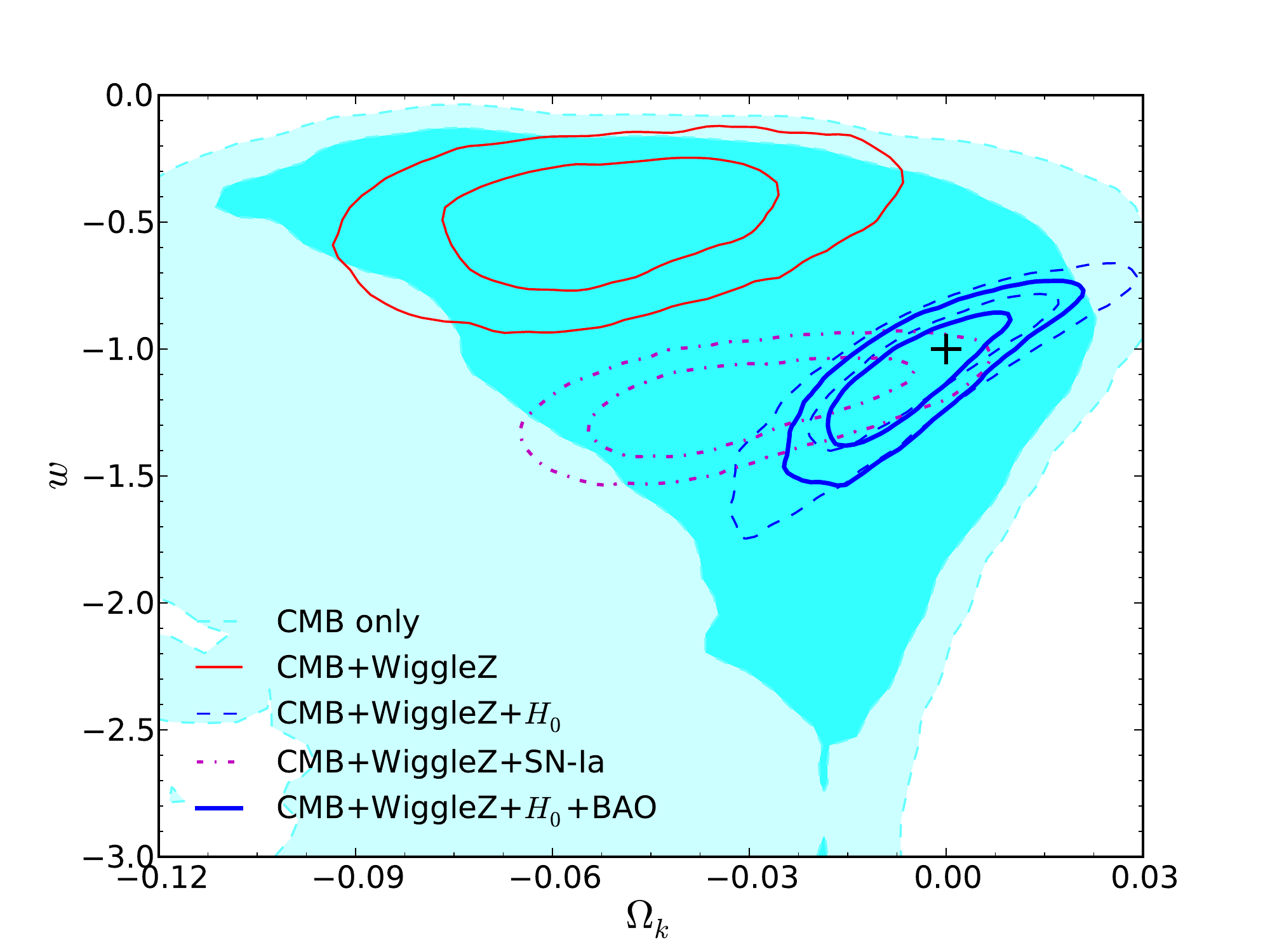}
	\includegraphics[width=0.99\columnwidth]{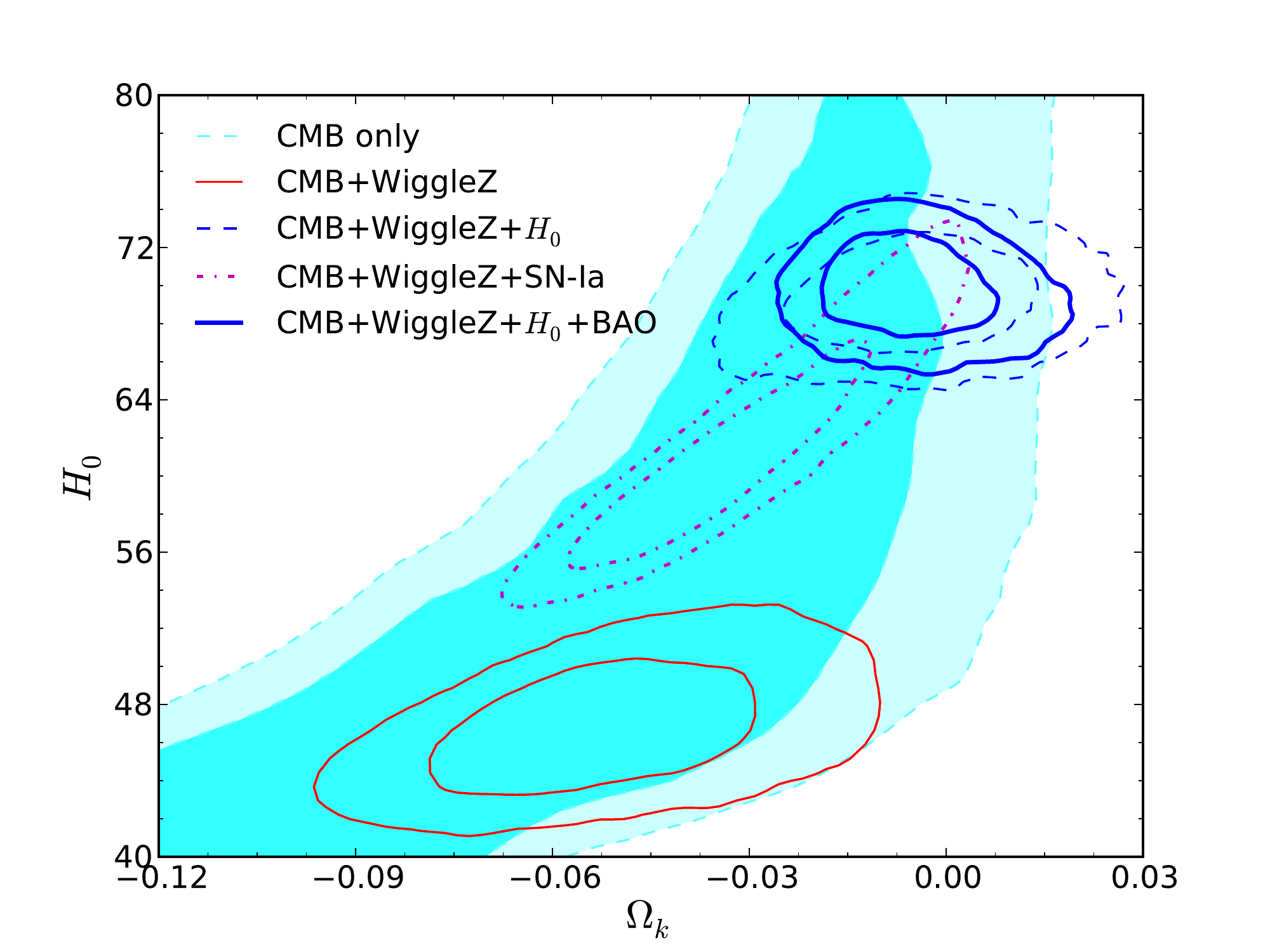}
	\caption{\label{fig:wkCDM2D} (color online). The two-dimensional constraints in the \{$\Omega_k$, $w$\} plane (left) and \{$\Omega_k$, $w$\} plane (right) for a $w$CDM model, using a number of different datasets (see legend). The solid black cross gives the location of Flat $\Lambda$CDM, i.e. a flat universe with the equation of state $w=-1$. }
	
\end{figure*}

Here we consider models where both the equation of state of the dark energy, $w$, is allowed to be different from $-1$ (but still constant with time) and also the curvature is allowed to be different from flat, allowing $\Omega_k$ to be non-zero. Dark energy dynamics (a constant equation of state different from $-1$ or an equation of state which varies with scale factor) has a somewhat degenerate signal with small curvature values \citep{Clarkson2007}. If we were to detect $w\neq-1$, it would be necessary to test both the equation of state and curvature simultaneously, to be sure that we had actually detected time-evolving dark energy (and so falsified the cosmological constant as the explanation for the late-time acceleration).

The marginalised constraints are given in table \ref{table:distanceconstraints}. The two-dimensional parameter constraint contours for  $\{\Omega_k,~w\}$ and $\{\Omega_k,~H_0\}$ are shown in Fig. \ref{fig:wkCDM2D}. We find that the CMB + WiggleZ data by themselves prefer a peak far away from the $w=-1$, $\Omega_k=0$ values of flat $\Lambda$CDM. This peak is similar to the low $H_0$, high $\Omega_m$ peak that appears in the analysis of the $\Lambda$CDM model, as it has the same similar best fit values for those parameters. As shown in Fig. \ref{fig:curved_models_pk}, the WiggleZ $P(k)$ data by itself cannot distinguish between these two sets of parameter values, and requires extra distance information in order to break the degeneracy. However, for this model, the $\Lambda$CDM point in the \{$\Omega_k$, $w$\} plane is disfavoured at more than 2-sigma. While we could consider this as marginal detection of some new physics, it is more likely to be a coincidence, randomly occurring due to the number of different data sets and models under consideration in this paper. It could also be a product of some systematic error, such as calibration of the selection function, population of haloes with WiggleZ galaxies in the GiggleZ simulations, or non-linear model fitting.  The option that it is a result of coincidence or systematic error rather seems especially true, given that most other datasets strongly disfavour this peak (the very low value of $H_0$ is devoured by the $H_0$ data for example). We discuss the goodness of fit improvement for this model, and consistency with other datasets, in section \ref{sec:modelcomparison}.

When the CMB + WiggleZ is combined with other distance data, the constraints on both $w$ and $\Omega_k$ are only twice as large as when the parameters are considered individually. The best constraints on both parameters come from the CMB + WiggleZ + BAO + $H_0$ combination, where $w=-1.11 \pm 0.17$ and $\Omega_k = -0.005 \pm 0.009$. This demonstrates that combining CMB, $P(k)$ and other distance data together provides reasonable constraints on these two extension parameters in combination.

\subsubsection{Massive neutrinos} \label{sec:mnu}

Recent particle physics experiments measuring neutrino oscillations have show that mass differences exist between the different species \citep{Fukuda:1998, Amsler:2008}, but there is no current experiment that can measure the absolute neutrino mass. The massive neutrinos provide a form of `Warm Dark Matter', suppressing structure formation on large scales. Measurements of large scale structure can constrain the total neutrino density, and so the sum of the particle masses. Though we fail to directly detect the neutrino mass, we find the total mass of all three species, $\sum m_\nu$, to be less than 0.58eV at 95\% confidence when combining WiggleZ with the CMB. This upper limit decreases to 0.34eV when we include distance data (BAO and $H_0$).

\begin{figure*}
\centering
	\includegraphics[width=0.99\columnwidth]{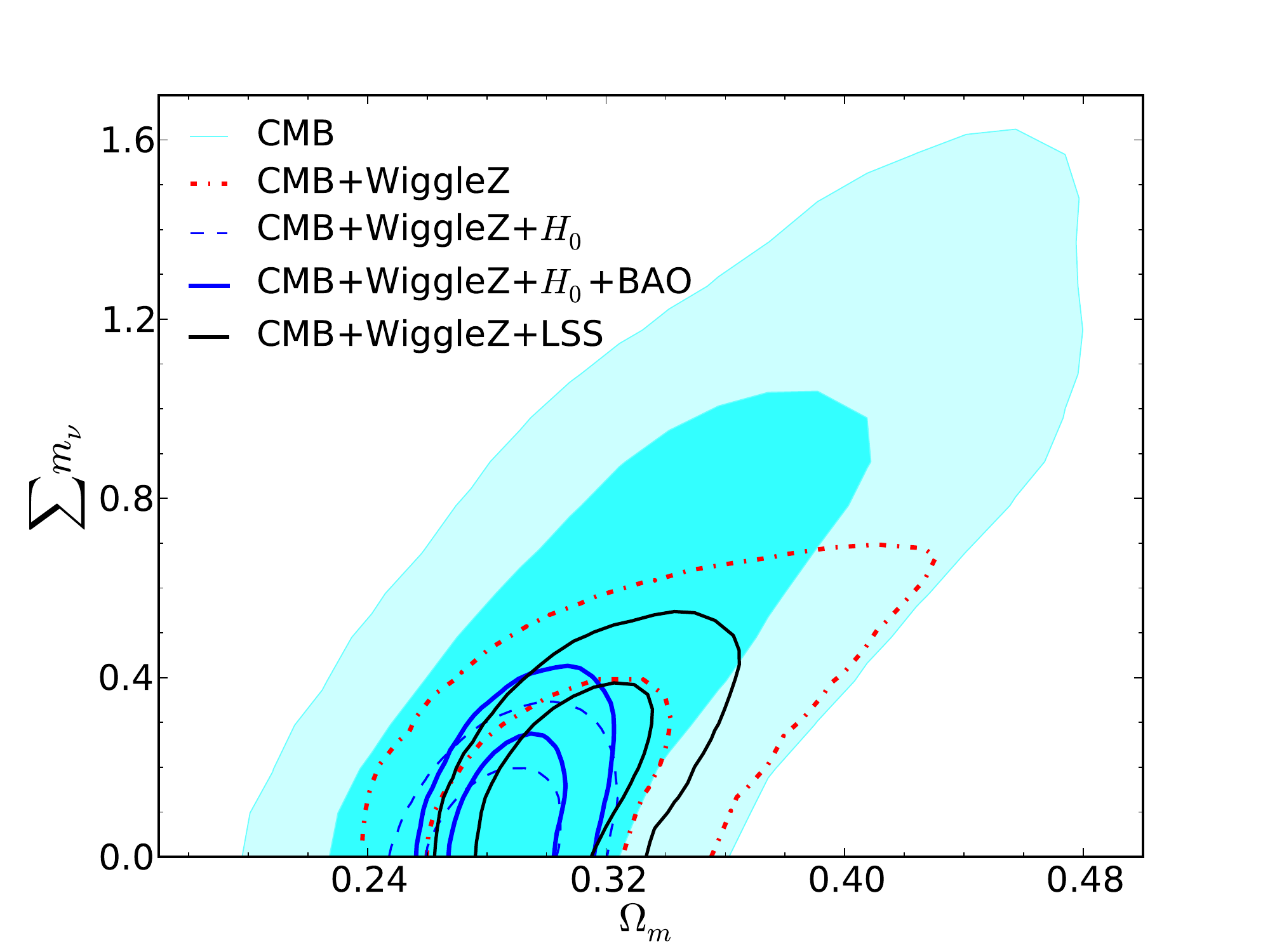}
	\includegraphics[width=0.99\columnwidth]{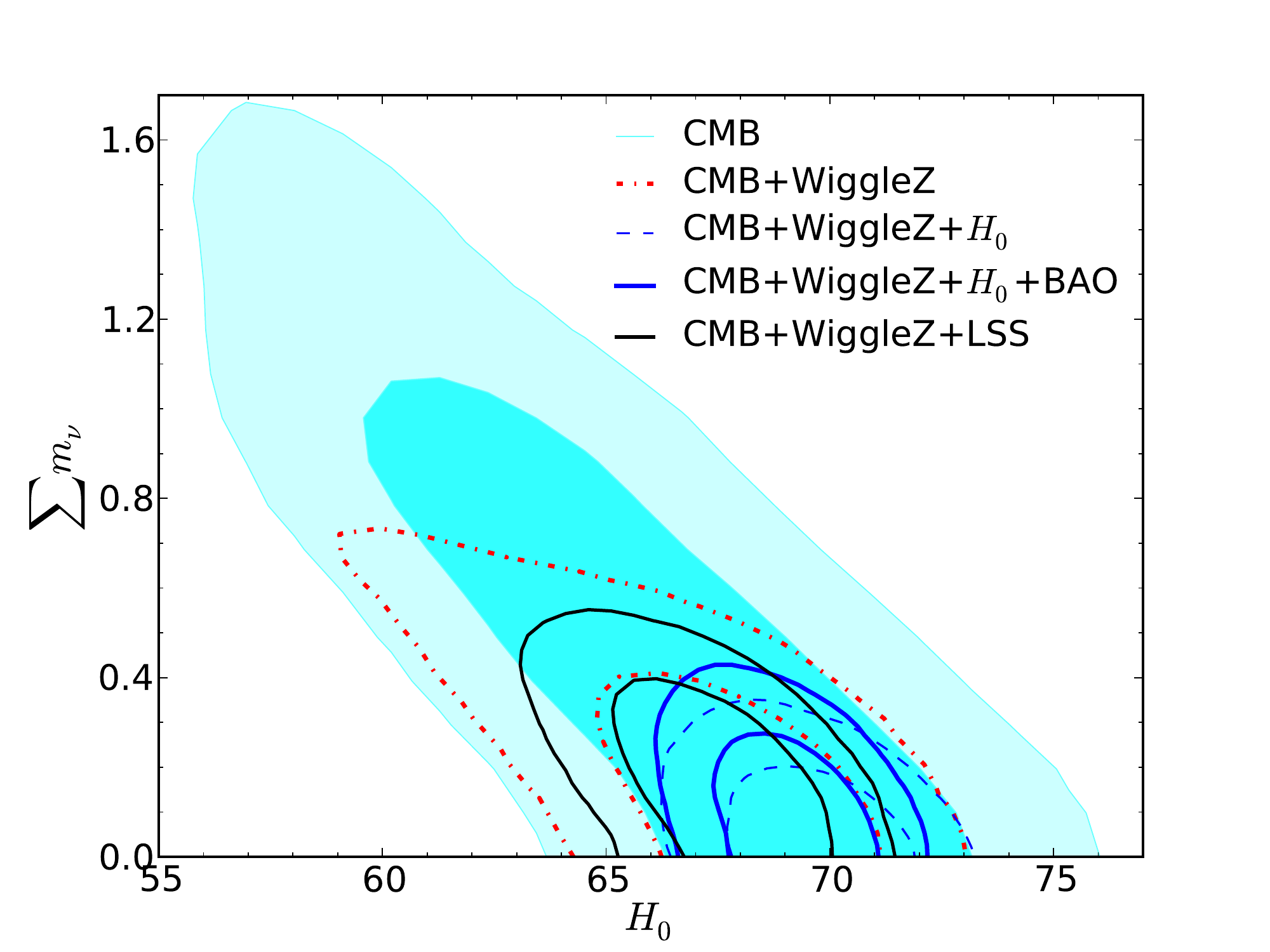}
	\caption{\label{fig:mnu2D} (color online). ({\it Left)} The two-dimensional constraints in the \{$\sum m_\nu$, $\Omega_m$\} ({\it Left}) and  \{$\sum m_\nu$, $H_0$\} ({\it Right}) planes for a massive neutrinos model, using a number of different data combinations (see legend).  }
	
\end{figure*}

Some of these results were first presented in \citet{riemersorensen:2011} using importance sampling of the WMAP MCMC chains. Here we present the output of the CosmoMC module and note that the two analyses give consistent results. The contours of the \{$H_0,~\sum m_\nu$\} and \{$\Omega_m,~\sum m_\nu$\} planes are shown in Fig. \ref{fig:mnu2D}. The marginalised constraints are given in table \ref{table:pertconstraints}.

\begin{table*}
\begin{tabular}{l l c c c c c} \hline
Model & Parameter 			& CMB + WiggleZ 			& + $H_0$					& + LSS 			& + BAO					& + $H_0$ + BAO \\ \hline
$\Lambda$CDM + $ m_\nu$& $ 100\Omega_b h^2 $ & $ 2.240 \pm 0.054 $ & $ 2.264 \pm 0.051 $ & $
2.244 \pm 0.052 $ & $ 2.250 \pm 0.051 $ & $ 2.266 \pm 0.051 $ \\
& $ \Omega_{DM} h^2 $ & $ 0.1183 \pm 0.0052 $ & $ 0.1157 \pm 0.0030 $
& $ 0.1179 \pm 0.0031 $ & $ 0.1161 \pm 0.0027 $ & $ 0.1153 \pm  0.0025
$ \\
& $ 100\theta $ & $ 1.039 \pm 0.003 $ & $ 1.040 \pm 0.002 $ & $ 1.039
\pm 0.002 $ & $ 1.039 \pm 0.002 $ & $ 1.040 \pm  0.002 $ \\
& $ \tau $ & $ 0.084 \pm 0.014 $ & $ 0.086 \pm 0.014 $ & $ 0.083 \pm
0.014 $ & $ 0.085 \pm 0.014 $ & $ 0.086 \pm  0.014 $ \\
& $ n_s $ & $ 0.964 \pm 0.013 $ & $ 0.971 \pm 0.012 $ & $ 0.965 \pm
0.012 $ & $ 0.967 \pm 0.012 $ & $ 0.971 \pm 0.012 $ \\
& $ log[10^{10} $ & $ 3.089 \pm 0.030 $ & $ 3.090 \pm 0.030 $ & $
3.088 \pm 0.030 $ & $ 3.087 \pm 0.030 $ & $ 3.089 \pm  0.030 $ \\
& $ f_\nu $ & $ 0.042 $ & $ 0.029 $ & $ 0.034 $ & $ 0.029 $ & $ 0.026$ \\
& $ \sum m_{\nu} $ & $ 0.58 $ & $ 0.32 $ & $ 0.45 $ & $ 0.38 $ & $ 0.34 $ \\
& $ \Omega_m $ & $ 0.316 \pm 0.042 $ & $ 0.293 \pm 0.018 $ & $ 0.311
\pm 0.021 $ & $ 0.298 \pm 0.016 $ & $ 0.291 \pm  0.014 $ \\
& $ H_0\, [\km\, \s^{-1}\, \Mpc ^{-1}] $ & $ 66.9 \pm 2.7 $ & $ 68.8 \pm 1.5 $ & $ 67.2 \pm
1.7 $ & $ 68.2 \pm 1.3 $ & $ 68.9 \pm  1.2 $ \\
& $ \sigma_8 $ & $ 0.787 \pm 0.030 $ & $ 0.798 \pm 0.026 $ & $ 0.791
\pm 0.027 $ & $ 0.794 \pm 0.026 $ & $ 0.799 \pm 0.025 $ \\

\hline
$\Lambda$CDM + $n_{\rm run}$& $100\Omega_{\rm b} h^2$  & $ 2.167 \pm 0.070 $ & $ 2.237 \pm 0.064 $ & $ 2.188 \pm 0.058 $ & $ 2.206 \pm 0.055 $ & $ 2.249 \pm 0.055 $ \\ 
& $\Omega_{\rm CDM} h^2$  & $ 0.1219 \pm 0.0056 $ & $ 0.1185 \pm 0.0050 $ & $ 0.1201 \pm 0.0035 $ & $ 0.1181 \pm 0.0032 $ & $ 0.1171 \pm  0.0035 $ \\ 
& $ 100\theta $ & $ 1.039 \pm 0.003 $ & $ 1.040 \pm 0.003 $ & $ 1.039 \pm 0.003 $ & $ 1.039 \pm 0.003 $ & $ 1.040 \pm  0.002 $ \\ 
& $ \tau $ & $ 0.090 \pm 0.016 $ & $ 0.091 \pm 0.016 $ & $ 0.090 \pm 0.016 $ & $ 0.091 \pm 0.016 $ & $ 0.092 \pm  0.016 $ \\ 
& $ n_s $ & $ 0.915 \pm 0.032 $ & $ 0.951 \pm 0.029 $ & $ 0.926 \pm 0.024 $ & $ 0.935 \pm 0.022 $ & $ 0.958 \pm  0.023 $ \\ 
& $ n_{run} $ & $ -0.040 \pm 0.022 $ & $ -0.016 \pm 0.021 $ & $ -0.033 \pm 0.018 $ & $ -0.027 \pm 0.017 $ & $ -0.011  \pm 0.018 $ \\ 
& $ \log(10^{10} A_s)$ & $ 3.113 \pm 0.035 $ & $ 3.113 \pm 0.035 $ & $ 3.108 \pm 0.034 $ & $ 3.106 \pm 0.034 $ & $ 3.110 \pm  0.035 $ \\ 
& $ \Omega_m $ & $ 0.332 \pm 0.039 $ & $ 0.304 \pm 0.030 $ & $ 0.318 \pm 0.022 $ & $ 0.305 \pm 0.018 $ & $ 0.295 \pm  0.019 $ \\ 
& $ H_0 \, [\km\, \s^{-1}\, \Mpc ^{-1}]$ & $ 66.0 \pm 2.6 $ & $ 68.2 \pm 2.3 $ & $ 66.9 \pm 1.7 $ & $ 67.9 \pm 1.5 $ & $ 68.9  \pm 1.6 $ \\ 
& $ \sigma_8 $ & $ 0.839 \pm 0.020 $ & $ 0.842 \pm 0.020 $ & $ 0.835 \pm 0.018 $ & $ 0.832 \pm 0.018 $ & $ 0.840  \pm 0.019 $ \\
\hline
$\Lambda$CDM + $r$ & $100\Omega_{\rm b} h^2$  		& $2.268  \pm 0.063 \,$	& $2.296\pm0.060\,$		& $2.257  \pm 0.00056 \,$		& $2.268 \pm 0.056 \,$ 		& $2.288\pm0.053\,$ \\
& $\Omega_{\rm CDM} h^2$  		& $0.1157 \pm 0.0036 \,$	& $0.1142\pm0.0033\,$		& $0.1168 \pm 0.0029 \,$		& $0.1158 \pm 0.0028\,$		& $0.1149\pm0.0027 \,$\\
& $100\theta$ 				& $1.039 \pm 0.003\,$		& $1.040\pm0.003 \,$		& $1.039 \pm 0.003 \,$			& $1.040 \pm 0.003\,$		& $1.04\pm0.002 \,$ \\
& $\tau$ 				& $0.086 \pm 0.014\,$		& $0.089\pm0.015\,$		& $0.084 \pm 0.014\,$			& $0.086 \pm 0.014\,$		& $0.0880\pm0.014\,$ \\
& $n_s$ 				& $0.974 \pm  0.017 \,$		& $0.982\pm0.016$			& $0.971 \pm  0.014 \,$			& $0.974 \pm 0.014\, $		& $0.979\pm0.014\,$ \\	
& $\log(10^{10} A_s)$		& $3.098\pm  0.030 \,$		& $3.101\pm0.030\,$			&  $3.096\pm  0.030 \,$			& $3.098\pm 0.030\,$		&$3.101\pm0.030\,$ \\
& $r$					& $<0.21 \,$				& $<0.23\,$		& $<0.18 \,$		& $<0.20\,$		&$<0.21\,$ \\
& $\Omega_m$ 			& $0.289 \pm 0.023 \,$		& $0.277\pm0.019\,$		& $0.297 \pm 0.018 \,$			& $0.289 \pm 0.016 \,$		& $0.282\pm0.015\,$\\
& $H_0\, [\km\, \s^{-1}\, \Mpc ^{-1}]$& $69.3 \pm  3.6 \,$ 	& $70.5\pm1.9 \,$		& $68.6 \pm  1.61 \,$ 			& $69.3 \pm 1.5 \,$		& $70.0\pm1.4\,$\\
& $\sigma_8$ 			& $0.835 \pm  0.018\,$		& $0.833\pm0.018 \,$		& $0.837 \pm  0.017\,$			& $0.836 \pm 0.018\,$		& $0.835\pm0.018\,$\\		
\hline
\end{tabular}
\caption{\label{table:pertconstraints}Cosmological parameter constraints combining WMAP + WiggleZ data with other cosmological datasets, assuming a flat $\Lambda$CDM model with some extra component. Uncertainties are $1\sigma$ and upper limits are $95\%$ confidence level.}
\end{table*}

It is worth noting how unconstrained $\Omega_m$ becomes from WMAP + WiggleZ when allowing the neutrino mass to vary. This is because there are now effectively two types of Dark Matter (ordinary CDM and massive neutrinos acting as Warm Dark matter), and the neutrinos can play some of the role in structure formation normally taken by CDM. This degeneracy is best broken by adding the BAO scale, as the extra distance data give better constraints on the combined matter density, and so reduce the size of the role neutrinos can play.

The best constraints come from the CMB + WiggleZ + $H_0$ data compilation, and adding the BAO data on top of this actually worsens the constraints, increasing the 95\% upper limit on $\sum m_{\nu}$ from 0.32eV to 0.34eV. This is due to the important role the measurement of the Hubble parameter plays in the analysis for this model, and the tension between these two datasets. The higher $H_0$ measurement from \citet{Riess:2009} gives a smaller best fit matter density, and such cosmologies generate less structure. As such the damping effect of massive neutrinos has to be smaller, and so the neutrino mass constraint becomes tighter. With the BAO constraint, the matter density is larger, and so more neutrino damping allowed, leading to a looser constraint. Combining the two datasets leads to a slightly worse fit than might be expected because of this tension, and consequently does not improve the mass constraint.

There are two possible effects of neutrinos we could consider. Here we have addressed the mass of the neutrino, but recently there has also been a slight preference for more than three neutrinos species \cite{Komatsu:2009,Komatsu:2010, Dunkley:2011}. We consider this subject in an upcoming paper.

\subsubsection{Running of the power spectrum} \label{sec:running}

The primordial power spectrum of density perturbations may not be a simple power law, where the spectral index, $n_s$, is independent of wavenumber, $k$. Instead the power spectrum may have some running \citep{KosowskyTurner95}, such that it is now parameterised as 
\begin{equation}
P(k) = A\left(\frac{k}{k_*}\right)^{(n_s-1)+n_{\rm run}\ln(k/k_*)} \,,
\end{equation}
where $n_{\rm run}= d\ln n_s/d\ln k$, and $k_*$ is some pivot scale. The presence of a large running of the spectral index, $n_{\rm run} \ge (n_s-1)^2$, is a departure from the predictions of ordinary single-field slow-roll inflation, and requires a more complex inflationary history (e.g., multiple scalar fields \citep{Dias2012} or a non-miminal coupling between the inflaton and Ricci tensor \citep{Li2006}).

We consider a model where the running is allowed to vary, rather than being held fixed at $n_{\rm run}=0$. The marginalised constraints are given in table \ref{table:pertconstraints}.

\begin{figure}
\centering
	\includegraphics[width=0.99\columnwidth]{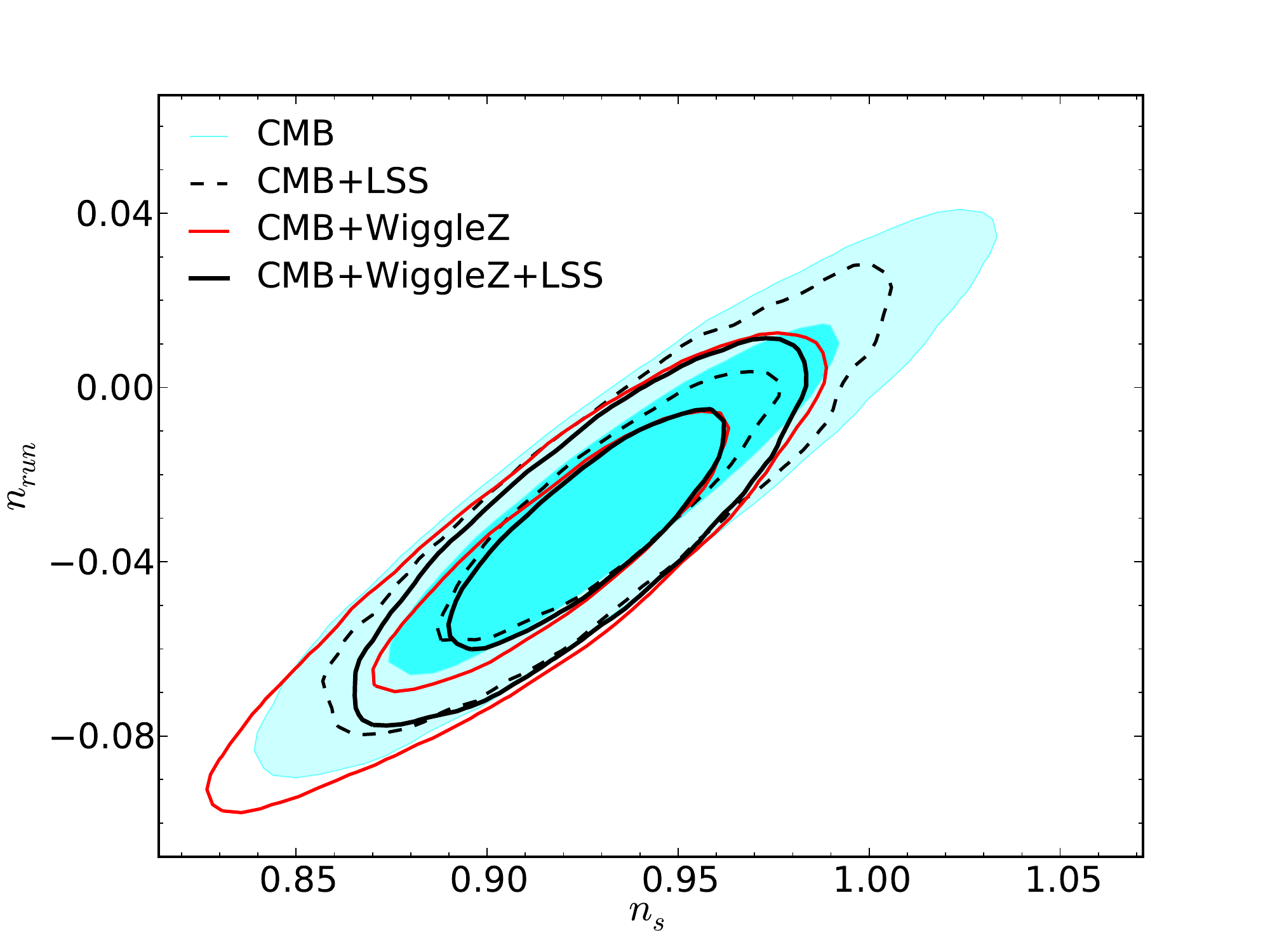}
	\caption{\label{fig:run2D}(color online). The two-dimensional constraints in the \{$n_s$, $n_{\rm run}$\} plane. Adding the WiggleZ data reduces the size of the error bars with a best fit value for the running, which is more negative than for WMAP alone ($n_{\rm run} =-0.033$ for CMB only, whereas it is  $n_{\rm run} =-0.040$ for CMB + WiggleZ).} 
	
\end{figure}

The two-dimensional parameter constraint contours for  $\{ n_s,~ n_{\rm run}\}$ using CMB data alone, and combining CMB with the WiggleZ and other LSS data are shown in Fig. \ref{fig:run2D}. The WiggleZ data drives the running to slightly more negative values, though CMB + WiggleZ is consistent with zero running within two-sigma ($n_{\rm run} = -0.04 \pm 0.022$). This result is driven by the large scale structure data, as the addition of this extra parameter allows the combination of CMB and WiggleZ to be consistent with a slightly larger value of the matter density, $\Omega_m=0.332\pm0.039$. This is similar to, but less extreme than an effect we have seen before, in section \ref{sec:kcdm}, when considering curvature. Combining with other datasets reduces the significance of this result, as the matter density is pushed back to lower values. Similarly, measurements of the matter power spectrum turnover would also allowed such large negative running models to be falsified.

\subsubsection{Tensors} \label{sec:grav}

We also consider models where the primordial power spectrum is composed of both scalar and tensor perturbations. Tensor perturbations are generated by primordial gravitational waves from cosmological inflation (for reviews see \citep{Linde2005,MFB1992,LiddleLyth,Bassett2006,LythLiddle}). We parameterise the tensor contribution through the tensor to scalar amplitude ratio, defined as
\be
r \equiv \frac{A_T}{A_S} \,,
\ee
where $A_s$ is the amplitude of scalar perturbations at some pivot scale $k_*$, and $A_T$ is the amplitude of tensor perturbations at that same scale. If inflation is the only mechanism for generating both the scalar and tensor perturbations (which is the case in single-field, slow-roll inflation), there is a consistency relation between the tensor to scalar ratio and the spectral index of the tensor power spectrum, $n_T$, given by
\be
r = -8 n_T\,.
\ee

In this analysis we have assumed the inflationary consistency relation, and only allowed the tensor to scalar ratio parameter to vary, as well as the other cosmological parameters in the $\Lambda$CDM model.
\begin{figure}
\centering
	\includegraphics[width=0.99\columnwidth]{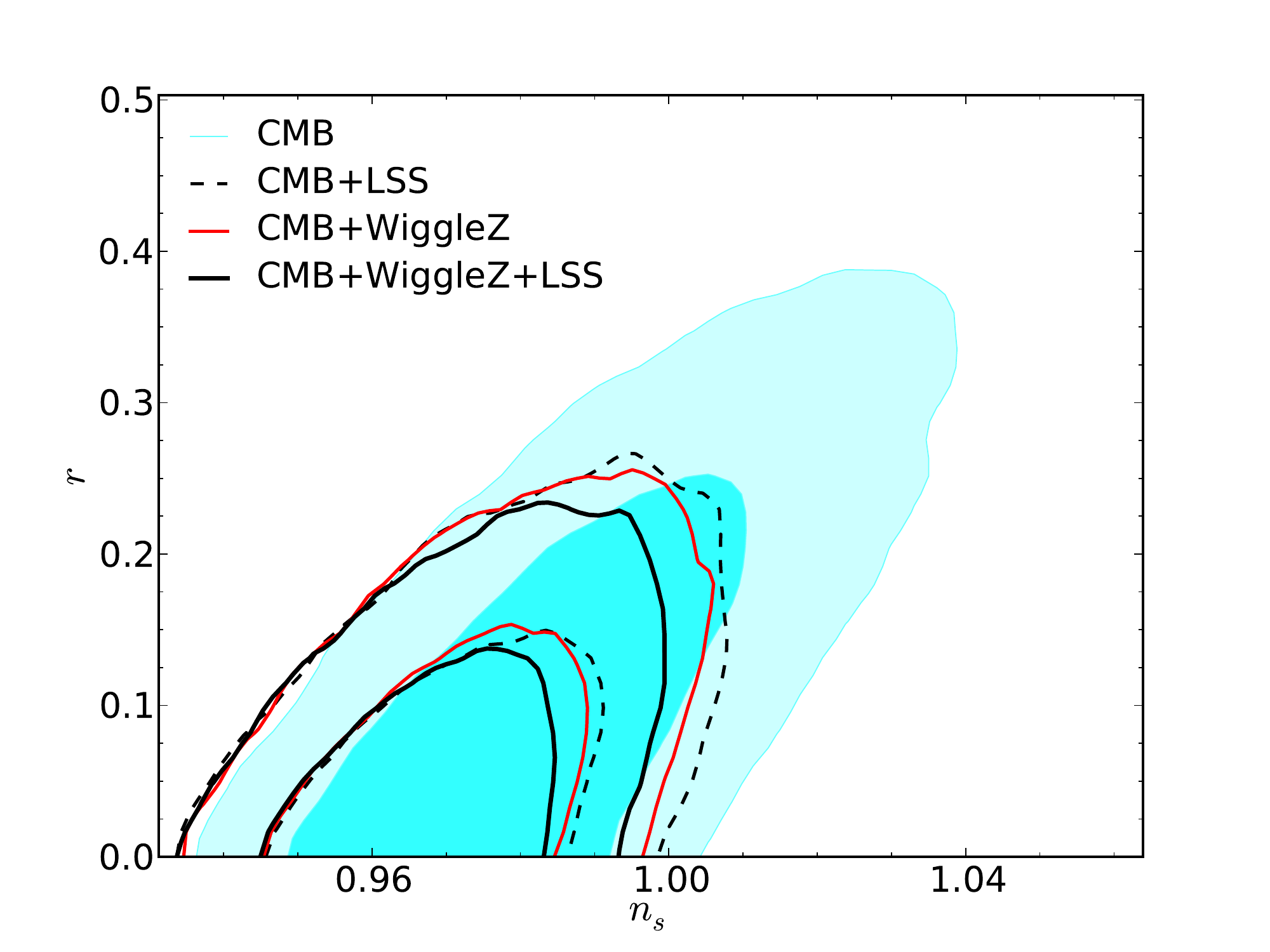}
	\caption{\label{fig:tensor2D}(color online). The two-dimensional constraints in the \{$n_s$, $r$\} plane, for a number of data combinations (see legend). We see that that the CMB + WiggleZ improves on the upper limit from CMB alone, and gives about the same limit as CMB + LSS. Combining all three, CMB + WiggleZ + LSS gives the best constraint.} 
	
\end{figure}

The marginalised constraints are given in table \ref{table:pertconstraints}. The two-dimensional parameter constraint contours for  $\{ n_s,~ r\}$ are shown in Fig. \ref{fig:tensor2D}. We do not detect the tensor contribution, but do improve the 95\% upper limit on it from $r < 0.36$ (CMB alone) to $r < 0.21$ (CMB + WiggleZ). This is very similar to constraints that come from CMB combined with other LSS data, and the combined CMB + WiggleZ + LSS data give the best constraints, of $r<0.18$.

\subsection{Comparison of all models}\label{sec:modelcomparison}

In table \ref{table:chisquarevalues}, we give the best fit log-likelihood ($\mathcal{L}$) values for the different models and dataset combinations considered in this paper. The increase in $-\log(\mathcal{L})$ when the extra datasets are added to the CMB are consistent with what is expected for Gaussian data-points with that many degrees of freedom (0.5 for $H_0$, 1 for BAO, 236 for SN-Ia, 406 for WiggleZ).  

While we do not use the log-likelihood as a model selection statistic, we notice that the WiggleZ data (in combination with the other data we have used) is equally likely when considered against all models. The only exception is the $w$CDM model (where the Universe is allowed to be nonflat), which provides a much better fit than any of the other models when compared against the CMB + WiggleZ data only. However, this improvement in fit is removed when another distance data set is added, and so is likely due to some undiagnosed systematic or simply some random coincidence.

Since the best fit log-likelihoods of all the models are approximately equal for a given data combination, any model selection power will come from the prior probability. A simple model selection criteria would be to approximate the prior by simply penalising each model best fit log-likelihood by the number of free parameters in the model (e.g., the Akaike Information Criteria, or AIC \citep{Akaike:1974}). In this case, the model with the least free parameters could be considered the best, which here would be flat $\Lambda$CDM. This conclusion is consistent with we see from the parameter constraints, where the limiting case of flat $\Lambda$CDM is always a good fit.

\begin{table*}
\begin{tabular}{l c c c c c c c c} \hline
Model 	& Number of & CMB only & CMB + WiggleZ 			& + $H_0$						& + BAO		& + $H_0$+ BAO & + SN-Ia  & + LSS \\
& Parameters & & & & & & \\  \hline
Best fit $-\log(\mathcal{L})$ values \\
Flat $\Lambda$CDM & 7 &  3734.1  & 4140.9  &  4141.8 & 4141.6  &  4142.6 & 4353.2  &  4164.4  \\
Flat $w$CDM  & 8 &  3734.0  & 4140.3  &  4141.6 & 4141.6  &  4141.9 & 4352.8  &  $\cdots$  \\
$\Lambda$CDM  & 8&  3733.9  & 4140.6  &  4141.5 & 4141.5  &  4142.6 & 4353.2  & $\cdots$  \\
$w$CDM & 9 &  3734.2  & 4136.5  &  4141.4 & 4141.4  &  4142.5 & 4351.4  &  $\cdots$ \\
Flat $\Lambda$CDM + $\sum m_\nu$ & 8 &  3734.0  & 4140.8  &  4141.9 & 4141.5  &  4142.6 & $\cdots$  &  4164.0  \\
Flat $\Lambda$CDM + $n_{\rm run}$ & 8&  3733.6  & 4139.6  &  4141.1 & 4140.5  &  4141.8 & $\cdots$  &  4162.9  \\
Flat $\Lambda$CDM + $r$ & 8 &  3734.0  & 4140.9  &  4141.9 & 4141.7  &  4142.7 & $\cdots$  &  4164.4  \\ \hline
\multicolumn{3}{l}{$\log(\mathcal{L})$ differences (to Flat $\Lambda$CDM)} &  \\
Flat $\Lambda$CDM   & 7&  0.00  & 0.00  &  0.00 & 0.00  &  0.00 & 0.0  &  0.0  \\
Flat $w$CDM   & 8&  -0.10  & -0.62  &  0.01 & -0.04  &  -0.67 & -0.48  &  $\cdots$ \\
$\Lambda$CDM  & 8&  -0.16  & -0.33  &  0.11 & -0.07  &  0.05 & -0.04  &  $\cdots$  \\
$w$CDM    & 9&  0.15  & -4.41  &  0.14 & -0.20  &  -0.03 & -1.86  &  $\cdots$ \\
Flat $\Lambda$CDM + $\sum m_\nu$   & 8&  -0.06  & -0.11  &  0.12 & -0.16  &  0.06 & $\cdots$  &  -0.34  \\
Flat $\Lambda$CDM + $n_{\rm run}$  & 8  &  -0.48  & -1.30  &  -0.75 & -1.06  &  -0.74 & $\cdots$  &  -1.43  \\
Flat $\Lambda$CDM + $r$  & 8 &  -0.05  & 0.04  &  0.07 & 0.05  &  0.10 &$\cdots$  &  0.05  \\ \hline
\end{tabular}
\caption{\label{table:chisquarevalues}Best fit $-\log(\mathcal{L})$ values for the different cosmological models and dataset combinations, and the differences between those best fit $-\log(\mathcal{L})$ values and the best fit Flat $\Lambda$CDM $-\log(\mathcal{L})$. For the second half of the table, a negative value represents a better fit of the data to the model, and a positive value a worse fit. }
\end{table*}

\section{Conclusions}
\label{sec:conclusions}

We have compared the measurement of the galaxy power spectrum from the  WiggleZ survey to predictions for the underlying matter power spectrum to conduct a likelihood analysis of a number of cosmological models. We have modelled the non-linear structure formation and observational effects that go into making these predictions, choosing between a number of different alternative models from the literature and testing against numerical simulations specifically created to match the galaxies measured by the WiggleZ survey. We found that the best performing model was one similar to that used for analysis of the SDSS LRG sample \citep{Reid:2010}, but where the damping of the BAO peak is removed.

We applied this non-linear approach to measurements of the WiggleZ power spectrum, and used data from the four redshift bins and seven regions to constrain the cosmological parameters. We performed the analysis using the CosmoMC analysis package, with an additional module to compute the likelihood of the WiggleZ $P(k)$. We make the module and necessary data products available with this paper.

The results from a number of different possible cosmological models are as follows:

{\it $\Lambda$CDM model:} In combination with the CMB, the WiggleZ data improves constraints on the matter density ($\Omega_m = 0.290 \pm 0.016$) and amplitude of fluctuations ($\sigma_8 = 0.825 \pm 0.017$).

{\it Dark energy and Curvature:} The CMB + WiggleZ data combination alone is not powerful at constraining the dark energy equation of state ($w=-0.53\pm0.29$) or the curvature parameter ($\Omega_k =  -0.051 \pm 0.028 $. This is because introducing an extra dynamical degree of freedom creates a degeneracy with the matter density,  $\Omega_m$, reducing the constraint on this extra parameter. Distance measurements from BAO and SN-Ia provide an independent and complimentary measurement that removes this degeneracy, reducing the errors to $\Omega_k =   -0.001 \pm 0.005 $ on the curvature and $w =  -1.008 \pm 0.085 $ on the equation of state (using CMB + WiggleZ + BAO + $H_0$ data).

{\it Massive neutrinos:} The CMB + WiggleZ data provides a powerful constraint on the neutrino mass, giving an upper limit on the neutrino mass (summed over the three species) of 0.58eV (95\% confidence). However this result is still somewhat degenerate with the matter density. Other distance measurements reduce this degeneracy, giving our best upper limit of 0.32 eV (CMB + WiggleZ + $H_0$).

{\it Primordial Power spectrum and gravitational waves}: The CMB + WiggleZ data does not provide any better constraint on the spectral index than is provided by the WMAP data by itself. However, by constraining the matter density, we can get better constraints on parameters such as the amplitude of fluctuations, $\sigma_8$. By combining with other large-scale structure data, we can also improve measurements on the running of the spectral index $n_{\rm run} =  -0.033 \pm 0.018 $ (CMB + WiggleZ + LSS) or the tensor to scalar amplitude ratio $r<0.18$ (CMB + WiggleZ + LSS).

To summarise, all our results were consistent with the standard flat $\Lambda$CDM cosmological model, though we have significantly improved constraints on a number of parameters.

We see that many of the models created degeneracies with $\Omega_m$ that the CMB + WiggleZ data combination cannot break, giving multiple or extended likelihood peaks, and best fit parameter values with high matter densities and low values of $H_0$. This is because the power spectrum cannot distinguish models with an extra degree of freedom if $\Omega_m$ takes a larger value (for example, curvature, massive neutrinos and running of spectral index). These high matter density peaks can be removed if extra data is added, but could also be falsified if the power spectrum were able to accurate measure the turnover at large scales. The WiggleZ data has been used to determine the turnover position and so measure the matter density $\Omega_m$ in a separate publication \citep{Poole2012}.

The analysis presented in this paper has demonstrated the importance of modelling non-linear structure formation, bias, and redshift space distortions. The importance of this modelling will only increase as future galaxy surveys probe the matter power spectrum at higher precision and down to smaller scales.
 

\section*{Acknowledgments}

DP thanks Antony Lewis, Andrew Liddle and Beth Reid for helpful discussions. We acknowledge financial support from the Australian Research Council through Discovery Project grants DP0772084 and DP1093738, funding the positions of SB, DP, MP, GP and TMD, and Linkage International travel grant LX0881951. DC and SC acknowledge the support of an Australian Research Council through QEII Fellowships. MJD and TMD thank the Gregg Thompson Dark Energy Travel Fund for financial support. 

GALEX (the Galaxy Evolution Explorer) is a NASA Small Explorer, launched in 2003. We gratefully acknowledge NASAs support for construction, operation and science analysis for the GALEX mission, developed in cooperation with the Centre National d'Etudes Spatiales of France and the Korean Ministry of Science and Technology. 

Finally, we thank the Anglo-Australian Telescope (AAT) Allocation Committee for supporting the WiggleZ survey over nine semesters, and we are very grateful for the dedicated work of the staff of the Australian Astronomical Observatory in the development and support of the AAOmega spectrograph, and the running of the AAT.



 
 \bibliography{references} 

\begin{thebibliography}{59}
\expandafter\ifx\csname natexlab\endcsname\relax\def\natexlab#1{#1}\fi
\expandafter\ifx\csname bibnamefont\endcsname\relax
  \def\bibnamefont#1{#1}\fi
\expandafter\ifx\csname bibfnamefont\endcsname\relax
  \def\bibfnamefont#1{#1}\fi
\expandafter\ifx\csname citenamefont\endcsname\relax
  \def\citenamefont#1{#1}\fi
\expandafter\ifx\csname url\endcsname\relax
  \def\url#1{\texttt{#1}}\fi
\expandafter\ifx\csname urlprefix\endcsname\relax\def\urlprefix{URL }\fi
\providecommand{\bibinfo}[2]{#2}
\providecommand{\eprint}[2][]{\url{#2}}

\bibitem[{\citenamefont{{Huchra} et~al.}(1983)\citenamefont{{Huchra}, {Davis},
  {Latham}, and {Tonry}}}]{Huchra:1983}
\bibinfo{author}{\bibfnamefont{J.}~\bibnamefont{{Huchra}}},
  \bibinfo{author}{\bibfnamefont{M.}~\bibnamefont{{Davis}}},
  \bibinfo{author}{\bibfnamefont{D.}~\bibnamefont{{Latham}}}, \bibnamefont{and}
  \bibinfo{author}{\bibfnamefont{J.}~\bibnamefont{{Tonry}}},
  \bibinfo{journal}{\apjs} \textbf{\bibinfo{volume}{52}}, \bibinfo{pages}{89}
  (\bibinfo{year}{1983}).

\bibitem[{\citenamefont{{Maddox} et~al.}(1990)\citenamefont{{Maddox},
  {Efstathiou}, {Sutherland}, and {Loveday}}}]{Maddox:1990}
\bibinfo{author}{\bibfnamefont{S.~J.} \bibnamefont{{Maddox}}},
  \bibinfo{author}{\bibfnamefont{G.}~\bibnamefont{{Efstathiou}}},
  \bibinfo{author}{\bibfnamefont{W.~J.} \bibnamefont{{Sutherland}}},
  \bibnamefont{and}
  \bibinfo{author}{\bibfnamefont{J.}~\bibnamefont{{Loveday}}},
  \bibinfo{journal}{\mnras} \textbf{\bibinfo{volume}{242}},
  \bibinfo{pages}{43P} (\bibinfo{year}{1990}).

\bibitem[{\citenamefont{{Le F{\`e}vre} et~al.}(2005)\citenamefont{{Le
  F{\`e}vre}, {Vettolani}, {Garilli}, {Tresse}, {Bottini}, {Le Brun},
  {Maccagni}, {Picat}, {Scaramella}, {Scodeggio} et~al.}}]{lefevre:2005}
\bibinfo{author}{\bibfnamefont{O.}~\bibnamefont{{Le F{\`e}vre}}},
  \bibinfo{author}{\bibfnamefont{G.}~\bibnamefont{{Vettolani}}},
  \bibinfo{author}{\bibfnamefont{B.}~\bibnamefont{{Garilli}}},
  \bibinfo{author}{\bibfnamefont{L.}~\bibnamefont{{Tresse}}},
  \bibinfo{author}{\bibfnamefont{D.}~\bibnamefont{{Bottini}}},
  \bibinfo{author}{\bibfnamefont{V.}~\bibnamefont{{Le Brun}}},
  \bibinfo{author}{\bibfnamefont{D.}~\bibnamefont{{Maccagni}}},
  \bibinfo{author}{\bibfnamefont{J.~P.} \bibnamefont{{Picat}}},
  \bibinfo{author}{\bibfnamefont{R.}~\bibnamefont{{Scaramella}}},
  \bibinfo{author}{\bibfnamefont{M.}~\bibnamefont{{Scodeggio}}},
  \bibnamefont{et~al.}, \bibinfo{journal}{\aa} \textbf{\bibinfo{volume}{439}},
  \bibinfo{pages}{845} (\bibinfo{year}{2005}), \eprint{arXiv:astro-ph/0409133}.

\bibitem[{\citenamefont{{Colless} et~al.}(2001)\citenamefont{{Colless},
  {Dalton}, {Maddox}, {Sutherland}, {Norberg}, {Cole}, {Bland-Hawthorn},
  {Bridges}, {Cannon}, {Collins} et~al.}}]{colless:2001}
\bibinfo{author}{\bibfnamefont{M.}~\bibnamefont{{Colless}}},
  \bibinfo{author}{\bibfnamefont{G.}~\bibnamefont{{Dalton}}},
  \bibinfo{author}{\bibfnamefont{S.}~\bibnamefont{{Maddox}}},
  \bibinfo{author}{\bibfnamefont{W.}~\bibnamefont{{Sutherland}}},
  \bibinfo{author}{\bibfnamefont{P.}~\bibnamefont{{Norberg}}},
  \bibinfo{author}{\bibfnamefont{S.}~\bibnamefont{{Cole}}},
  \bibinfo{author}{\bibfnamefont{J.}~\bibnamefont{{Bland-Hawthorn}}},
  \bibinfo{author}{\bibfnamefont{T.}~\bibnamefont{{Bridges}}},
  \bibinfo{author}{\bibfnamefont{R.}~\bibnamefont{{Cannon}}},
  \bibinfo{author}{\bibfnamefont{C.}~\bibnamefont{{Collins}}},
  \bibnamefont{et~al.}, \bibinfo{journal}{\mnras}
  \textbf{\bibinfo{volume}{328}}, \bibinfo{pages}{1039} (\bibinfo{year}{2001}),
  \eprint{arXiv:astro-ph/0106498}.

\bibitem[{\citenamefont{{York} et~al.}(2000)\citenamefont{{York}, {Adelman},
  {Anderson}, {Anderson}, {Annis}, {Bahcall}, {Bakken}, {Barkhouser},
  {Bastian}, {Berman} et~al.}}]{York:2000}
\bibinfo{author}{\bibfnamefont{D.~G.} \bibnamefont{{York}}},
  \bibinfo{author}{\bibfnamefont{J.}~\bibnamefont{{Adelman}}},
  \bibinfo{author}{\bibfnamefont{J.~E.} \bibnamefont{{Anderson}},
  \bibfnamefont{Jr.}}, \bibinfo{author}{\bibfnamefont{S.~F.}
  \bibnamefont{{Anderson}}},
  \bibinfo{author}{\bibfnamefont{J.}~\bibnamefont{{Annis}}},
  \bibinfo{author}{\bibfnamefont{N.~A.} \bibnamefont{{Bahcall}}},
  \bibinfo{author}{\bibfnamefont{J.~A.} \bibnamefont{{Bakken}}},
  \bibinfo{author}{\bibfnamefont{R.}~\bibnamefont{{Barkhouser}}},
  \bibinfo{author}{\bibfnamefont{S.}~\bibnamefont{{Bastian}}},
  \bibinfo{author}{\bibfnamefont{E.}~\bibnamefont{{Berman}}},
  \bibnamefont{et~al.}, \bibinfo{journal}{\aj} \textbf{\bibinfo{volume}{120}},
  \bibinfo{pages}{1579} (\bibinfo{year}{2000}),
  \eprint{arXiv:astro-ph/0006396}.

\bibitem[{\citenamefont{{Drinkwater} et~al.}(2010)\citenamefont{{Drinkwater},
  {Jurek}, {Blake}, {Woods}, {Pimbblet}, {Glazebrook}, {Sharp}, {Pracy},
  {Brough}, {Colless} et~al.}}]{Drinkwater:2010}
\bibinfo{author}{\bibfnamefont{M.~J.} \bibnamefont{{Drinkwater}}},
  \bibinfo{author}{\bibfnamefont{R.~J.} \bibnamefont{{Jurek}}},
  \bibinfo{author}{\bibfnamefont{C.}~\bibnamefont{{Blake}}},
  \bibinfo{author}{\bibfnamefont{D.}~\bibnamefont{{Woods}}},
  \bibinfo{author}{\bibfnamefont{K.~A.} \bibnamefont{{Pimbblet}}},
  \bibinfo{author}{\bibfnamefont{K.}~\bibnamefont{{Glazebrook}}},
  \bibinfo{author}{\bibfnamefont{R.}~\bibnamefont{{Sharp}}},
  \bibinfo{author}{\bibfnamefont{M.~B.} \bibnamefont{{Pracy}}},
  \bibinfo{author}{\bibfnamefont{S.}~\bibnamefont{{Brough}}},
  \bibinfo{author}{\bibfnamefont{M.}~\bibnamefont{{Colless}}},
  \bibnamefont{et~al.}, \bibinfo{journal}{\mnras}
  \textbf{\bibinfo{volume}{401}}, \bibinfo{pages}{1429} (\bibinfo{year}{2010}).

\bibitem[{\citenamefont{{Blake}
  et~al.}(2011{\natexlab{a}})\citenamefont{{Blake}, {Davis}, {Poole},
  {Parkinson}, {Brough}, {Colless}, {Contreras}, {Couch}, {Croom}, {Drinkwater}
  et~al.}}]{blakedavis11}
\bibinfo{author}{\bibfnamefont{C.}~\bibnamefont{{Blake}}},
  \bibinfo{author}{\bibfnamefont{T.}~\bibnamefont{{Davis}}},
  \bibinfo{author}{\bibfnamefont{G.~B.} \bibnamefont{{Poole}}},
  \bibinfo{author}{\bibfnamefont{D.}~\bibnamefont{{Parkinson}}},
  \bibinfo{author}{\bibfnamefont{S.}~\bibnamefont{{Brough}}},
  \bibinfo{author}{\bibfnamefont{M.}~\bibnamefont{{Colless}}},
  \bibinfo{author}{\bibfnamefont{C.}~\bibnamefont{{Contreras}}},
  \bibinfo{author}{\bibfnamefont{W.}~\bibnamefont{{Couch}}},
  \bibinfo{author}{\bibfnamefont{S.}~\bibnamefont{{Croom}}},
  \bibinfo{author}{\bibfnamefont{M.~J.} \bibnamefont{{Drinkwater}}},
  \bibnamefont{et~al.}, \bibinfo{journal}{\mnras}
  \textbf{\bibinfo{volume}{415}}, \bibinfo{pages}{2892}
  (\bibinfo{year}{2011}{\natexlab{a}}), \eprint{1105.2862}.

\bibitem[{\citenamefont{{Blake}
  et~al.}(2011{\natexlab{b}})\citenamefont{{Blake}, {Kazin}, {Beutler},
  {Davis}, {Parkinson}, {Brough}, {Colless}, {Contreras}, {Couch}, {Croom}
  et~al.}}]{blakekazin11}
\bibinfo{author}{\bibfnamefont{C.}~\bibnamefont{{Blake}}},
  \bibinfo{author}{\bibfnamefont{E.~A.} \bibnamefont{{Kazin}}},
  \bibinfo{author}{\bibfnamefont{F.}~\bibnamefont{{Beutler}}},
  \bibinfo{author}{\bibfnamefont{T.~M.} \bibnamefont{{Davis}}},
  \bibinfo{author}{\bibfnamefont{D.}~\bibnamefont{{Parkinson}}},
  \bibinfo{author}{\bibfnamefont{S.}~\bibnamefont{{Brough}}},
  \bibinfo{author}{\bibfnamefont{M.}~\bibnamefont{{Colless}}},
  \bibinfo{author}{\bibfnamefont{C.}~\bibnamefont{{Contreras}}},
  \bibinfo{author}{\bibfnamefont{W.}~\bibnamefont{{Couch}}},
  \bibinfo{author}{\bibfnamefont{S.}~\bibnamefont{{Croom}}},
  \bibnamefont{et~al.}, \bibinfo{journal}{\mnras}
  \textbf{\bibinfo{volume}{418}}, \bibinfo{pages}{1707}
  (\bibinfo{year}{2011}{\natexlab{b}}), \eprint{1108.2635}.

\bibitem[{\citenamefont{{Blake}
  et~al.}(2011{\natexlab{c}})\citenamefont{{Blake}, {Brough}, {Colless},
  {Contreras}, {Couch}, {Croom}, {Davis}, {Drinkwater}, {Forster}, {Gilbank}
  et~al.}}]{Blake:2011growth}
\bibinfo{author}{\bibfnamefont{C.}~\bibnamefont{{Blake}}},
  \bibinfo{author}{\bibfnamefont{S.}~\bibnamefont{{Brough}}},
  \bibinfo{author}{\bibfnamefont{M.}~\bibnamefont{{Colless}}},
  \bibinfo{author}{\bibfnamefont{C.}~\bibnamefont{{Contreras}}},
  \bibinfo{author}{\bibfnamefont{W.}~\bibnamefont{{Couch}}},
  \bibinfo{author}{\bibfnamefont{S.}~\bibnamefont{{Croom}}},
  \bibinfo{author}{\bibfnamefont{T.}~\bibnamefont{{Davis}}},
  \bibinfo{author}{\bibfnamefont{M.~J.} \bibnamefont{{Drinkwater}}},
  \bibinfo{author}{\bibfnamefont{K.}~\bibnamefont{{Forster}}},
  \bibinfo{author}{\bibfnamefont{D.}~\bibnamefont{{Gilbank}}},
  \bibnamefont{et~al.}, \bibinfo{journal}{\mnras}
  \textbf{\bibinfo{volume}{415}}, \bibinfo{pages}{2876}
  (\bibinfo{year}{2011}{\natexlab{c}}), \eprint{1104.2948}.

\bibitem[{\citenamefont{{Blake}
  et~al.}(2011{\natexlab{d}})\citenamefont{{Blake}, {Glazebrook}, {Davis},
  {Brough}, {Colless}, {Contreras}, {Couch}, {Croom}, {Drinkwater}, {Forster}
  et~al.}}]{Blake:2011ap}
\bibinfo{author}{\bibfnamefont{C.}~\bibnamefont{{Blake}}},
  \bibinfo{author}{\bibfnamefont{K.}~\bibnamefont{{Glazebrook}}},
  \bibinfo{author}{\bibfnamefont{T.~M.} \bibnamefont{{Davis}}},
  \bibinfo{author}{\bibfnamefont{S.}~\bibnamefont{{Brough}}},
  \bibinfo{author}{\bibfnamefont{M.}~\bibnamefont{{Colless}}},
  \bibinfo{author}{\bibfnamefont{C.}~\bibnamefont{{Contreras}}},
  \bibinfo{author}{\bibfnamefont{W.}~\bibnamefont{{Couch}}},
  \bibinfo{author}{\bibfnamefont{S.}~\bibnamefont{{Croom}}},
  \bibinfo{author}{\bibfnamefont{M.~J.} \bibnamefont{{Drinkwater}}},
  \bibinfo{author}{\bibfnamefont{K.}~\bibnamefont{{Forster}}},
  \bibnamefont{et~al.}, \bibinfo{journal}{\mnras}
  \textbf{\bibinfo{volume}{418}}, \bibinfo{pages}{1725}
  (\bibinfo{year}{2011}{\natexlab{d}}), \eprint{1108.2637}.

\bibitem[{\citenamefont{{Riemer--S{\o}rensen}
  et~al.}(2012)\citenamefont{{Riemer--S{\o}rensen}, {Blake}, {Parkinson},
  {Davis}, {Brough}, {Colless}, {Contreras}, {Couch}, {Croom}, {Croton}
  et~al.}}]{riemersorensen:2011}
\bibinfo{author}{\bibfnamefont{S.}~\bibnamefont{{Riemer--S{\o}rensen}}},
  \bibinfo{author}{\bibfnamefont{C.}~\bibnamefont{{Blake}}},
  \bibinfo{author}{\bibfnamefont{D.}~\bibnamefont{{Parkinson}}},
  \bibinfo{author}{\bibfnamefont{T.~M.} \bibnamefont{{Davis}}},
  \bibinfo{author}{\bibfnamefont{S.}~\bibnamefont{{Brough}}},
  \bibinfo{author}{\bibfnamefont{M.}~\bibnamefont{{Colless}}},
  \bibinfo{author}{\bibfnamefont{C.}~\bibnamefont{{Contreras}}},
  \bibinfo{author}{\bibfnamefont{W.}~\bibnamefont{{Couch}}},
  \bibinfo{author}{\bibfnamefont{S.}~\bibnamefont{{Croom}}},
  \bibinfo{author}{\bibfnamefont{D.}~\bibnamefont{{Croton}}},
  \bibnamefont{et~al.}, \bibinfo{journal}{\prd} \textbf{\bibinfo{volume}{85}},
  \bibinfo{eid}{081101} (\bibinfo{year}{2012}).

\bibitem[{\citenamefont{{Jennings} et~al.}(2011)\citenamefont{{Jennings},
  {Baugh}, and {Pascoli}}}]{Jennings:2010}
\bibinfo{author}{\bibfnamefont{E.}~\bibnamefont{{Jennings}}},
  \bibinfo{author}{\bibfnamefont{C.~M.} \bibnamefont{{Baugh}}},
  \bibnamefont{and}
  \bibinfo{author}{\bibfnamefont{S.}~\bibnamefont{{Pascoli}}},
  \bibinfo{journal}{\mnras} \textbf{\bibinfo{volume}{410}},
  \bibinfo{pages}{2081} (\bibinfo{year}{2011}).

\bibitem[{\citenamefont{{Bird} et~al.}(2012)\citenamefont{{Bird}, {Viel}, and
  {Haehnelt}}}]{Bird:2011}
\bibinfo{author}{\bibfnamefont{S.}~\bibnamefont{{Bird}}},
  \bibinfo{author}{\bibfnamefont{M.}~\bibnamefont{{Viel}}}, \bibnamefont{and}
  \bibinfo{author}{\bibfnamefont{M.~G.} \bibnamefont{{Haehnelt}}},
  \bibinfo{journal}{\mnras} \textbf{\bibinfo{volume}{420}},
  \bibinfo{pages}{2551} (\bibinfo{year}{2012}), \eprint{1109.4416}.

\bibitem[{\citenamefont{{Marulli} et~al.}(2011)\citenamefont{{Marulli},
  {Carbone}, {Viel}, {Moscardini}, and {Cimatti}}}]{Marulli:2011}
\bibinfo{author}{\bibfnamefont{F.}~\bibnamefont{{Marulli}}},
  \bibinfo{author}{\bibfnamefont{C.}~\bibnamefont{{Carbone}}},
  \bibinfo{author}{\bibfnamefont{M.}~\bibnamefont{{Viel}}},
  \bibinfo{author}{\bibfnamefont{L.}~\bibnamefont{{Moscardini}}},
  \bibnamefont{and}
  \bibinfo{author}{\bibfnamefont{A.}~\bibnamefont{{Cimatti}}},
  \bibinfo{journal}{\mnras} \textbf{\bibinfo{volume}{418}},
  \bibinfo{pages}{346} (\bibinfo{year}{2011}), \eprint{1103.0278}.

\bibitem[{\citenamefont{{Smith} et~al.}(2003)\citenamefont{{Smith}, {Peacock},
  {Jenkins}, {White}, {Frenk}, {Pearce}, {Thomas}, {Efstathiou}, and
  {Couchman}}}]{Smith:2003}
\bibinfo{author}{\bibfnamefont{R.~E.} \bibnamefont{{Smith}}},
  \bibinfo{author}{\bibfnamefont{J.~A.} \bibnamefont{{Peacock}}},
  \bibinfo{author}{\bibfnamefont{A.}~\bibnamefont{{Jenkins}}},
  \bibinfo{author}{\bibfnamefont{S.~D.~M.} \bibnamefont{{White}}},
  \bibinfo{author}{\bibfnamefont{C.~S.} \bibnamefont{{Frenk}}},
  \bibinfo{author}{\bibfnamefont{F.~R.} \bibnamefont{{Pearce}}},
  \bibinfo{author}{\bibfnamefont{P.~A.} \bibnamefont{{Thomas}}},
  \bibinfo{author}{\bibfnamefont{G.}~\bibnamefont{{Efstathiou}}},
  \bibnamefont{and} \bibinfo{author}{\bibfnamefont{H.~M.~P.}
  \bibnamefont{{Couchman}}}, \bibinfo{journal}{\mnras}
  \textbf{\bibinfo{volume}{341}}, \bibinfo{pages}{1311} (\bibinfo{year}{2003}).

\bibitem[{\citenamefont{{Kaiser}}(1984)}]{Kaiser:1984}
\bibinfo{author}{\bibfnamefont{N.}~\bibnamefont{{Kaiser}}},
  \bibinfo{journal}{\apjl} \textbf{\bibinfo{volume}{284}}, \bibinfo{pages}{L9}
  (\bibinfo{year}{1984}).

\bibitem[{\citenamefont{{Lewis} and {Bridle}}(2002)}]{Lewis:2002}
\bibinfo{author}{\bibfnamefont{A.}~\bibnamefont{{Lewis}}} \bibnamefont{and}
  \bibinfo{author}{\bibfnamefont{S.}~\bibnamefont{{Bridle}}},
  \bibinfo{journal}{\prd} \textbf{\bibinfo{volume}{66}},
  \bibinfo{pages}{103511} (\bibinfo{year}{2002}).

\bibitem[{\citenamefont{{Gilbank} et~al.}(2011)\citenamefont{{Gilbank},
  {Gladders}, {Yee}, and {Hsieh}}}]{Gilbank:2011}
\bibinfo{author}{\bibfnamefont{D.~G.} \bibnamefont{{Gilbank}}},
  \bibinfo{author}{\bibfnamefont{M.~D.} \bibnamefont{{Gladders}}},
  \bibinfo{author}{\bibfnamefont{H.~K.~C.} \bibnamefont{{Yee}}},
  \bibnamefont{and} \bibinfo{author}{\bibfnamefont{B.~C.}
  \bibnamefont{{Hsieh}}}, \bibinfo{journal}{\aj}
  \textbf{\bibinfo{volume}{141}}, \bibinfo{eid}{94} (\bibinfo{year}{2011}),
  \eprint{1012.3470}.

\bibitem[{\citenamefont{{Sharp} et~al.}(2006)\citenamefont{{Sharp}, {Saunders},
  {Smith}, {Churilov}, {Correll}, {Dawson}, {Farrel}, {Frost}, {Haynes},
  {Heald} et~al.}}]{Sharp:2006}
\bibinfo{author}{\bibfnamefont{R.}~\bibnamefont{{Sharp}}},
  \bibinfo{author}{\bibfnamefont{W.}~\bibnamefont{{Saunders}}},
  \bibinfo{author}{\bibfnamefont{G.}~\bibnamefont{{Smith}}},
  \bibinfo{author}{\bibfnamefont{V.}~\bibnamefont{{Churilov}}},
  \bibinfo{author}{\bibfnamefont{D.}~\bibnamefont{{Correll}}},
  \bibinfo{author}{\bibfnamefont{J.}~\bibnamefont{{Dawson}}},
  \bibinfo{author}{\bibfnamefont{T.}~\bibnamefont{{Farrel}}},
  \bibinfo{author}{\bibfnamefont{G.}~\bibnamefont{{Frost}}},
  \bibinfo{author}{\bibfnamefont{R.}~\bibnamefont{{Haynes}}},
  \bibinfo{author}{\bibfnamefont{R.}~\bibnamefont{{Heald}}},
  \bibnamefont{et~al.}, in \emph{\bibinfo{booktitle}{Society of Photo-Optical
  Instrumentation Engineers (SPIE) Conference Series}} (\bibinfo{year}{2006}),
  vol. \bibinfo{volume}{6269} of \emph{\bibinfo{series}{Society of
  Photo-Optical Instrumentation Engineers (SPIE) Conference Series}},
  \eprint{arXiv:astro-ph/0606137}.

\bibitem[{\citenamefont{{Feldman} et~al.}(1994)\citenamefont{{Feldman},
  {Kaiser}, and {Peacock}}}]{Feldman:1994}
\bibinfo{author}{\bibfnamefont{H.~A.} \bibnamefont{{Feldman}}},
  \bibinfo{author}{\bibfnamefont{N.}~\bibnamefont{{Kaiser}}}, \bibnamefont{and}
  \bibinfo{author}{\bibfnamefont{J.~A.} \bibnamefont{{Peacock}}},
  \bibinfo{journal}{\apj} \textbf{\bibinfo{volume}{426}}, \bibinfo{pages}{23}
  (\bibinfo{year}{1994}).

\bibitem[{\citenamefont{{Blake} et~al.}(2010)\citenamefont{{Blake}, {Brough},
  {Colless}, {Couch}, {Croom}, {Davis}, {Drinkwater}, {Forster}, {Glazebrook},
  {Jelliffe} et~al.}}]{Blake:2010a}
\bibinfo{author}{\bibfnamefont{C.}~\bibnamefont{{Blake}}},
  \bibinfo{author}{\bibfnamefont{S.}~\bibnamefont{{Brough}}},
  \bibinfo{author}{\bibfnamefont{M.}~\bibnamefont{{Colless}}},
  \bibinfo{author}{\bibfnamefont{W.}~\bibnamefont{{Couch}}},
  \bibinfo{author}{\bibfnamefont{S.}~\bibnamefont{{Croom}}},
  \bibinfo{author}{\bibfnamefont{T.}~\bibnamefont{{Davis}}},
  \bibinfo{author}{\bibfnamefont{M.~J.} \bibnamefont{{Drinkwater}}},
  \bibinfo{author}{\bibfnamefont{K.}~\bibnamefont{{Forster}}},
  \bibinfo{author}{\bibfnamefont{K.}~\bibnamefont{{Glazebrook}}},
  \bibinfo{author}{\bibfnamefont{B.}~\bibnamefont{{Jelliffe}}},
  \bibnamefont{et~al.}, \bibinfo{journal}{\mnras}
  \textbf{\bibinfo{volume}{406}}, \bibinfo{pages}{803} (\bibinfo{year}{2010}).

\bibitem[{\citenamefont{{Poole} and {The WiggleZ
  Collaboration}}(2012)}]{Poole:2012}
\bibinfo{author}{\bibfnamefont{G.~B.} \bibnamefont{{Poole}}} \bibnamefont{and}
  \bibinfo{author}{\bibnamefont{{The WiggleZ Collaboration}}},
  \bibinfo{journal}{In preparation}  (\bibinfo{year}{2012}).

\bibitem[{\citenamefont{{Springel} et~al.}(2001)\citenamefont{{Springel},
  {White}, {Tormen}, and {Kauffmann}}}]{Springel:2001}
\bibinfo{author}{\bibfnamefont{V.}~\bibnamefont{{Springel}}},
  \bibinfo{author}{\bibfnamefont{S.~D.~M.} \bibnamefont{{White}}},
  \bibinfo{author}{\bibfnamefont{G.}~\bibnamefont{{Tormen}}}, \bibnamefont{and}
  \bibinfo{author}{\bibfnamefont{G.}~\bibnamefont{{Kauffmann}}},
  \bibinfo{journal}{\mnras} \textbf{\bibinfo{volume}{328}},
  \bibinfo{pages}{726} (\bibinfo{year}{2001}), \eprint{arXiv:astro-ph/0012055}.

\bibitem[{\citenamefont{{Tegmark} et~al.}(2006)\citenamefont{{Tegmark},
  {Eisenstein}, {Strauss}, {Weinberg}, {Blanton}, {Frieman}, {Fukugita},
  {Gunn}, {Hamilton}, {Knapp} et~al.}}]{Tegmark:2006}
\bibinfo{author}{\bibfnamefont{M.}~\bibnamefont{{Tegmark}}},
  \bibinfo{author}{\bibfnamefont{D.~J.} \bibnamefont{{Eisenstein}}},
  \bibinfo{author}{\bibfnamefont{M.~A.} \bibnamefont{{Strauss}}},
  \bibinfo{author}{\bibfnamefont{D.~H.} \bibnamefont{{Weinberg}}},
  \bibinfo{author}{\bibfnamefont{M.~R.} \bibnamefont{{Blanton}}},
  \bibinfo{author}{\bibfnamefont{J.~A.} \bibnamefont{{Frieman}}},
  \bibinfo{author}{\bibfnamefont{M.}~\bibnamefont{{Fukugita}}},
  \bibinfo{author}{\bibfnamefont{J.~E.} \bibnamefont{{Gunn}}},
  \bibinfo{author}{\bibfnamefont{A.~J.~S.} \bibnamefont{{Hamilton}}},
  \bibinfo{author}{\bibfnamefont{G.~R.} \bibnamefont{{Knapp}}},
  \bibnamefont{et~al.}, \bibinfo{journal}{\prd} \textbf{\bibinfo{volume}{74}},
  \bibinfo{pages}{123507} (\bibinfo{year}{2006}).

\bibitem[{\citenamefont{{Reid} et~al.}(2010{\natexlab{a}})\citenamefont{{Reid},
  {Percival}, {Eisenstein}, {Verde}, {Spergel}, {Skibba}, {Bahcall},
  {Budavari}, {Frieman}, {Fukugita} et~al.}}]{Reid:2009}
\bibinfo{author}{\bibfnamefont{B.~A.} \bibnamefont{{Reid}}},
  \bibinfo{author}{\bibfnamefont{W.~J.} \bibnamefont{{Percival}}},
  \bibinfo{author}{\bibfnamefont{D.~J.} \bibnamefont{{Eisenstein}}},
  \bibinfo{author}{\bibfnamefont{L.}~\bibnamefont{{Verde}}},
  \bibinfo{author}{\bibfnamefont{D.~N.} \bibnamefont{{Spergel}}},
  \bibinfo{author}{\bibfnamefont{R.~A.} \bibnamefont{{Skibba}}},
  \bibinfo{author}{\bibfnamefont{N.~A.} \bibnamefont{{Bahcall}}},
  \bibinfo{author}{\bibfnamefont{T.}~\bibnamefont{{Budavari}}},
  \bibinfo{author}{\bibfnamefont{J.~A.} \bibnamefont{{Frieman}}},
  \bibinfo{author}{\bibfnamefont{M.}~\bibnamefont{{Fukugita}}},
  \bibnamefont{et~al.}, \bibinfo{journal}{\mnras}
  \textbf{\bibinfo{volume}{404}}, \bibinfo{pages}{60}
  (\bibinfo{year}{2010}{\natexlab{a}}).

\bibitem[{\citenamefont{{Swanson} et~al.}(2010)\citenamefont{{Swanson},
  {Percival}, and {Lahav}}}]{Swanson:2010}
\bibinfo{author}{\bibfnamefont{M.~E.~C.} \bibnamefont{{Swanson}}},
  \bibinfo{author}{\bibfnamefont{W.~J.} \bibnamefont{{Percival}}},
  \bibnamefont{and} \bibinfo{author}{\bibfnamefont{O.}~\bibnamefont{{Lahav}}},
  \bibinfo{journal}{\mnras} \textbf{\bibinfo{volume}{409}},
  \bibinfo{pages}{1100} (\bibinfo{year}{2010}).

\bibitem[{\citenamefont{{Percival} et~al.}(2010)\citenamefont{{Percival},
  {Reid}, {Eisenstein}, {Bahcall}, {Budavari}, {Frieman}, {Fukugita}, {Gunn},
  {Ivezi{\'c}}, {Knapp} et~al.}}]{Percival:2009}
\bibinfo{author}{\bibfnamefont{W.~J.} \bibnamefont{{Percival}}},
  \bibinfo{author}{\bibfnamefont{B.~A.} \bibnamefont{{Reid}}},
  \bibinfo{author}{\bibfnamefont{D.~J.} \bibnamefont{{Eisenstein}}},
  \bibinfo{author}{\bibfnamefont{N.~A.} \bibnamefont{{Bahcall}}},
  \bibinfo{author}{\bibfnamefont{T.}~\bibnamefont{{Budavari}}},
  \bibinfo{author}{\bibfnamefont{J.~A.} \bibnamefont{{Frieman}}},
  \bibinfo{author}{\bibfnamefont{M.}~\bibnamefont{{Fukugita}}},
  \bibinfo{author}{\bibfnamefont{J.~E.} \bibnamefont{{Gunn}}},
  \bibinfo{author}{\bibfnamefont{{\v Z}.}~\bibnamefont{{Ivezi{\'c}}}},
  \bibinfo{author}{\bibfnamefont{G.~R.} \bibnamefont{{Knapp}}},
  \bibnamefont{et~al.}, \bibinfo{journal}{\mnras}
  \textbf{\bibinfo{volume}{401}}, \bibinfo{pages}{2148} (\bibinfo{year}{2010}).

\bibitem[{\citenamefont{{Reid} et~al.}(2010{\natexlab{b}})\citenamefont{{Reid},
  {Verde}, {Jimenez}, and {Mena}}}]{Reid:2010}
\bibinfo{author}{\bibfnamefont{B.~A.} \bibnamefont{{Reid}}},
  \bibinfo{author}{\bibfnamefont{L.}~\bibnamefont{{Verde}}},
  \bibinfo{author}{\bibfnamefont{R.}~\bibnamefont{{Jimenez}}},
  \bibnamefont{and} \bibinfo{author}{\bibfnamefont{O.}~\bibnamefont{{Mena}}},
  \bibinfo{journal}{\jcap} \textbf{\bibinfo{volume}{2010}},
  \bibinfo{pages}{003} (\bibinfo{year}{2010}{\natexlab{b}}).

\bibitem[{\citenamefont{{Peacock} and {Dodds}}(1994)}]{Peacock:1994}
\bibinfo{author}{\bibfnamefont{J.~A.} \bibnamefont{{Peacock}}}
  \bibnamefont{and} \bibinfo{author}{\bibfnamefont{S.~J.}
  \bibnamefont{{Dodds}}}, \bibinfo{journal}{\mnras}
  \textbf{\bibinfo{volume}{267}}, \bibinfo{pages}{1020} (\bibinfo{year}{1994}).

\bibitem[{\citenamefont{{Scoccimarro}}(2004)}]{Scoccimarro:2004}
\bibinfo{author}{\bibfnamefont{R.}~\bibnamefont{{Scoccimarro}}},
  \bibinfo{journal}{\prd} \textbf{\bibinfo{volume}{70}},
  \bibinfo{pages}{083007} (\bibinfo{year}{2004}).

\bibitem[{\citenamefont{{Komatsu}
  et~al.}(2011{\natexlab{a}})\citenamefont{{Komatsu}, {Smith}, {Dunkley},
  {Bennett}, {Gold}, {Hinshaw}, {Jarosik}, {Larson}, {Nolta}, {Page}
  et~al.}}]{Komatsu:2011}
\bibinfo{author}{\bibfnamefont{E.}~\bibnamefont{{Komatsu}}},
  \bibinfo{author}{\bibfnamefont{K.~M.} \bibnamefont{{Smith}}},
  \bibinfo{author}{\bibfnamefont{J.}~\bibnamefont{{Dunkley}}},
  \bibinfo{author}{\bibfnamefont{C.~L.} \bibnamefont{{Bennett}}},
  \bibinfo{author}{\bibfnamefont{B.}~\bibnamefont{{Gold}}},
  \bibinfo{author}{\bibfnamefont{G.}~\bibnamefont{{Hinshaw}}},
  \bibinfo{author}{\bibfnamefont{N.}~\bibnamefont{{Jarosik}}},
  \bibinfo{author}{\bibfnamefont{D.}~\bibnamefont{{Larson}}},
  \bibinfo{author}{\bibfnamefont{M.~R.} \bibnamefont{{Nolta}}},
  \bibinfo{author}{\bibfnamefont{L.}~\bibnamefont{{Page}}},
  \bibnamefont{et~al.}, \bibinfo{journal}{\apjs}
  \textbf{\bibinfo{volume}{192}}, \bibinfo{eid}{18}
  (\bibinfo{year}{2011}{\natexlab{a}}), \eprint{1001.4538}.

\bibitem[{\citenamefont{{Percival} et~al.}(2007)\citenamefont{{Percival},
  {Cole}, {Eisenstein}, {Nichol}, {Peacock}, {Pope}, and
  {Szalay}}}]{Percival:2007}
\bibinfo{author}{\bibfnamefont{W.~J.} \bibnamefont{{Percival}}},
  \bibinfo{author}{\bibfnamefont{S.}~\bibnamefont{{Cole}}},
  \bibinfo{author}{\bibfnamefont{D.~J.} \bibnamefont{{Eisenstein}}},
  \bibinfo{author}{\bibfnamefont{R.~C.} \bibnamefont{{Nichol}}},
  \bibinfo{author}{\bibfnamefont{J.~A.} \bibnamefont{{Peacock}}},
  \bibinfo{author}{\bibfnamefont{A.~C.} \bibnamefont{{Pope}}},
  \bibnamefont{and} \bibinfo{author}{\bibfnamefont{A.~S.}
  \bibnamefont{{Szalay}}}, \bibinfo{journal}{\mnras}
  \textbf{\bibinfo{volume}{381}}, \bibinfo{pages}{1053} (\bibinfo{year}{2007}),
  \eprint{0705.3323}.

\bibitem[{\citenamefont{{Guy} et~al.}(2010)\citenamefont{{Guy}, {Sullivan},
  {Conley}, {Regnault}, {Astier}, {Balland}, {Basa}, {Carlberg}, {Fouchez},
  {Hardin} et~al.}}]{Guy:2010}
\bibinfo{author}{\bibfnamefont{J.}~\bibnamefont{{Guy}}},
  \bibinfo{author}{\bibfnamefont{M.}~\bibnamefont{{Sullivan}}},
  \bibinfo{author}{\bibfnamefont{A.}~\bibnamefont{{Conley}}},
  \bibinfo{author}{\bibfnamefont{N.}~\bibnamefont{{Regnault}}},
  \bibinfo{author}{\bibfnamefont{P.}~\bibnamefont{{Astier}}},
  \bibinfo{author}{\bibfnamefont{C.}~\bibnamefont{{Balland}}},
  \bibinfo{author}{\bibfnamefont{S.}~\bibnamefont{{Basa}}},
  \bibinfo{author}{\bibfnamefont{R.~G.} \bibnamefont{{Carlberg}}},
  \bibinfo{author}{\bibfnamefont{D.}~\bibnamefont{{Fouchez}}},
  \bibinfo{author}{\bibfnamefont{D.}~\bibnamefont{{Hardin}}},
  \bibnamefont{et~al.}, \bibinfo{journal}{Astronomy \& Astrophysics}
  \textbf{\bibinfo{volume}{523}}, \bibinfo{eid}{A7} (\bibinfo{year}{2010}),
  \eprint{1010.4743}.

\bibitem[{\citenamefont{{Conley} et~al.}(2011)\citenamefont{{Conley}, {Guy},
  {Sullivan}, {Regnault}, {Astier}, {Balland}, {Basa}, {Carlberg}, {Fouchez},
  {Hardin} et~al.}}]{Conley:2011}
\bibinfo{author}{\bibfnamefont{A.}~\bibnamefont{{Conley}}},
  \bibinfo{author}{\bibfnamefont{J.}~\bibnamefont{{Guy}}},
  \bibinfo{author}{\bibfnamefont{M.}~\bibnamefont{{Sullivan}}},
  \bibinfo{author}{\bibfnamefont{N.}~\bibnamefont{{Regnault}}},
  \bibinfo{author}{\bibfnamefont{P.}~\bibnamefont{{Astier}}},
  \bibinfo{author}{\bibfnamefont{C.}~\bibnamefont{{Balland}}},
  \bibinfo{author}{\bibfnamefont{S.}~\bibnamefont{{Basa}}},
  \bibinfo{author}{\bibfnamefont{R.~G.} \bibnamefont{{Carlberg}}},
  \bibinfo{author}{\bibfnamefont{D.}~\bibnamefont{{Fouchez}}},
  \bibinfo{author}{\bibfnamefont{D.}~\bibnamefont{{Hardin}}},
  \bibnamefont{et~al.}, \bibinfo{journal}{\apjs}
  \textbf{\bibinfo{volume}{192}}, \bibinfo{eid}{1} (\bibinfo{year}{2011}),
  \eprint{1104.1443}.

\bibitem[{\citenamefont{{Sullivan} et~al.}(2011)\citenamefont{{Sullivan},
  {Guy}, {Conley}, {Regnault}, {Astier}, {Balland}, {Basa}, {Carlberg},
  {Fouchez}, {Hardin} et~al.}}]{Sullivan:2011}
\bibinfo{author}{\bibfnamefont{M.}~\bibnamefont{{Sullivan}}},
  \bibinfo{author}{\bibfnamefont{J.}~\bibnamefont{{Guy}}},
  \bibinfo{author}{\bibfnamefont{A.}~\bibnamefont{{Conley}}},
  \bibinfo{author}{\bibfnamefont{N.}~\bibnamefont{{Regnault}}},
  \bibinfo{author}{\bibfnamefont{P.}~\bibnamefont{{Astier}}},
  \bibinfo{author}{\bibfnamefont{C.}~\bibnamefont{{Balland}}},
  \bibinfo{author}{\bibfnamefont{S.}~\bibnamefont{{Basa}}},
  \bibinfo{author}{\bibfnamefont{R.~G.} \bibnamefont{{Carlberg}}},
  \bibinfo{author}{\bibfnamefont{D.}~\bibnamefont{{Fouchez}}},
  \bibinfo{author}{\bibfnamefont{D.}~\bibnamefont{{Hardin}}},
  \bibnamefont{et~al.}, \bibinfo{journal}{The Astrophysical Journal}
  \textbf{\bibinfo{volume}{737}}, \bibinfo{eid}{102} (\bibinfo{year}{2011}),
  \eprint{1104.1444}.

\bibitem[{\citenamefont{{Riess} et~al.}(2009)\citenamefont{{Riess}, {Macri},
  {Casertano}, {Sosey}, {Lampeitl}, {Ferguson}, {Filippenko}, {Jha}, {Li},
  {Chornock} et~al.}}]{Riess:2009}
\bibinfo{author}{\bibfnamefont{A.~G.} \bibnamefont{{Riess}}},
  \bibinfo{author}{\bibfnamefont{L.}~\bibnamefont{{Macri}}},
  \bibinfo{author}{\bibfnamefont{S.}~\bibnamefont{{Casertano}}},
  \bibinfo{author}{\bibfnamefont{M.}~\bibnamefont{{Sosey}}},
  \bibinfo{author}{\bibfnamefont{H.}~\bibnamefont{{Lampeitl}}},
  \bibinfo{author}{\bibfnamefont{H.~C.} \bibnamefont{{Ferguson}}},
  \bibinfo{author}{\bibfnamefont{A.~V.} \bibnamefont{{Filippenko}}},
  \bibinfo{author}{\bibfnamefont{S.~W.} \bibnamefont{{Jha}}},
  \bibinfo{author}{\bibfnamefont{W.}~\bibnamefont{{Li}}},
  \bibinfo{author}{\bibfnamefont{R.}~\bibnamefont{{Chornock}}},
  \bibnamefont{et~al.}, \bibinfo{journal}{\apj} \textbf{\bibinfo{volume}{699}},
  \bibinfo{pages}{539} (\bibinfo{year}{2009}).

\bibitem[{\citenamefont{{Beltr{\'a}n} et~al.}(2005)\citenamefont{{Beltr{\'a}n},
  {Garc{\'{\i}}a-Bellido}, {Lesgourgues}, {Liddle}, and
  {Slosar}}}]{Beltran:2005}
\bibinfo{author}{\bibfnamefont{M.}~\bibnamefont{{Beltr{\'a}n}}},
  \bibinfo{author}{\bibfnamefont{J.}~\bibnamefont{{Garc{\'{\i}}a-Bellido}}},
  \bibinfo{author}{\bibfnamefont{J.}~\bibnamefont{{Lesgourgues}}},
  \bibinfo{author}{\bibfnamefont{A.~R.} \bibnamefont{{Liddle}}},
  \bibnamefont{and} \bibinfo{author}{\bibfnamefont{A.}~\bibnamefont{{Slosar}}},
  \bibinfo{journal}{\prd} \textbf{\bibinfo{volume}{71}}, \bibinfo{eid}{063532}
  (\bibinfo{year}{2005}), \eprint{arXiv:astro-ph/0501477}.

\bibitem[{\citenamefont{{Bridges} et~al.}(2006)\citenamefont{{Bridges},
  {Lasenby}, and {Hobson}}}]{Bridges:2006}
\bibinfo{author}{\bibfnamefont{M.}~\bibnamefont{{Bridges}}},
  \bibinfo{author}{\bibfnamefont{A.~N.} \bibnamefont{{Lasenby}}},
  \bibnamefont{and} \bibinfo{author}{\bibfnamefont{M.~P.}
  \bibnamefont{{Hobson}}}, \bibinfo{journal}{\mnras}
  \textbf{\bibinfo{volume}{369}}, \bibinfo{pages}{1123} (\bibinfo{year}{2006}),
  \eprint{arXiv:astro-ph/0511573}.

\bibitem[{\citenamefont{{Mukherjee} et~al.}(2006)\citenamefont{{Mukherjee},
  {Parkinson}, and {Liddle}}}]{Mukherjee:2006}
\bibinfo{author}{\bibfnamefont{P.}~\bibnamefont{{Mukherjee}}},
  \bibinfo{author}{\bibfnamefont{D.}~\bibnamefont{{Parkinson}}},
  \bibnamefont{and} \bibinfo{author}{\bibfnamefont{A.~R.}
  \bibnamefont{{Liddle}}}, \bibinfo{journal}{\apjl}
  \textbf{\bibinfo{volume}{638}}, \bibinfo{pages}{L51} (\bibinfo{year}{2006}),
  \eprint{arXiv:astro-ph/0508461}.

\bibitem[{\citenamefont{{Parkinson} et~al.}(2006)\citenamefont{{Parkinson},
  {Mukherjee}, and {Liddle}}}]{Parkinson:2006}
\bibinfo{author}{\bibfnamefont{D.}~\bibnamefont{{Parkinson}}},
  \bibinfo{author}{\bibfnamefont{P.}~\bibnamefont{{Mukherjee}}},
  \bibnamefont{and} \bibinfo{author}{\bibfnamefont{A.~R.}
  \bibnamefont{{Liddle}}}, \bibinfo{journal}{\prd}
  \textbf{\bibinfo{volume}{73}}, \bibinfo{eid}{123523} (\bibinfo{year}{2006}),
  \eprint{arXiv:astro-ph/0605003}.

\bibitem[{\citenamefont{{Trotta}}(2007)}]{Trotta:2007}
\bibinfo{author}{\bibfnamefont{R.}~\bibnamefont{{Trotta}}},
  \bibinfo{journal}{\mnras} \textbf{\bibinfo{volume}{375}},
  \bibinfo{pages}{L26} (\bibinfo{year}{2007}), \eprint{arXiv:astro-ph/0608116}.

\bibitem[{\citenamefont{{V{\"a}liviita} and
  {Giannantonio}}(2009)}]{Valiviita:2009}
\bibinfo{author}{\bibfnamefont{J.}~\bibnamefont{{V{\"a}liviita}}}
  \bibnamefont{and}
  \bibinfo{author}{\bibfnamefont{T.}~\bibnamefont{{Giannantonio}}},
  \bibinfo{journal}{\prd} \textbf{\bibinfo{volume}{80}}, \bibinfo{eid}{123516}
  (\bibinfo{year}{2009}), \eprint{0909.5190}.

\bibitem[{\citenamefont{{Kilbinger} et~al.}(2010)\citenamefont{{Kilbinger},
  {Wraith}, {Robert}, {Benabed}, {Capp{\'e}}, {Cardoso}, {Fort}, {Prunet}, and
  {Bouchet}}}]{Kilbinger:2010}
\bibinfo{author}{\bibfnamefont{M.}~\bibnamefont{{Kilbinger}}},
  \bibinfo{author}{\bibfnamefont{D.}~\bibnamefont{{Wraith}}},
  \bibinfo{author}{\bibfnamefont{C.~P.} \bibnamefont{{Robert}}},
  \bibinfo{author}{\bibfnamefont{K.}~\bibnamefont{{Benabed}}},
  \bibinfo{author}{\bibfnamefont{O.}~\bibnamefont{{Capp{\'e}}}},
  \bibinfo{author}{\bibfnamefont{J.-F.} \bibnamefont{{Cardoso}}},
  \bibinfo{author}{\bibfnamefont{G.}~\bibnamefont{{Fort}}},
  \bibinfo{author}{\bibfnamefont{S.}~\bibnamefont{{Prunet}}}, \bibnamefont{and}
  \bibinfo{author}{\bibfnamefont{F.~R.} \bibnamefont{{Bouchet}}},
  \bibinfo{journal}{\mnras} \textbf{\bibinfo{volume}{405}},
  \bibinfo{pages}{2381} (\bibinfo{year}{2010}), \eprint{0912.1614}.

\bibitem[{\citenamefont{{Poole} et~al.}(2012)}]{Poole2012}
\bibinfo{author}{\bibfnamefont{G.~B.} \bibnamefont{{Poole}}}
  \bibnamefont{et~al.} (\bibinfo{year}{2012}).

\bibitem[{\citenamefont{{Clarkson} et~al.}(2007)\citenamefont{{Clarkson},
  {Cort{\^e}s}, and {Bassett}}}]{Clarkson2007}
\bibinfo{author}{\bibfnamefont{C.}~\bibnamefont{{Clarkson}}},
  \bibinfo{author}{\bibfnamefont{M.}~\bibnamefont{{Cort{\^e}s}}},
  \bibnamefont{and}
  \bibinfo{author}{\bibfnamefont{B.}~\bibnamefont{{Bassett}}},
  \bibinfo{journal}{\jcap} \textbf{\bibinfo{volume}{8}}, \bibinfo{pages}{11}
  (\bibinfo{year}{2007}), \eprint{arXiv:astro-ph/0702670}.

\bibitem[{\citenamefont{{Fukuda} et~al.}(1998)\citenamefont{{Fukuda},
  {Hayakawa}, {Ichihara}, {Inoue}, {Ishihara}, {Ishino}, {Itow}, {Kajita},
  {Kameda}, {Kasuga} et~al.}}]{Fukuda:1998}
\bibinfo{author}{\bibfnamefont{Y.}~\bibnamefont{{Fukuda}}},
  \bibinfo{author}{\bibfnamefont{T.}~\bibnamefont{{Hayakawa}}},
  \bibinfo{author}{\bibfnamefont{E.}~\bibnamefont{{Ichihara}}},
  \bibinfo{author}{\bibfnamefont{K.}~\bibnamefont{{Inoue}}},
  \bibinfo{author}{\bibfnamefont{K.}~\bibnamefont{{Ishihara}}},
  \bibinfo{author}{\bibfnamefont{H.}~\bibnamefont{{Ishino}}},
  \bibinfo{author}{\bibfnamefont{Y.}~\bibnamefont{{Itow}}},
  \bibinfo{author}{\bibfnamefont{T.}~\bibnamefont{{Kajita}}},
  \bibinfo{author}{\bibfnamefont{J.}~\bibnamefont{{Kameda}}},
  \bibinfo{author}{\bibfnamefont{S.}~\bibnamefont{{Kasuga}}},
  \bibnamefont{et~al.}, \bibinfo{journal}{\prl} \textbf{\bibinfo{volume}{81}},
  \bibinfo{pages}{1562} (\bibinfo{year}{1998}).

\bibitem[{\citenamefont{Amsler et~al.}(2008)}]{Amsler:2008}
\bibinfo{author}{\bibfnamefont{C.}~\bibnamefont{Amsler}} \bibnamefont{et~al.}
  (\bibinfo{collaboration}{Particle Data Group}), \bibinfo{journal}{Phys.
  Lett.} \textbf{\bibinfo{volume}{B667}}, \bibinfo{pages}{1}
  (\bibinfo{year}{2008}).

\bibitem[{\citenamefont{{Komatsu} et~al.}(2009)\citenamefont{{Komatsu},
  {Dunkley}, {Nolta}, {Bennett}, {Gold}, {Hinshaw}, {Jarosik}, {Larson},
  {Limon}, {Page} et~al.}}]{Komatsu:2009}
\bibinfo{author}{\bibfnamefont{E.}~\bibnamefont{{Komatsu}}},
  \bibinfo{author}{\bibfnamefont{J.}~\bibnamefont{{Dunkley}}},
  \bibinfo{author}{\bibfnamefont{M.~R.} \bibnamefont{{Nolta}}},
  \bibinfo{author}{\bibfnamefont{C.~L.} \bibnamefont{{Bennett}}},
  \bibinfo{author}{\bibfnamefont{B.}~\bibnamefont{{Gold}}},
  \bibinfo{author}{\bibfnamefont{G.}~\bibnamefont{{Hinshaw}}},
  \bibinfo{author}{\bibfnamefont{N.}~\bibnamefont{{Jarosik}}},
  \bibinfo{author}{\bibfnamefont{D.}~\bibnamefont{{Larson}}},
  \bibinfo{author}{\bibfnamefont{M.}~\bibnamefont{{Limon}}},
  \bibinfo{author}{\bibfnamefont{L.}~\bibnamefont{{Page}}},
  \bibnamefont{et~al.}, \bibinfo{journal}{\apjs}
  \textbf{\bibinfo{volume}{180}}, \bibinfo{pages}{330} (\bibinfo{year}{2009}),
  \eprint{0803.0547}.

\bibitem[{\citenamefont{{Komatsu}
  et~al.}(2011{\natexlab{b}})\citenamefont{{Komatsu}, {Smith}, {Dunkley},
  {Bennett}, {Gold}, {Hinshaw}, {Jarosik}, {Larson}, {Nolta}, L.
  et~al.}}]{Komatsu:2010}
\bibinfo{author}{\bibfnamefont{E.}~\bibnamefont{{Komatsu}}},
  \bibinfo{author}{\bibfnamefont{K.~M.} \bibnamefont{{Smith}}},
  \bibinfo{author}{\bibfnamefont{J.}~\bibnamefont{{Dunkley}}},
  \bibinfo{author}{\bibfnamefont{C.~L.} \bibnamefont{{Bennett}}},
  \bibinfo{author}{\bibfnamefont{B.}~\bibnamefont{{Gold}}},
  \bibinfo{author}{\bibfnamefont{G.}~\bibnamefont{{Hinshaw}}},
  \bibinfo{author}{\bibfnamefont{N.}~\bibnamefont{{Jarosik}}},
  \bibinfo{author}{\bibfnamefont{D.}~\bibnamefont{{Larson}}},
  \bibinfo{author}{\bibfnamefont{M.~R.} \bibnamefont{{Nolta}}},
  \bibinfo{author}{\bibfnamefont{P.}~\bibnamefont{L.}}, \bibnamefont{et~al.},
  \bibinfo{journal}{\apjs} \textbf{\bibinfo{volume}{192}}, \bibinfo{eid}{18}
  (\bibinfo{year}{2011}{\natexlab{b}}).

\bibitem[{\citenamefont{Dunkley et~al.}({2011})\citenamefont{Dunkley, Hlozek,
  Sievers, Acquaviva, Ade, Aguirre, Amiri, Appel, Barrientos, Battistelli
  et~al.}}]{Dunkley:2011}
\bibinfo{author}{\bibfnamefont{J.}~\bibnamefont{Dunkley}},
  \bibinfo{author}{\bibfnamefont{R.}~\bibnamefont{Hlozek}},
  \bibinfo{author}{\bibfnamefont{J.}~\bibnamefont{Sievers}},
  \bibinfo{author}{\bibfnamefont{V.}~\bibnamefont{Acquaviva}},
  \bibinfo{author}{\bibfnamefont{P.~A.~R.} \bibnamefont{Ade}},
  \bibinfo{author}{\bibfnamefont{P.}~\bibnamefont{Aguirre}},
  \bibinfo{author}{\bibfnamefont{M.}~\bibnamefont{Amiri}},
  \bibinfo{author}{\bibfnamefont{J.~W.} \bibnamefont{Appel}},
  \bibinfo{author}{\bibfnamefont{L.~F.} \bibnamefont{Barrientos}},
  \bibinfo{author}{\bibfnamefont{E.~S.} \bibnamefont{Battistelli}},
  \bibnamefont{et~al.}, \bibinfo{journal}{{\apj}}
  \textbf{\bibinfo{volume}{{739}}} (\bibinfo{year}{{2011}}), ISSN
  \bibinfo{issn}{{0004-637X}}.

\bibitem[{\citenamefont{{Kosowsky} and {Turner}}(1995)}]{KosowskyTurner95}
\bibinfo{author}{\bibfnamefont{A.}~\bibnamefont{{Kosowsky}}} \bibnamefont{and}
  \bibinfo{author}{\bibfnamefont{M.~S.} \bibnamefont{{Turner}}},
  \bibinfo{journal}{\prd} \textbf{\bibinfo{volume}{52}}, \bibinfo{pages}{1739}
  (\bibinfo{year}{1995}), \eprint{arXiv:astro-ph/9504071}.

\bibitem[{\citenamefont{{Dias} et~al.}(2012)\citenamefont{{Dias}, {Frazer}, and
  {Liddle}}}]{Dias2012}
\bibinfo{author}{\bibfnamefont{M.}~\bibnamefont{{Dias}}},
  \bibinfo{author}{\bibfnamefont{J.}~\bibnamefont{{Frazer}}}, \bibnamefont{and}
  \bibinfo{author}{\bibfnamefont{A.~R.} \bibnamefont{{Liddle}}},
  \bibinfo{journal}{ArXiv e-prints}  (\bibinfo{year}{2012}),
  \eprint{1203.3792}.

\bibitem[{\citenamefont{{Li}}(2006)}]{Li2006}
\bibinfo{author}{\bibfnamefont{M.}~\bibnamefont{{Li}}},
  \bibinfo{journal}{\jcap} \textbf{\bibinfo{volume}{10}}, \bibinfo{pages}{3}
  (\bibinfo{year}{2006}), \eprint{arXiv:astro-ph/0607525}.

\bibitem[{\citenamefont{{Linde}}(2005)}]{Linde2005}
\bibinfo{author}{\bibfnamefont{A.}~\bibnamefont{{Linde}}},
  \bibinfo{journal}{ArXiv High Energy Physics - Theory e-prints}
  (\bibinfo{year}{2005}), \eprint{arXiv:hep-th/0503203}.

\bibitem[{\citenamefont{{Mukhanov} et~al.}(1992)\citenamefont{{Mukhanov},
  {Feldman}, and {Brandenberger}}}]{MFB1992}
\bibinfo{author}{\bibfnamefont{V.~F.} \bibnamefont{{Mukhanov}}},
  \bibinfo{author}{\bibfnamefont{H.~A.} \bibnamefont{{Feldman}}},
  \bibnamefont{and} \bibinfo{author}{\bibfnamefont{R.~H.}
  \bibnamefont{{Brandenberger}}}, \bibinfo{journal}{\physrep}
  \textbf{\bibinfo{volume}{215}}, \bibinfo{pages}{203} (\bibinfo{year}{1992}).

\bibitem[{\citenamefont{{Liddle} and {Lyth}}(2000)}]{LiddleLyth}
\bibinfo{author}{\bibfnamefont{A.~R.} \bibnamefont{{Liddle}}} \bibnamefont{and}
  \bibinfo{author}{\bibfnamefont{D.~H.} \bibnamefont{{Lyth}}},
  \emph{\bibinfo{title}{{Cosmological Inflation and Large-Scale Structure}}}
  (\bibinfo{publisher}{Cambridge University Press}, \bibinfo{year}{2000}).

\bibitem[{\citenamefont{{Bassett} et~al.}(2006)\citenamefont{{Bassett},
  {Tsujikawa}, and {Wands}}}]{Bassett2006}
\bibinfo{author}{\bibfnamefont{B.~A.} \bibnamefont{{Bassett}}},
  \bibinfo{author}{\bibfnamefont{S.}~\bibnamefont{{Tsujikawa}}},
  \bibnamefont{and} \bibinfo{author}{\bibfnamefont{D.}~\bibnamefont{{Wands}}},
  \bibinfo{journal}{Reviews of Modern Physics} \textbf{\bibinfo{volume}{78}},
  \bibinfo{pages}{537} (\bibinfo{year}{2006}), \eprint{arXiv:astro-ph/0507632}.

\bibitem[{\citenamefont{{Lyth} and {Liddle}}(2009)}]{LythLiddle}
\bibinfo{author}{\bibfnamefont{D.~H.} \bibnamefont{{Lyth}}} \bibnamefont{and}
  \bibinfo{author}{\bibfnamefont{A.~R.} \bibnamefont{{Liddle}}},
  \emph{\bibinfo{title}{{The Primordial Density Perturbation: Cosmology,
  Inflation and the Origin of Structure}}} (\bibinfo{publisher}{Cambridge
  University Press}, \bibinfo{year}{2009}).

\bibitem[{\citenamefont{Akaike}(1974)}]{Akaike:1974}
\bibinfo{author}{\bibfnamefont{H.}~\bibnamefont{Akaike}},
  \bibinfo{journal}{Automatic Control, IEEE Transactions on}
  \textbf{\bibinfo{volume}{19}}, \bibinfo{pages}{716 } (\bibinfo{year}{1974}),
  ISSN \bibinfo{issn}{0018-9286}.

\end{thebibliography}


\appendix

\section{Fitting fomulae for GiggleZ Power Spectum}
\label{sec:gigglezfitting}

We fit the simulated power spectra from GiggleZ using a fifth-order polynomial in $k$ (in units of $h^{-1}{\rm Mpc}$), for the logarithm of the power spectrum, $P(k)$ (in units of $h^{-3}~{\rm Mpc}^3$), in each of the four redshift bins. In order to generate mock WiggleZ catalogues from the GiggleZ simulations at a particular redshift, we selected the range of halo groupings with large-scale clustering bias closest to the WiggleZ sample under analysis, and applied the WiggleZ selection function to the mock dataset.

\begin{table}
\begin{tabular}{ccccccc} \hline
$z$ & $A_0$ &$A_1$ &$A_2$ &$A_3$ &$A_4$ & $A_5$ \\ \hline
0.22 & 4.619d0 & -13.7787 & 58.94 & -175.24 & 284.321 & -187.284 \\
 0.41 & 4.63079 & -12.6293 & 42.9265 & -91.8068 &  97.808 & -37.633 \\
 0.60 & 4.69659 & -12.7287 & 42.5681 & -89.5578 & 96.664 &  -41.2564 \\
 0.78 & 4.6849 & -13.4747 & 53.7172 & -145.832 & 216.638 & -132.782 \\\hline
\end{tabular}
\caption{\label{table:gigglezcoefficients} Coefficients for the polynomial fitting formula for the GiggleZ power spectrum.}

\end{table}

The resulting fitting formula is:
\begin{eqnarray}
\log_{10}(P(k)) &=& A_0 + A_1 k + A_2 k^2 \\ \nonumber
&& + A_3 k^3 + A_4 k^4 + A_5 k^5
\end{eqnarray}
with different coefficients for each of the four redshift bins. The coefficients are given in table \ref{table:gigglezcoefficients}, and built-in in the CosmoMC module.

\end{document}